\documentclass[12pt,english]{article}
\usepackage{avant}
\usepackage[T1]{fontenc}
\usepackage[latin9]{inputenc}
\usepackage{geometry}
\usepackage{babel}
\usepackage{amsmath}
\usepackage{amssymb}
\usepackage{stmaryrd}
\usepackage{graphicx}
\usepackage{esint}
\usepackage[authoryear]{natbib}
\usepackage[unicode=true,
 bookmarks=false,
 breaklinks=false,pdfborder={0 0 1},backref=false,colorlinks=false]
 {hyperref}
\usepackage{breakurl}

\makeatletter
\@ifundefined{date}{}{\date{}}
\usepackage{babel}
\usepackage{amsthm}
\usepackage{bm}\usepackage{graphicx}\usepackage{rotating}\usepackage{array}\usepackage{booktabs}\usepackage{color}\usepackage{algorithmic}
\usepackage{algorithm}\usepackage{multirow}\usepackage{enumerate}\usepackage{graphicx}\usepackage{caption}\usepackage{subcaption}

\topmargin by -\headheight \advance
\topmargin by -\headsep

\evensidemargin
\oddsidemargin
\floatstyle{ruled}
\newfloat{algorithm}{tbp}{loa}
\providecommand{\algorithmname}{Algorithm}
\floatname{algorithm}{\protect\algorithmname}

\@ifundefined{definecolor}{color}{}
\RequirePackage[displaymath]{lineno}

\@ifundefined{definecolor}{color}{}
\usepackage{amsfonts}\usepackage{latexsym}

\usepackage{mathrsfs}\usepackage{amsthm}\usepackage{latexsym}\usepackage{booktabs}\@ifundefined{definecolor}{color}{}

\usepackage{times}\@ifundefined{definecolor}{color}{}

\newtheorem{lemma}{Lemma}
\newtheorem{cy}{Corollary}

\newtheorem{theorem}{Theorem}
\newtheorem{definition}{Definition}

\newcommand{\T}{^{\mbox{\scriptsize {\sf T}}}} 

\DeclareMathOperator*{\argmax}{argmax}
\DeclareMathOperator*{\argmin}{argmin}

\usepackage{babel}

\makeatother

\begin{document}
{\setlength{\baselineskip}{1.0\baselineskip}

\global\long\def\mbA{\mathbf{A}}
\global\long\def\mbf{\mathbf{f}}
 \global\long\def\hatmbA{\widehat{\mathbf{A}}}
 \global\long\def\mbB{\mathbf{B}}
 \global\long\def\mbD{\mathbf{D}}
 \global\long\def\mbN{\mathbf{N}}
 \global\long\def\mbV{\mathbf{V}}
 \global\long\def\mbZ{\mathbf{Z}}
 \global\long\def\hatmbV{\widehat{\mathbf{V}}}
 \global\long\def\mbv{\mathbf{v}}
 \global\long\def\mbx{\mathbf{x}}
 \global\long\def\hatmbu{\widehat{\mathbf{u}}}
 \global\long\def\mbu{\mathbf{u}}
 \global\long\def\hatmbv{\widehat{\mathbf{v}}}
 \global\long\def\tilmbv{\widetilde{\mbv}}
 \global\long\def\mbX{\mathbf{X}}
 \global\long\def\mbY{\mathbf{Y}}
 \global\long\def\mbU{\mathbf{U}}
 \global\long\def\mbW{\mathbf{W}}
 \global\long\def\mbP{\mathbf{P}}
 \global\long\def\mbQ{\mathbf{Q}}
 \global\long\def\mbO{\mathbf{O}}
 \global\long\def\mbM{\mathbf{M}}
 \global\long\def\mbR{\mathbf{R}}
 \global\long\def\mbS{\mathbf{S}}
 \global\long\def\mbG{\mathbf{G}}
 \global\long\def\mbI{\mathbf{I}}
 \global\long\def\hatmbG{\widehat{\mbG}}
 \global\long\def\mbg{\mathbf{g}}
 \global\long\def\hatmbg{\widehat{\mbg}}
 \global\long\def\hatmbM{\widehat{\mbM}}
 \global\long\def\hatmbW{\widehat{\mbW}}
 \global\long\def\hatmbU{\widehat{\mbU}}
 \global\long\def\mbw{\mathbf{w}}
 \global\long\def\hatmbw{\widehat{\mbw}}
 \global\long\def\bolGamma{\boldsymbol{\Gamma}}
 \global\long\def\hatbolGamma{\widehat{\bolGamma}}
 \global\long\def\bolOmega{\boldsymbol{\Omega}}
 \global\long\def\bolLambda{\boldsymbol{\Lambda}}
 \global\long\def\hatbolLambda{\widehat{\boldsymbol{\Lambda}}}
 \global\long\def\bolPhi{\boldsymbol{\Phi}}
 \global\long\def\mbbR{\mathbb{R}}
 \global\long\def\mbbS{\mathbb{S}}
 \global\long\def\calE{\mathcal{E}}
 \global\long\def\calB{\mathcal{B}}
 \global\long\def\boltheta{\boldsymbol{\theta}}
 \global\long\def\boleta{\boldsymbol{\eta}}
 \global\long\def\bolphi{\boldsymbol{\phi}}
 \global\long\def\bolpsi{\boldsymbol{\psi}}
 \global\long\def\hatboltheta{\widehat{\boltheta}}
 \global\long\def\bolepsilon{\boldsymbol{\epsilon}}
 \global\long\def\bolbeta{\boldsymbol{\beta}}
 \global\long\def\hatbolbeta{\widehat{\bolbeta}}
 \global\long\def\bolgamma{\boldsymbol{\gamma}}
 \global\long\def\hatbolgamma{\widehat{\bolgamma}}
 \global\long\def\hatboleta{\widehat{\boleta}}
 \global\long\def\hatmbP{\widehat{\mbP}}
 \global\long\def\hatmbbP{\widehat{\mathbb{P}}}
 \global\long\def\hatbolpsi{\widehat{\bolpsi}}
 \global\long\def\hatlambda{\widehat{\lambda}}
 \global\long\def\hatbolOmega{\widehat{\bolOmega}}
 \global\long\def\bolvarepsilon{\boldsymbol{\varepsilon}}
 \global\long\def\bolalpha{\boldsymbol{\alpha}}
 \global\long\def\bolmu{\boldsymbol{\mu}}
 \global\long\def\bolTheta{\boldsymbol{\Theta}}
 \global\long\def\bolSigma{\boldsymbol{\Sigma}}
 \global\long\def\hatbolSigma{\widehat{\bolSigma}}
 \global\long\def\calR{{\cal R}}
 \global\long\def\calRp{\calR^{\perp}}
 \global\long\def\calG{\mathcal{G}}
 \global\long\def\calD{\mathcal{D}}
 \global\long\def\calU{\mathcal{U}}
 \global\long\def\hatphi{\widehat{\phi}}
 \global\long\def\hatcalE{\widehat{\calE}}
 \global\long\def\hatsigma{\widehat{\sigma}}
 \global\long\def\hatmbY{\widehat{\mathbf{Y}}}

\global\long\def\vecc{\mathrm{vec}}
 \global\long\def\Prob{\mathrm{Pr}}
 \global\long\def\E{\mathrm{E}}
 \global\long\def\Cov{\mathrm{cov}}
 \global\long\def\Corr{\mathrm{corr}}
 \global\long\def\F{\mathrm{F}}
 \global\long\def\J{\mathrm{J}}
 \global\long\def\I{\mathcal{I}_{n}}
 \global\long\def\II{\mathcal{I}_{n}^{\mathrm{1D}}}
 \global\long\def\Jn{\mathrm{J}_{n}}
 \global\long\def\Ln{\ell_{n}}
 \global\long\def\Var{\mathrm{var}}
 \global\long\def\dimension{\mathrm{dim}}
 \global\long\def\spn{\mathrm{span}}
 \global\long\def\vech{\mathrm{vech}}

\global\long\def\Env{\mathrm{Env}}
 \global\long\def\tr{\mathrm{trace}}
 \global\long\def\dg{\mathrm{diag}}
 \global\long\def\asyVar{\mathrm{avar}}

\global\long\def\MenvU{\calE_{\mbM}(\mbU)}
\global\long\def\hatMenvU{\widehat{\calE}_{\mbM}(\mbU)}
\newcommand{\ra}[1]{\renewcommand{\arraystretch}{#1}}

\title{Envelopes and principal component regression}
\author{
\bigskip
Xin Zhang\thanks{Correspondence: Xin Zhang (xzhang8@fsu.edu)}, Kai Deng, and Qing Mai \\\normalsize{\textit{Florida State University}}
}
\date{}
\maketitle

\begin{abstract}
Envelope methods offer targeted dimension reduction for various models. The overarching goal is to improve efficiency in multivariate parameter estimation by projecting the data onto a lower-dimensional subspace known as the envelope. Envelope approaches have advantages in analyzing data with highly correlated variables, but their iterative Grassmannian optimization algorithms do not scale very well with ultra high-dimensional data. 
While the connections between envelopes and partial least squares in multivariate linear regression have promoted recent progress in high-dimensional studies of envelopes, we propose a more straightforward way of envelope modeling from a novel principal components regression perspective.
The proposed procedure, Non-Iterative Envelope Component Estimation (NIECE), has excellent computational advantages over the iterative Grassmannian optimization alternatives in high dimensions.
We develop a unified NIECE theory that bridges the gap between envelope methods  and principal components in regression.
The new theoretical insights also shed light on the envelope subspace estimation error as a function of eigenvalue gaps of two symmetric positive definite matrices used in envelope modeling.
We apply the new theory and algorithm to several envelope models, including response and predictor reduction in multivariate linear models, logistic regression, and Cox proportional hazard model. Simulations and illustrative data analysis show the potential for NIECE to improve standard methods in linear and generalized linear models significantly.\\


\end{abstract}

\newpage

\section{Introduction}
	The idea of envelope modeling is to exploit the covariance structure in variables and identify and eliminate the part of data tangent to the parameter space of interest but only brings extraneous variability to model fitting.
	Firstly introduced in \citet{Cook2010envelope}, envelope methods have demonstrated promising performances in various multivariate statistical problems. Different envelope structures are proposed in the multivariate linear regression model, for example, partial envelope \citep{SuCook2011partial}, inner envelope \citep{su2012inner}, scaled envelope \citep{cook2013scaled}, and simultaneous envelope \citep{CookZhang2015simultaneous}.
Connections between envelopes and classical multivariate analysis methods are intensively studied: envelope and partial least squares \citep{Cook2013PLS}, envelope and reduced-rank regression \citep{CFZ2015RRE}, envelope and Bayesian statistics \citep{khare2017bayesian}, envelope and discriminant analysis \citep{zhang2019efficient}, and among others.	
Recent extensions of envelopes beyond the standard linear models is also an emerging field of research. \citet{CookZhang2015foundation} proposed a general constructive framework for adapting envelope methods to any estimation procedure, and applied this to generalized linear models and Cox regression. More recently, envelope methodology has been extended to quantile regression \citep{envelopeQR}, Huber regression \citep{zhou2020enveloped}, matrix-variate \citep{ding2018matrix} and tensor-variate regressions \citep{li2017parsimonious}. See \citet{cook2018introduction} for a detailed introduction, and \citet{cook2020envelope} and \citet{lee2020review} for recent reviews, on envelopes.

	In this paper, we will study two challenging and fundamental questions in high-dimensional envelopes: (i) How to overcome the computational bottleneck of envelope estimation in ultra high-dimensions; and (ii) How to quantify the envelope subspace estimation error in a generic model-free setting, especially when the dimension $p$ diverges much faster than the sample size $n$.
	To address the computational issue (i) and the theoretical issue (ii) of high-dimensional envelopes, we propose a scalable and straightforward envelope estimation procedure based on a novel principal components regression formulation. Computationally, the proposed novel procedure, Non-Iterative Envelope Component Estimation (NIECE), is a perfect complement to the more delicate and much more expensive iterative Grassmannian optimization approaches that were used in the literature (i.e.~all the envelope methods as mentioned earlier). 
	Theoretically, a unified theory for NIECE shows that we can estimate the envelope subspace consistently when the dimension diverges exponentially fast as the sample size in a wide range of models and applications. 
	To the best of our knowledge, this paper is also the first in (a) establishing the finite sample connections between envelope estimation and principal components; (b) providing non-asymptotic analysis of envelope subspace estimation error; (c) devising a novel ``eigenvalue gap'' argument for theoretical analysis on two positive semi-definite matrices $\mbM$ and $\mbU$ in a model-free setting of envelopes: For NIECE, we extract the eigenvectors $\mbv_i$ of $\mbM$ and examine the ``eigen-gaps'' of the quadratic form $\mbv_i\T\mbU\mbv_i$. 
		
	In what follows, we outline in Section \ref{sub:intro1pop} the population construct of an envelope $\MenvU$ following the same notation in \cite{CookZhang2015foundation} for a general-purpose multivariate parameter estimation. We provide a literature review on the computational aspects, algorithms, and existing high-dimensional theory of envelopes in Section \ref{sub:intro2est}. The population-level connections between envelopes and principal components are given in Section \ref{sub:intro3pca} to motivate our methodology. The specific goals and organization of this article are outlined in Section \ref{sub:intro4notation}.
		
	\subsection{Envelopes: definition and working mechanism}\label{sub:intro1pop}
	For a matrix $\mbB \in \mbbR^{p\times d}$ with full column rank $d$, let $\calB = \spn(\mbB) \subseteq\mbbR^{p}$ denote the subspace spanned by the columns of $\mbB$. Let $\mbP_{\calB} = \mbP_{\mbB} = \mbB(\mbB^{\T}\mbB)^{-1}\mbB^{\T}$ denote the projection onto $\calB$, and $\mbQ_{\calB} = \mbQ_{\mbB} = \mbI_p - \mbP_{\mbB}$ denote the projection onto the orthogonal complement of $\calB$, where $\mbI_{p}$ is the $p\times p$ identity matrix. Formally, an envelope is defined as follows \citep[see,][for example]{CookZhang2015foundation}, where the notion of reducing subspace is  from functional analysis \citep[e.g.]{Conway1990}. 
	\begin{definition}\label{defi:env}
	A reducing subspace of a symmetric matrix $\mbM \in \mbbR^{p\times p}$ is defined as the subspace $\calR \subseteq \mbbR^{p}$ such that $\mbM=\mbP_{\calR}\mbM\mbP_{\calR}+\mbQ_{\calR}\mbM\mbQ_{\calR}$. An $\mbM$-envelope of a subspace $\calU=\spn(\mbU) \subseteq \mbbR^{p}$, denoted as $\MenvU$ or $\calE_{\mbM}(\calU)$, is the intersection of all reducing subspaces of $\mbM$ that contain $\calU$. 
	\end{definition}

	 \begin{figure}[ht!]
	\centering
	\includegraphics[scale=0.66]{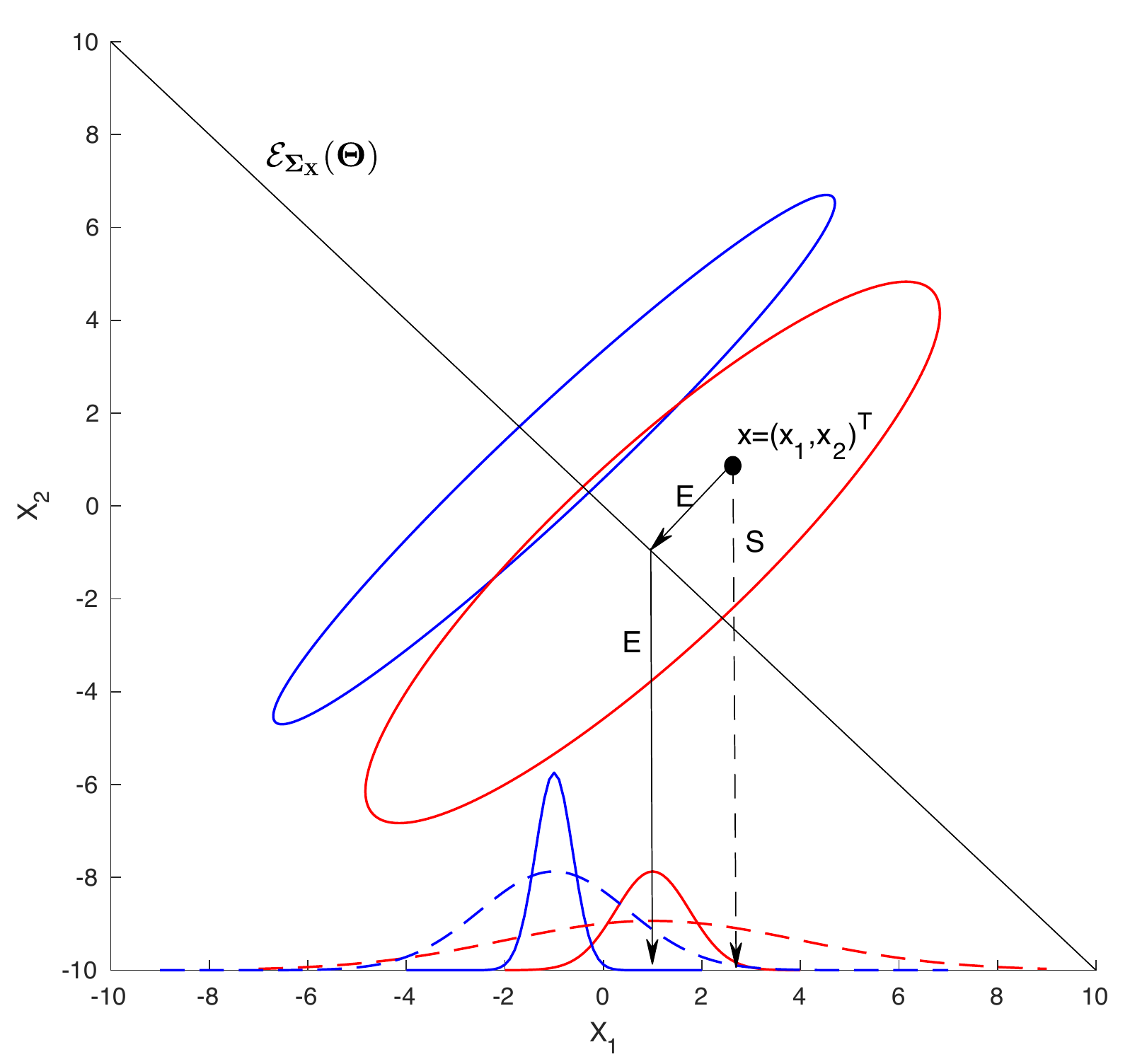}
	\caption{The working mechanism of envelope methodology when  estimating the mean difference of $\mbX$ across two groups (indicated by red and blue colors). The plot illustrates the efficiency gain by envelope methods in estimating the mean difference $\boltheta=\bolmu_2-\bolmu_1$; representative projection paths, labeled `E' for envelope analysis and `S' for standard analysis. }
	\label{fig: intuition}
\end{figure} 

The construction and estimation of envelope $\calE_{\mbM}(\mbU)$  is the center of our future development. \citet{Cook2010envelope} showed that the envelope $\MenvU$ is unique and always exists. In multivariate analysis, $\mbU$ comes from the  parameter of interest; and $\mbM$ represents a helpful nuisance parameter. The specific choices of $\mbM$ and $\mbU$ depend on the context of applications and the goals of studies. They may vary from one model to another, as we see later in analyzing several envelope models in later sections (including response and predictor envelopes in linear regression, envelopes in logistic regression and Cox model).

Figure~\ref{fig: intuition} illustrates the working mechanism of envelope methods. 
Consider a simple case with group indicator $G\in\{1,2\}$ and predictor $\mbX=(X_1,X_2)\T\in\mbbR^2$. The target parameter is the mean difference $\boltheta = \bolmu_2-\bolmu_1=\E(\mbX\mid G=2)-\E(\mbX\mid G=1)$ or equivalently the matrix form $\bolTheta = \boltheta\boltheta\T$ (i.e. the $\mbU$ matrix in Definition~\ref{defi:env}). The nuisance parameter is the marginal covariance $\bolSigma_{\mbX}=\Cov(\mbX)$ (i.e. the $\mbM$ matrix in Definition~\ref{defi:env}); and it helps improving the estimation efficiency in the mean difference as illustrated in Figure~\ref{fig: intuition}. 
The two ellipses represent the contours of the conditional distributions of $\mbX$ within each of the two classes. We can see that the mean difference lies in the shorter axis of the ellipses, while the long axis brings large variability but does not contribute to comparing the means. Therefore, if we calculate the difference in the sample means, the differences will also be blurred. For $X_1$, as shown in Figure~\ref{fig: intuition}, we can see that the two empirical distributions of $X_1\mid (G=1)$ and $X_1\mid(G=2)$, as the two lower flatten curves represented, are almost indistinguishable. 
In contrast, the envelope method can identify the ellipses' longer axis as \emph{immaterial variation}. The envelope estimation procedure will project all the data first onto the \emph{envelope}, which contains all the \emph{material variation}, and then onto each axis of $\mbX$ to compare the two classes. 
The elimination of immaterial variation leads to much well-separated two distributions as shown in Figure~\ref{fig: intuition}, and therefore massive gain in estimation accuracy of $\boltheta=\bolmu_2-\bolmu_1$. 
	
\subsection{A brief overview of envelope estimation}\label{sub:intro2est}
	Without loss of generality, we assume $\mbU$ is symmetric positive semi-definite henceforth because we can always replace $\mbU$ with $\mbU\mbU\T\geq0$ without changing the column subspace $\spn(\mbU)$ and the envelope $\MenvU$. The dimension of $\calE_{\mbM}(\mbU)$, denoted by $u$, $0\leq u\leq p$, is important for all envelope methods and can be estimated consistently \citep{EnvDim}. In this paper, we assume the envelope dimension $u\geq1$ is known or pre-specified.

	Given the dimension of an envelope, envelope estimation generally reduces to solving for $\hatbolGamma\in\mbbR^{p\times u}$ from the following constrained optimization and letting $\hatMenvU=\spn(\hatbolGamma)$,
	\begin{equation}
	\hatbolGamma=\argmin_{\bolGamma\T\bolGamma=\mbI_u}\J_n(\bolGamma),\quad \J_n(\bolGamma)=\log\vert\bolGamma\T\hatmbM\bolGamma\vert+\log\vert\bolGamma\T(\hatmbM+\hatmbU)^{-1}\bolGamma\vert,\label{obj_pop}
	\end{equation}
where the symmetric matrices $\hatmbM>0$ and $\hatmbU\geq 0$ are finite sample estimators of their population counterparts $\mbM$ and $\mbU$. From the orthogonality constraint $\bolGamma\T\bolGamma=\mbI_u$, we note that \eqref{obj_pop} is a non-convex optimization on Stiefel manifolds. Equivalently, if we consider the subspace as the argument in the above optimization, then \eqref{obj_pop} is also equivalent to a non-convex optimization on Grassmann manifolds (aka Grassmannian). 

Almost all envelope methods are connected to this type of optimization. 
For a broad class of envelope estimators, it was shown in \citet{CookZhang2015foundation} that the partially maximized log-likelihood function is $(-n/2)$ times certain sample version of \eqref{obj_pop}. Therefore, normal likelihood-based envelope methods are usually required to solve a version of \eqref{obj_pop}. Furthermore, given a parameter vector of interests, $\boltheta\in\mbbR^p$, and some standard $\sqrt{n}$-consistent estimator $\hatboltheta$, the particular choice of $\mbU=\boltheta\boltheta\T$ and $\mbM$ being the asymptotic covariance of $\hatboltheta$ reproduces these likelihood-based envelope methods in the literature. The envelope estimator $\hatboltheta_{\Env}=\hatbolGamma\hatbolGamma\T\hatboltheta$ is asymptotically more efficient than the standard estimator $\hatboltheta$ in various contexts such as linear and generalized linear models. Such envelope estimation solves for $\hatbolGamma$ based on \eqref{obj_pop} and then plugs-in $\hatboltheta_{\Env}=\hatbolGamma\hatbolGamma\T\hatboltheta$, is essentially a two-stage projection pursuit multivariate parameter estimation relying on this generic objective function of envelope basis. In this paper, we focus on the envelope subspace estimation problem and characterizing the estimation error in terms of the distance between the true and estimated envelope subspaces when $p$ is allowed to diverge much faster than $n$.

Most computational methods for manifold optimizations can be directly used to solve \eqref{obj_pop}, see \citet{edelman1998geometry,absil2009optimization} and \citet{wen2013feasible} for more background. By exploiting the geometry of envelopes, \citet{CookZhang2014algorithm} developed the 1D algorithm that approximately solves \eqref{obj_pop} by sequential optimization over $u$ one-dimensional vectors than over $p\times u$ matrices. Relatedly, \citet{cook2018fast} developed the envelope coordinate descent algorithm that sequentially solves the same 1D algorithm objective functions and is shown to be even faster than the original 1D algorithm. Another approach for solving \eqref{obj_pop} is proposed by \citet{cook2016note}. The authors suggest to remove the orthogonality constraint by rotating $\bolGamma$ to as $\bolGamma = (\mbI_u,\mbA\T)\T \mbG$ for some coordinates $\mbA\in\mbbR^{(p-u)\times u}$ and a full rank matrix $\mbG\in\mbbR^{u\times u}$. Then the iterative Grassmannian optimization is transformed into unconstrained iterative optimization over matrix $\mbA$. All such computational methods adopt gradient-based iterative schemes for a highly non-convex objective function, which involves log-determinant of symmetric positive definite matrices, and are quite sensitive to initialization when dimensions $(p,u)$ are high. 


Extending envelope methods to high dimensional data analysis is challenging, mainly due to the delicate and complicated optimization in \eqref{obj_pop} that provides no closed-form for $\hatbolGamma$ and thus makes the envelope estimator $\hatboltheta_{\Env}=\hatbolGamma\hatbolGamma\T\hatboltheta$ almost mysterious. 
Nevertheless, some progress on this problem has been made. \citet{Su2016} proposed a sparse envelope estimator for response reduction and variable selection; and \citet{zhu2020envelope} proposed a sparse envelope estimator for predictor reduction and variable selection. Both methods rely on a penalized version of \eqref{obj_pop} with sparsity inducing penalties on rows of $\bolGamma$. This type of coordinate-free penalty \citep{chen2010coordinate} takes the form of $P_\lambda(\bolGamma)=\lambda_i\Vert\bolgamma_i\Vert_2$, $i=1,\dots,p$, where $\bolgamma_i\in\mbbR^u$ is the $i$-th row of $\bolGamma$. Their approaches come at the cost of assuming that the dimension $p$ (i.e.~the total number of response or predictor variables) can not diverge too quickly: $(p+s)\log(p)/n\rightarrow0$, where $s$ is the sparsity level. Moreover, the $p$ tuning parameters are assumed to satisfy $\sqrt{n}\lambda_i\rightarrow0$ for the $s$ truly active variables and $\sqrt{n}\lambda_i\rightarrow\infty$ for the $(p-s)$ inactive variables.

To overcome the computational bottleneck of envelope estimation in the ultra high-dimensional regimes, i.e.~$\log(p)/n \rightarrow 0$, we need alternative objectives of envelope estimation. Indeed, we propose a sparse principal component approach to envelope estimation that does not involve the usual manifold optimizations. The new approach aims not to replace the existing methods when applicable but to provide a feasible approach when they are not. It also fills the gap in theoretical analysis and provides valuable insights about principal components in regression.

\subsection{Envelope and principal components }\label{sub:intro3pca}

Principal component analysis  \citep{jolliffe2002principal,jolliffe2016principal} is routinely used in exploratory data analysis as a dimension reduction and visualization method and in regression to improve prediction. In some cases, the PCs with the smallest variances (also known as the minor components, see \citet{oja1992principal} for example) are kept for subsequent analysis in addition to the PCs with the largest variances. Using latent variable and normal likelihood, \citet{tipping1999probabilistic} provided a model-based probabilistic formulation for principal component analysis where only the leading PCs are relevant in the analysis; \citet{welling2004extreme} generalized such approach to probabilistic models where the maximum likelihood estimation for the dimension reduction subspace is a combination of PCs with largest and smallest variances; \citet{zhang2020principal} discussed scenarios that the maximum likelihood is attained by any combinations of PCs. 

On principal component regression, \citet{jolliffe1982note} noted that the PCs with smaller variability could be as useful as PCs with the largest variability; more recently, \citet{langsimple} provided a response-guided formulation for regression with PCs that unifies the usual principal component regression with ridge regression \citep{hoerl1970ridge}.

Our approach is conceptually closely related to principal components regression but is formulated in more general settings (i.e.~beyond linear regression) with rigorous theoretical justification using envelopes. From Definition~\ref{defi:env}, the subspace spanned by any set of eigenvectors of $\mbM$ is a reducing subspace of $\mbM$. In this article, to ensure identifiability, we assume that the first $d\lesssim n$ eigenvalues of $\mbM$ are distinct and their span contains $\spn(\mbU)$. In practice, we will specify $d\simeq n$ in the proposed algorithm so that we lose little information by focusing on the first $d$ eigenvectors. When some of the eigenvalues coincide, we will be targeting an upper bound of the envelope $\MenvU$, similar to Proposition 4 in \citet{cook2018fast}. 

In many applications, $\mbM=\bolSigma_{\mbX}\equiv\Cov(\mbX)>0$. Then the population level connection between envelope $\MenvU$ and the principal components (PCs) of $\mbX$ can be easily shown as follows,
\begin{equation}\label{pop_pcaEnv}
\MenvU = \sum_{\substack{i=1,\\ \mbv_i\T\mbU\mbv_i\neq0}}^{d}\spn(\mbv_i),
\end{equation}
where $\mbv_i\in\mbbR^p$ is the $i$-th eigenvector of $\mbM=\bolSigma_{\mbX}$ and hence the $i$-th PC. Our algorithm is an efficient realization of the above characteristics of envelopes and is thus straightforward to implement based on the penalized matrix decomposition \citep[PMD]{witten2009pmd}. In our theoretical study, we first establish the consistency of PMD, which is of independent interest. It is also straightforward to adopt other sparse PC methods in our algorithm \citep{ZHT06,shen2008sparse,AW08,Ma13,vu2013minimax}.

Several recent studies have connected envelopes with PCs. In particular, \citet{li2016supervised} studied the supervised dimension reduction problem, where the envelope is formed as the leading PCs in a latent variable model and estimated by EM algorithm. The covariance can be written as $\bolSigma_{\mbX} = \mbV\bolSigma_{\mbf}\mbV\T + \sigma^2\mbI_p$, where $\bolSigma_{\mbf}\in\mbbR^{u\times u}$ is the covariance of the $u$ latent variables $\mbf$ and $\mbV\in\mbbR^{p\times u}$ is the basis of the $u$-dimensional target subspace. \citet{franks2019shared} studied a shared subspace spiked model that $\bolSigma_{\mbX\mid Y=k}=\sigma^2_k(\mbV\boldsymbol{\Psi}_k\mbV\T+\mbI_p)$ for  groups of observations indexed by $k=1,\dots,K$, where $\boldsymbol{\Psi}_k$ symmetric positive definite matrix and $\mbV\in\mbbR^{p\times u}$ is the basis of the $u$-dimensional target subspace. More recently, \citet{franks2020reducing} extended the model of \citet{li2016supervised} to large-$p$-small-$n$ setting by adopting the Monte Carlo EM algorithm similar to that in \citet{franks2019shared}. These models, as well as the common principal component analysis models \citep{flury1984,flury1988book,schott1999partial}, can be viewed as special forms of the more general envelope structure in this paper. However, none of these methods has established consistency in subspace estimation when $p$ diverges with $n$. 
Although the population level connections between the envelope and principal components, including \eqref{pop_pcaEnv}, are noticed by several recent studies, we provide for the first time a unified computational and theoretical approach for ultra high-dimensional envelope estimation.

It is also worth mentioning that envelope methods in multivariate linear regression are regarded as a likelihood-based alternative to the partial least squares (PLS) regression: \citet{Cook2013PLS} showed that the sequence of Krylov subspaces in PLS algorithm \citep{de1993simpls,Helland1990} converges in population to the same envelope subspace. 
This connection between envelopes and partial least squares has promoted recent progress in high-dimensional studies of envelopes and PLS \citep{cook2019partial,zhu2020envelope}. Our method can thus also be viewed as an extension of the sparse PLS methods \citep{chun2010sparse,chun2011identification} because it applies to a wide range of multivariate analysis problems beyond linear regression.

	\subsection{Notation and organization}\label{sub:intro4notation}

For any matrix $\mbM$, we will let $\lambda_{j}(\mbM)$ be the $j$-th
eigenvalue of $\mbM$, e.g.~$\lambda_{1}(\mbM)$ is the largest eigenvalue
of $\mbM$. We use the Frobenius norm,
operator norm, $\ell_{1}$ norm and the maximum norm of matrices.
For a matrix $\bolOmega\in\mathbb{R}^{p_{1}\times p_{2}}$, let $\omega_{ij}$ denote its $(i,j)$-th element, then its Frobenius
norm is $\Vert\bolOmega\Vert_{F}=\sqrt{\sum_{i,j}\omega_{ij}^{2}}$;
its operator norm $\Vert\bolOmega\Vert_{op}$ is its largest eigenvalue;
its $\ell_{1}$ norm is $\Vert\bolOmega\Vert_{1}=\max_{j}\sum_{i}|\omega_{ij}|$;
and its maximum norm is $\Vert\bolOmega\Vert_{\max}=\max_{i,j}|\omega_{ij}|$.

We will study the estimation error of envelopes, characterized by the distance between two subspaces $\calE=\spn(\bolGamma)$ and $\hatcalE=\spn(\hatbolGamma)$
For $\bolGamma,\hatbolGamma\in\mbbR^{p\times u}$ such that $\bolGamma\T\bolGamma=\mbI_{u}=\hatbolGamma\T\hatbolGamma$,
we let $\bolgamma_{i},\hatbolgamma_{i}\in\mbbR^{p}$, $i=1,\dots,u$,
be the $i$-th columns of $\bolGamma$ and $\hatbolGamma$, respectively.
Similar to \citet{YWS15}, we define the $u\times u$ diagonal matrix $\bolTheta(\hatbolGamma,\bolGamma)$
such that the $j$-th principal angle between the two subspace $\spn(\bolGamma)$
and $\spn(\hatbolGamma)$ is at the $j$-th diagonal element of $\bolTheta(\hatbolGamma,\bolGamma)$,
and let $\sin\bolTheta(\hatbolGamma,\bolGamma)\in\mbbR^{u\times u}$
be defined entrywise. The principal angles between the two subspace
$\spn(\bolGamma)$ and $\spn(\hatbolGamma)$ are denoted as $\theta_{j}$,
$j=1,\dots,u$, then $\cos(\theta_{j})=\sqrt{1-\sin^{2}(\theta_{j})}=\lambda_{j}(\bolGamma\T\hatbolGamma)$, which is the $j$-th eigenvalue of $\bolGamma\T\hatbolGamma$. 
In sufficient dimension reduction literature, distance between two subspace $\calE=\spn(\bolGamma)$ and $\hatcalE=\spn(\hatbolGamma)$
are also commonly characterized as $\Vert\mbP_{\widehat{\calE}}-\mbP_{\calE}\Vert_F$
instead of $\sin\bolTheta(\hatbolGamma,\bolGamma)$. A simple (and somewhat well-known) equivalence is provided in the following.

\begin{lemma} \label{lem: norm_vs_angle} $\Vert\mbP_{\widehat{\calE}}-\mbP_{\calE}\Vert_{F}=\sqrt{2}\Vert\sin\bolTheta(\hatbolGamma,\bolGamma)\Vert_{F}$.
\end{lemma}

Unless otherwise specified, we will be using $\Vert\mbP_{\widehat{\calE}}-\mbP_{\calE}\Vert_{F}$ as the natural measure for the distance between the true and estimated envelope subspaces. 

The rest of the article is organized as follows. Section \ref{sec:NIECE} introduces the proposed Non-Iterative Envelope Component Estimation (NIECE) and a general approach to connect the subspace estimation error with the newly introduced ``envelope scores''. The general theory is applied to the multivariate linear model in Section \ref{subsec:lowdimMLM}, where we allow both the numbers of predictors and responses to diverge with the sample size. Section \ref{sec:SNIECE} provides the sparse NIECE for high-dimensional data analysis, using the penalized matrix decomposition \citep[PMD]{witten2009pmd}. The general theory for the sparse NIECE is established first, including the high-dimensional consistency result of PMD, and then applied to the linear and generalized linear models and the Cox model. Section~\ref{sec:imple} discusses some practical considerations of using NIECE in high-dimensional regression as an alternative to principal component regression. Sections \ref{sec:sim} and \ref{sec:real}, respectively, contain a simulation study and a real data example. The Supplementary Materials contains all technical details and additional numerical results.

\section{Non-Iterative Envelope Component Estimation}\label{sec:NIECE}

\subsection{Population algorithm and envelope scores}

We propose the following NIECE procedure in population. The sample algorithm is readily available by replacing $\mbM$ and $\mbU$ with their sample counterparts and is shown to be consistent in terms of envelope subspace estimation when $p$ grows at a relatively slow rate of the sample size $n$. 

\begin{enumerate}
\item Input: symmetric $p\times p$ matrices $\mbM>0$ and $\mbU\geq0$;
number of principal components $d$; envelope dimension
$u$. Note that $0\leq u\leq d\leq p$.
\item Obtain the first $d$ eigenvectors of $\mbM$: $\mbV_d=(\mbv_{1},\dots,\mbv_{d})\in\mbbR^{p\times d}$ that satisfies $\mbV_d\T\mbV_d=\mbI_{d}$. The corresponding $d$ eigenvalues are $\lambda_{1}(\mbM)\equiv\mbv_1\T\mbM\mbv_1>\cdots>\lambda_{d}(\mbM)\equiv\mbv_d\T\mbM\mbv_d>0$. 
\item Calculate the \emph{envelope scores}: ${\phi}_{j}\equiv\mbv_{j}\T\mbU\mbv_{j}$
for $j=1,\dots,d$, and organize them in descending order ${\phi}_{(1)}\geq\cdots\geq{\phi}_{(d)}$ and define $\mbv_{(j)}$ such that $\phi_{(j)}=\mbv_{(j)}\T\mbU\mbv_{(j)}$. 
\item Output: envelope is $\MenvU=\spn(\mbv_{(1)},\cdots,\mbv_{(u)})$. 
\end{enumerate}

Formally, we provide the following fundamental property of envelopes and, more importantly, the newly introduced envelope scores. 

\begin{lemma}\label{lem:assumption} Suppose that $\MenvU\subseteq\spn(\mbV_d)$ for some $d\leq p$ and that $\lambda_j(\mbM)\neq\lambda_k(\mbM)$ for any $j,k\in\{1,\dots,d\}$ and $j\neq k$. Then \eqref{pop_pcaEnv} holds, the envelope can be written as $\MenvU=\spn(\mbv_{(1)},\cdots,\mbv_{(u)})$, and the envelope scores satisfy $\phi_{(1)}\geq\dots\geq\phi_{(u)}>\phi_{(u+1)}=\cdots=\phi_{(d)}=0$.
\end{lemma}

Lemma~\ref{lem:assumption} is a consequence of Proposition~2.2 in \citet{Cook2010envelope}, but it allows easy construction of envelope estimation in a non-iterative manner. We make the following remarks regarding the assumptions in Lemma~\ref{lem:assumption}.

First of all, the assumption that $\MenvU\subseteq\spn(\mbV_d)$ for some $d\leq p$ is trivially true if we take $d=p$, which is a reasonable choice in low-dimensional settings. However, when $p> n$, we have to make such an assumption for some $d<n$; otherwise, the envelope $\MenvU$ is not estimable. Intuitively, this is because there is only a limited number of eigenvectors of $\mbM$ estimable in high dimensions. In all our theoretical studies, we treat $d$ as fixed for simplicity and allow both $n,p\rightarrow\infty$. 

Second, the distinct eigenvalue assumption that $\lambda_j(\mbM)\neq\lambda_k(\mbM)$ for any $j,k\in\{1,\dots,d\}$ and $j\neq k$ is to ensure the identifiability of each eigenvectors. Moreover, this assumption is crucial to the high-dimensional theoretical analysis of the penalized matrix decomposition \citep[PMD,][]{witten2009pmd} method, which sequentially obtains the sparse estimates for the eigenvectors and is adopted in our sparse NIECE approach. The distinct eigenvalue assumption may be relaxed if we estimate the principal subspaces \citep[e.g.,][]{vu2013minimax} instead of vectors. We define the minimal eigen-gap of the first $d$ eigenvalues of $\mbM$ as follows,
\begin{equation}\label{Delta}
0<\Delta=\min_{j=1,\dots,d} \{\lambda_{j}(\mbM) - \lambda_{j+1}(\mbM)\}.
\end{equation}

Unless otherwise noted, we will make the assumptions as in Lemma~\ref{lem:assumption}. The envelope is thus constructed by the eigenvectors $\mbv_{(1)},\dots,\mbv_{(u)}$.
Let $\pi(\cdot): \{1,\dots,d\}\mapsto \{1,\dots,d\}$ be the permutation of indices according the envelope scores (cf. Step 3 of the NIECE algorithm), so that $\mbv_{(j)}$ is $\pi(j)$-th eigenvector of $\mbM$ and $\lambda_{\pi(j)}(\mbM)=\mbv_{(j)}\T\mbM\mbv_{(j)}$.


Finally, we note that the envelope scores, which satisfy $\phi_{(1)}\geq\dots\geq\phi_{(u)}>\phi_{(u+1)}=\cdots=\phi_{(d)}=0$, do not need to be distinct from each other. We only need a gap between the $u$-th and $(u+1)$-th envelope scores. In our theoretical analysis, the following envelope score gap plays a crucial role
\begin{equation}\label{DeltaU}
\Delta_{\mbU}\equiv\phi_{(u)}-\phi_{(u+1)}=\mbv_{(u)}\T\mbU\mbv_{(u)}>0.
\end{equation} 
In the next section, we show how the minimal eigen-gap $\Delta$ and the envelope score gap $\Delta_{\mbU}$ would affect the subspace estimation in finite sample. 

\subsection{Theory for the sample NIECE algorithm}

For the sample NIECE procedure, we simply replace $\mbM$ and $\mbU$ with their sample counterparts (e.g.~sample covariance matrices) and let $\widehat{\mbv}_{(j)}$,
$j=1,\dots,u,$ be the $\widehat{\pi}(j)$-th eigenvectors
of  $\hatmbM$, where $\widehat\pi(\cdot)$ is the permutation of indices in $\{1,\dots,d\}$ according the the estimated envelope scores. Specifically, we can write  $\lambda_{\widehat{\pi}(j)}(\hatmbM)=\hatmbv_{(j)}\T\hatmbM\hatmbv_{(j)}$.
One critical condition that is necessary for the consistency of the
envelope estimation is that the index set $\{\widehat{\pi}(j)\mid j=1,\dots,u\}$
converges to the true set $\{\pi(j)\mid j=1,\dots,u\}$. In fact,
we prove a slightly stronger condition $\widehat{\pi}(j)\rightarrow\pi(j)$
for all $j=1,\dots,u$ in our theoretical studies. 

We first present a generic theory to characterize how the envelope subspace estimation error is affected by: (i) the envelope dimension; (ii) the estimation errors, measured in operator norms, of sample matrices $\hatmbU$ and $\hatmbM$; (iii) the minimal eigen-gap $\Delta$ defined in \eqref{Delta}; (iv) the envelope score gap $\Delta_\mbU$ defined in \eqref{DeltaU}; and (v) the largest eigenvalue of $\mbU$ denoted as $\nu=\Vert\mbU\Vert_{op}$. The following theory is for the analysis in relatively low-dimensional data, and it is not easy to find estimates satisfying \eqref{op_epsilon} in high dimensions.

\begin{theorem} \label{thm: basic_niece} For any $\epsilon>0$ and $\epsilon\leq\max\left\{ \dfrac{\Delta}{4},\!\dfrac{\Delta_{\mbU}}{2}\left(1+\dfrac{10\nu}{\Delta}\right)^{-1}\right\}$, if we have
\begin{equation}\label{op_epsilon}
\Vert\hatmbM-\mbM\Vert_{op}\leq\epsilon,\quad\Vert\hatmbU-\mbU\Vert_{op}\leq\epsilon,
\end{equation} then $
\Vert\mbP_{\hatcalE}-\mbP_{\calE}\Vert_{F}\leq\dfrac{2\sqrt{2u}}{\Delta}\epsilon$, where $\hatcalE$ and $\calE$ are the envelope $\hatMenvU$ and $\MenvU$, respectively.
\end{theorem}

In Theorem~\ref{thm: basic_niece} and all our theoretical analysis later, the envelope subspace estimation error is measured by $\Vert\mbP_{\hatcalE}-\mbP_{\calE}\Vert_{F}$, which is bounded between 0 and $\sqrt{2u}$ by definition. Therefore, it is natural to see $\Vert\mbP_{\hatcalE}-\mbP_{\calE}\Vert_{F}$ proportional to $\sqrt{2u}$ in our results. From Theorem~\ref{thm: basic_niece}, the subspace estimation error is proportional to the estimation error of the matrices $\mbM$ and $\hatmbU$, which is indicated by $\epsilon$; and is inversely proportional to the minimal eigen-gap $\Delta$, which directly affects the estimation of eigenvectors $\mbv_{(1)},\dots,\mbv_{(u)}$ that lie within the envelope. On the other hand, the envelope score gap $\Delta_\mbU$ together $\nu$, which indicates the magnitude of $\mbU$, affects the envelope subspace estimation error indirectly because $\hatmbU$ is only used in Step 3 of NIECE procedure to determine the index ordering. When the matrices $\hatmbM$ and $\hatmbU$ are close enough to their population counterparts, i.e.~$\epsilon\leq \dfrac{\Delta_{\mbU}}{2}\left(1+\dfrac{10\nu}{\Delta}\right)^{-1}$, the estimated index set $\{\widehat{\pi}(j)\mid j=1,\dots,u\}$
converges to the true set $\{\pi(j)\mid j=1,\dots,u\}$. 

In the Supplementary Materials, Lemma~\ref{lem:eigenspace}, we consider the ideal setting where $\{\widehat{\pi}(j)\mid j=1,\dots,u\}=\{\pi(j)\mid j=1,\dots,u\}$. Then we may replace the minimal eigen-gap $\Delta$ of the $d$ eigenvalues of $\mbM$ to the following quantity that only involves $u\leq d$ eigenvalues. For simplicity, suppose the eigenvalues $\lambda_{\pi(j)}$'s, $j=1,\dots,u$, are all
distinct, then we may re-define $\Delta=\min_{j=1,\dots,u}\Delta_{j}$, where $\Delta_{j}=\min\{\lambda_{\pi(j)-1}(\mbM)-\lambda_{\pi(j)}(\mbM),\ \lambda_{\pi(j)}(\mbM)-\lambda_{\pi(j)+1}(\mbM)\}$ and $\lambda_{-1}\equiv-\infty$, $\lambda_{p+1}\equiv\infty$. In the Supplementary Materials, we show that the similar results of Theorem~\ref{thm: basic_niece} still holds by re-defining $\Delta$, even when some of these $\lambda_{\pi(j)}$'s are not distinct.

To apply the sample NIECE in different contexts, we only need to calculate the probability of \eqref{op_epsilon} to verify the consistency of envelope subspace estimation. This theory allows us to study the un-penalized envelopes with (slowly) diverging dimensions. We next demonstrate that such results hold for the simultaneous predictor and response reduction.

\subsection{Simultaneous envelopes in multivariate linear regression}\label{subsec:lowdimMLM}
To illustrate the application of Theorem~\ref{thm: basic_niece}, we consider the simultaneous envelopes model \citep{CookZhang2015simultaneous}. The simultaneous envelopes are constructed, by jointly estimating the response envelope \citep{Cook2010envelope} and the predictor envelope \citep{Cook2013PLS}, to simultaneous reduce the multivariate response $\mbY\in\mbbR^{r}$ and predictor $\mbX\in\mbbR^{p}$ in the following regression model, 
\begin{equation}
\mbY=\bolalpha+\bolbeta\mbX+\boldsymbol{\varepsilon},\label{lm}
\end{equation}
where $\bolalpha\in\mbbR^r$ is the intercept vector, $\bolbeta\in\mbbR^{r\times p}$ is the regression coefficient matrix, and $\boldsymbol{\varepsilon}\in\mbbR^r$ is independent of $\mbX$. 

In the classical likelihood-based envelope methods, the response is reduced by projecting onto the response envelope $\calE_{\bolSigma}(\bolbeta)$, whose likelihood-based estimation is derived from the normal error $\boldsymbol{\varepsilon}\sim N(0,\bolSigma)$. On the other hand, the predictor reduction is achieved by projection onto the predictor envelope $\calE_{\bolSigma_{\mbX}}(\bolbeta\T)$, whose likelihood-based estimation is derived from further assuming $\mbX\sim N(0,\bolSigma_{\mbX})$. Thus, for predictor reduction and simultaneous reduction, the joint normality of $(\mbX\T,\mbY\T)\T$ is required. We consider the following sub-Gaussian tail distribution assumption that is weaker than the joint normality assumption.

We assume that there exists $\sigma>0$ such that for any $\mbw\in\mbbR^{p+r}$,
$\Vert\mbw\Vert_{2}=1$, we have
\begin{equation}\label{XYsubG}
P(|\mbw\T\left(\begin{array}{c}
\mbX\\
\mbY
\end{array}\right)|\ge t)\le2\exp(-\dfrac{t^{2}}{\sigma^{2}}).
\end{equation}

While alternating updates obtain the likelihood-based simultaneous envelopes estimators, we apply the NIECE to the predictor and response envelopes separately. 
For response envelope, note that $\calE_{\bolSigma}(\bolbeta)=\calE_{\bolSigma_{\mbY}}(\bolSigma_{\mbY\mbX})$, where $\bolSigma_\mbY\in\mbbR^{r\times r}$ is the covariance matrix of $\mbY$ and $\bolSigma_{\mbY\mbX}\equiv\bolSigma_{\mbX\mbY}\T\in\mbbR^{r\times p}$ is the cross-covariance matrix. For predictor envelope, we exploit the symmetry by noticing $\calE_{\bolSigma_{\mbX}}(\bolbeta\T)=\calE_{\bolSigma_{\mbX}}(\bolSigma_{\mbX\mbY})$. Given the data, the NIECE estimator is obtained by using the sample covariances $\hatmbM=\widehat\bolSigma_{\mbY}$ and $\hatmbU=\hatbolSigma_{\mbY\mbX}\hatbolSigma_{\mbX\mbY}$ for response reduction, and $\hatmbM=\widehat\bolSigma_{\mbX}$ and $\hatmbU=\hatbolSigma_{\mbX\mbY}\hatbolSigma_{\mbY\mbX}$ for predictor reduction.

We have the following result that applies to both response and predictor envelopes (i.e.~simultaneous envelopes), where $\MenvU=\calE_{\bolSigma_{\mbY}}(\bolSigma_{\mbY\mbX})\subseteq\mbbR^r$ for response envelope and is re-defined as $\MenvU=\calE_{\bolSigma_{\mbX}}(\bolSigma_{\mbX\mbY})\subseteq\mbbR^p$ for predictor envelope. Other quantities such as $\hatcalE$, $\Delta$, $\Delta_{\mbU}$, $u$ and $\nu$ also alters when we switch from response envelope to predictor envelope.

\begin{theorem}\label{thm.res.low} For any $\epsilon>0$ such that
$\epsilon\leq\max\left\{ \dfrac{\Delta}{4},\!\dfrac{\Delta_{\mbU}}{2}\left(1+\dfrac{10\nu}{\Delta}\right)^{-1}\right\} $
and $\epsilon\leq\Vert\bolSigma_{\mbX\mbY}\Vert_{1}+\Vert\bolSigma_{\mbY\mbX}\Vert_{1}<c_{1}$
for some constant $c_{1}$, we have 
\begin{equation}
\Vert \mbP_{\hatcalE} - \mbP_{\calE}\Vert_{F}\le\dfrac{2\sqrt{2u}\epsilon}{\Delta},
\end{equation}
with a probability greater than $1-Cr^{2}\exp(-C\dfrac{n\epsilon^{2}}{r^{2}})-Cpr\exp(-C\dfrac{n\epsilon^{2}}{r^{2}})-Cpr\exp(-C\dfrac{n\epsilon^{2}}{p^{2}})$.
\end{theorem}
To the best of our knowledge, the above theorem is the first result in establishing non-asymptotic properties of any envelope estimators without sparsity. It also leads to the first known consistency result of envelope subspace estimation with diverging $n$, $p$, and $r$.
Specifically, we have the following Corollary by letting $\epsilon=pr\sqrt{{(\log p+\log r)}/{n}}\rightarrow0$.
\begin{cy}\label{cy.res.low}
Suppose $\max\left(pr\sqrt{{\log p+\log r}},{pr}{\Delta^{-1}_{\mbU}}\sqrt{{\log p+\log r}}\right)=o(\sqrt{n})$
and $\Vert\bolSigma_{\mbX\mbY}\Vert_{1}+\Vert\bolSigma_{\mbY\mbX}\Vert_{1}<c_{1}$
for some constant $c_{1}$, then  
$\Vert \mbP_{\hatcalE}-\mbP_{\calE}\Vert_{F}\rightarrow0
$ in probability as $n,p,r\rightarrow\infty$.
\end{cy}
The above result holds for both predictor and response envelopes. We have thus established the consistency of the simultaneous envelopes.
Analogous to Theorem~\ref{thm.res.low} and Corollary~\ref{cy.res.low}, it is straightforward to show similar results for other types of envelopes by applying Theorem~\ref{thm: basic_niece}. We omit details for such extensions and study some of them (envelopes in the generalized linear model and Cox model) in the more challenging high-dimensional settings in the following sections.

\subsection{Comparing Envelope Algorithms}\label{subsec:simalg}

	\begin{figure}[t!]
		\includegraphics[scale=0.42]{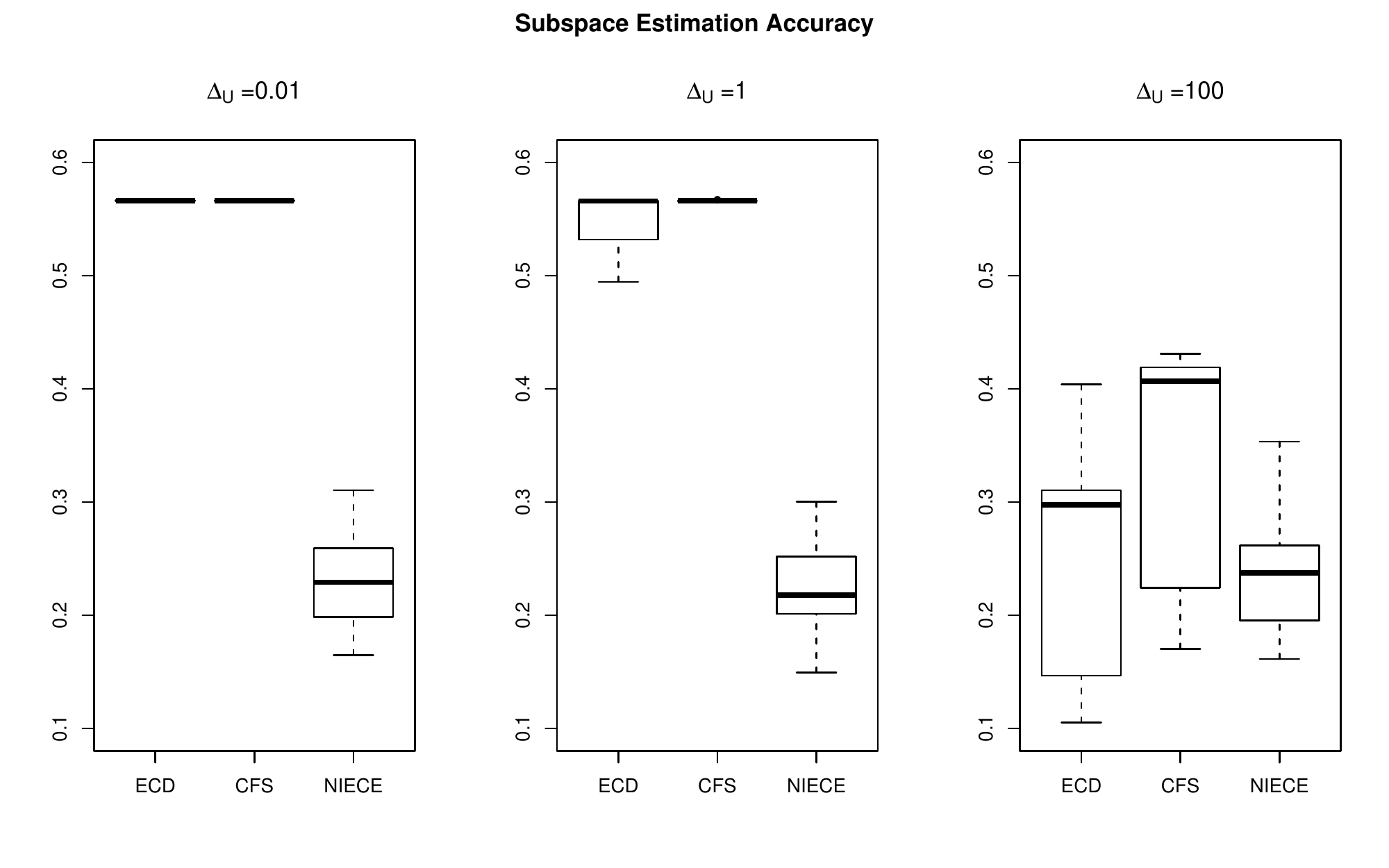}\includegraphics[scale=0.42]{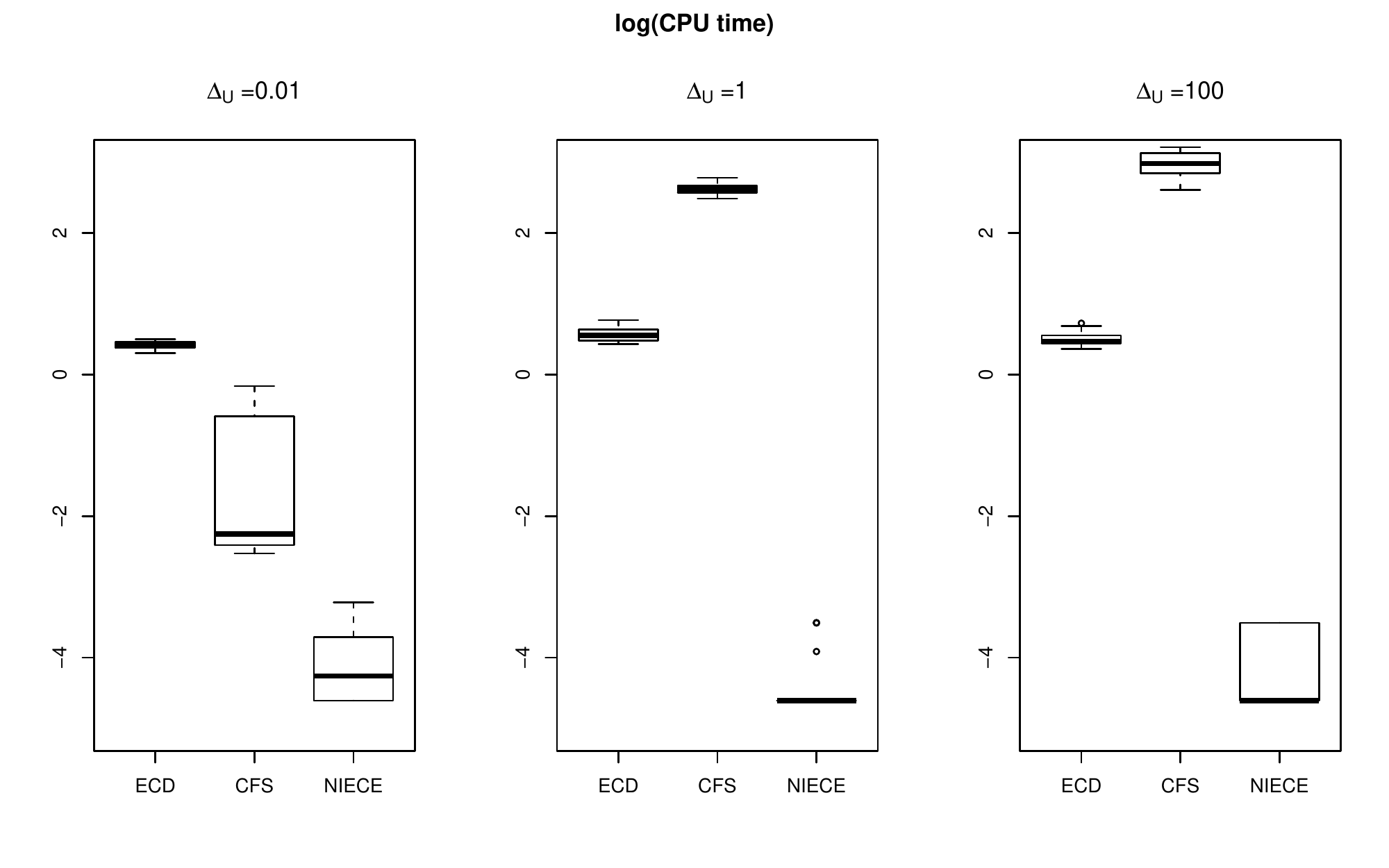}
		\caption{Advantages of NIECE over the state-of-the-art envelope algorithms in both accuracy (left three panels; smaller value $\calD$ indicates more accurate subspace estimation) and time (right three panels; logarithm of CPU time in seconds), and across a wide range of signal strengths: weak signal (left, $\Delta_{\mbU}=0.01$), moderate signal (middle, $\Delta_{\mbU}=1$), and strong signal (right, $\Delta_{\mbU}=100$). \label{fig1} }
	\end{figure}

We first compare NIECE with two fast envelope algorithms, CFS \citep{cook2016note} and ECD \citep{cook2018fast} that are both recently proposed state-of-the-art algorithms and were shown to be much faster and more stable than the full Grassmannian optimization \citep{edelman1998geometry,absil2009optimization} and sequential 1D algorithm \citep{CookZhang2014algorithm} in the literature. We consider the moderately high-dimensions, where $(n,p,u)=(200,100,5)$, because the CFS and ECD are not applicable to high-dimensional settings. 
	We set $\mbM=\bolGamma\bolOmega\bolGamma\T+\bolGamma_{0}\bolOmega_{0}\bolGamma_{0}\T$,
	$\mbU=\Delta_{\mbU}\cdot\bolGamma\bolPhi\bolGamma\T$, where $\bolGamma=(\mbv_{2},\mbv_{3},\mbv_{10},\mbv_{11},\mbv_{19})$ and $\bolOmega$ and $\bolOmega_{0}$ are diagonal matrix of eigenvalues $\lambda_{k}=k^{3}$ for $k=1,\dots,20$ and $\lambda_{k}=0.05$ for
	$k>20$, $\bolPhi=\mbO_{u}\mbD\mbO_{u}\T$ with orthonormal matrix $\mbO_{u}\in\mbbR^{u\times u}$ and $\mbD=\mathrm{diag}(1,\dots,u)$. We tried three different signal strength settings with $\Delta_{\mbU}=0.01$, $1$, and $100$. For each setting, we randomly generated 100 replicates of $\hatmbM$ and $\hatmbU$ from Wishart distributions with degrees of freedom $n$ and means $\mbM$ and $\mbU$ to mimic the sample covariances in regression models. We reported the subspace estimation error $\calD(\hatbolGamma,\bolGamma)=\calD(\hatcalE,\calE)=\Vert\mbP_{\calE}-\mbP_{\hatcalE}\Vert_{F}/\sqrt{2u}$, which is a number between $0$ and $1$, and the logarithm of CPU time. 
	
	Figure~\ref{fig1} shows the advantages of NIECE in both subspace estimation accuracy and computational speed. This numerical study is an illustration of our theoretical analysis in the previous section. When the eigenvalue gap $\Delta$ in $\mbM$ is large enough, the NIECE procedure is very fast (due to the computational advantage of SVD in NIECE versus manifold optimizations in others) and also very accurate for a wide range of signal strengths measured by $\Delta_{\mbU}$.	

\section{NIECE for high-dimensional data analysis}\label{sec:SNIECE}

\subsection{High-dimensional sparse NIECE}

In high-dimensional sparse envelope settings, we only need to modify Step 2 of NIECE sample algorithm, by replacing the eigen-decomposition of sample matrix $\hatmbM$ with the penalized eigen-decomposition or penalized principal component analysis methods that are more suitable for such scenarios. Without loss of generality, we assume $d< n$ and obtain the $d$ sparse eigenvectors. For almost all envelope problems, we can write $\hatmbM=\mbX_{n}\T\mbX_{n}$ from some data matrix $\mbX_{n}\in\mbbR^{n\times p}$. As we have seen from the previous Section, the data matrix $\mbX_{n}$ can be either the $n$ samples of $\mbX$ or $\mbY$. For envelopes in GLM and Cox model, $\mbX_{n}$ is also the $n$ samples of $\mbX$. Another example is the weighted least squares envelope \citep{CookZhang2015foundation}, where the $i$-th row of $\mbX_{n}$ is the squared-root weight $\sqrt{W_i}$ times the predictor vector $\mbX_i$. 

For simplicity, the data matrix $\mbX_{n}\in\mbbR^{n\times p}$ is henceforth defined as the $n$ i.i.d.~samples of mean zero random variable $\mbX\in\mbbR^p$.
We obtain the $d$ sparse eigenvectors of $\hatmbM\geq0$ based
on the penalized matrix decomposition (PMD) method from \citet{witten2009pmd}. Because $n\ll p$ is very common in high-dimensional data analysis, we adopt the special type of PMD named PMD($\cdot$, $L_{1}$) on $\mbX_{n}$ instead of the much bigger matrix $\hatmbM$ as follows. 

Let $\mbX^{1}=\mbX_{n}$ be the original data matrix. For $k=1,\dots,d$, we sequentially solve for
\begin{equation}
(\hatmbu_{k},\hatmbv_{k})=\argmax_{\mbu\in\mbbR^n,\mbv\in\mbbR^p}\mbu\T\mbX^{k}\mbv\quad\mathrm{subject\ to}\ \Vert\mbv\Vert_{1}\leq c,\ \Vert\mbv\Vert_{2}^{2}\leq1,\ \Vert\mbu\Vert_{2}^{2}\leq1;\label{SPC}
\end{equation}
and then deflate the data matrix as $\mbX^{k+1}=\mbX^{k}-\widehat{\sigma}_{k}\hatmbu_{k}\hatmbv_{k}\T$, where $\widehat{\sigma}_{k}=\hatmbu_{k}\T\mbX^{k}\hatmbv_{k}$.

To solve \eqref{SPC}, we need alternating updates between $\mbu\in\mbbR^{n}$
and $\mbv\in\mbbR^{p}$. Since there is no penalty on $\mbu$, we
can see that the solution for $\hatmbu_{k}$ has to be
\begin{equation}
\hatmbu_{k}=\frac{\mbX^{k}\hatmbv_{k}}{\Vert\mbX^{k}\hatmbv_{k}\Vert_{2}},\label{SPC_uhat_k}
\end{equation}
which leads to the following equivalent presentation.

\begin{lemma} \label{lemma: SPC} The optimization in \eqref{SPC}
is equivalent to \eqref{SPC_uhat_k} and 
\begin{equation}
\hatmbv_{k}=\arg\max_{\mbv}\mbv\T\hatmbM_{k}\mbv\quad\mathrm{subject\ to}\ \Vert\mbv\Vert_{1}\leq\tau\ \mathrm{and}\ \Vert\mbv\Vert_{2}^{2}\leq1,\label{SPC_vhat_k}
\end{equation}
where $\widehat{\mbM}_{k}=(\mbX^{k})\T\mbX^{k}$ and $\mbX^{k}=\mbX^{k-1}(\mbI_{p}-\hatmbv_{k-1}\hatmbv_{k-1}\T)$.
\end{lemma}
As noted in \citet{witten2009pmd}, by the Karush-Kuhn-Tucker conditions
in convex optimization, the solution to \eqref{SPC_vhat_k} is also the
solution to 
\begin{equation}
\hatmbv_{k}=\arg\max_{\mbv}\mbv\T\hatmbM_{k}\mbv\quad\mathrm{subject\ to}\ \Vert\mbv\Vert_{1}\leq c,\ \mathrm{and}\ \Vert\mbv\Vert_{2}^{2}=1,
\end{equation}
if $c$ is chosen so that the maximizer of $\mbv\T\hatmbM_{k}\mbv$
subject to only one constraint $\Vert\mbv\Vert_{1}\leq c$ has $L_{2}$
norm greater than or equals to $1$.

\subsection{A general high-dimensional theory}
Notice that we have re-defined $\hatmbv_k$ as the PMD solution for the high-dimensional sparse NIECE. Moreover, we re-define $\hatmbV_{d}=(\hatmbv_{1},\dots,\hatmbv_{d})$ and other estimates for ease of presentation. However, it is important to note that $\hatmbv_k$ is no longer the eigenvector of $\hatmbM$ and that orthogonality among eigenvectors no longer holds, i.e.~$\hatmbV_d\T\hatmbV_d\neq \mbI_{d}$. Furthermore, $\widehat{\sigma}_{k}^{2}=\hatmbv_k\T\hatmbM\hatmbv_k$ is different from the eigenvalue $\lambda_{k}(\hatmbM)$. Due to the non-orthogonality among $\hatmbv_k$'s and related issues, the deflation of data matrix is not easy to handle. Nevertheless, we are able to show consistency of $\hatmbv_1,\dots,\hatmbv_d$ by analyzing $\mbX^k=\mbX_n\prod_{j=1}^{k-1}(\mbI_d - \hatmbv_{j}\hatmbv_{j}\T)$ based on Lemma~\ref{lemma: SPC}. 
Because we assume that the eigenvalues are distinct, PMD offers a computationally efficient way to estimate the eigenvectors. If we use principal subspace estimation \citep[e.g.,][]{cai2013sparse}, the rate may potentially be improved but with higher computation costs.

We first show a general theory of PMD, which is of independent interest. The sparsity in the true eigenvectors $\mbv_k$ (e.g.~sparsity in principal component loadings) is imposed as follows.
\begin{equation}\label{sparsityPMD}
\tau_{0} =  \max_{1\leq k\leq d}\Vert\mbv_{k}\Vert_{1},
\end{equation}
where a small value of $\tau_0$ implies that the first $d$ eigenvectors of $\mbM$ are all reasonably sparse in high dimensions. 

\begin{theorem} \label{thm:SPC} If $\tau>\tau_{0}$ and $\Vert\hatmbM-\mbM\Vert_{\max}\leq\epsilon$,
then there exists a constant $c_{0}>0$ that does not depend on $n$,
$p$ or $\epsilon$, such that,
\begin{equation}
\sin^{2}\Theta(\mbv_{k},\hatmbv_{k})\leq c_{0}\epsilon\tau^{2},\quad k=1,\ldots,d.\label{eq:thmspc}
\end{equation}
\end{theorem}

Hence, to show the consistency of NIECE in high-dimensional settings,
we only need to verify $\Vert\hatmbM-\mbM\Vert_{\max}\leq\epsilon$,
which is weaker than the assumption of $\Vert\hatmbM-\mbM\Vert_{op}\leq\epsilon$
in the non-sparse settings. This is because we apply regularization to obtain accurate estimates of $\mbv_{k}$ in high dimensions.
Also, note that $\Vert\hatmbM-\mbM\Vert_{op}\leq\epsilon$ generally
cannot hold in high dimensions, while later we will show that $\Vert\hatmbM-\mbM\Vert_{\max}\leq\epsilon$
holds with a probability tending to 1 under mild conditions.

\begin{theorem}\label{thm: NIECE_hd} Assume that $\tau>\tau_{0}$,
$\Vert\hatmbM-\mbM\Vert_{\max}\leq\epsilon$ and $\Vert\hatmbU-\mbU\Vert_{\max}\leq\epsilon$,
then if $\Delta_{\mbU}>\sqrt{8c_{0}\nu^{2}\tau^{2}\epsilon}+(2c_{0}\nu+1)\tau^{2}\epsilon$
with constant $c_{0}$ defined in \eqref{eq:thmspc}, then there exists
a constant $C>0$ that does not depend on $n$, $p$ or $\epsilon$
such that $
\Vert\mbP_{\hatcalE}-\mbP_{\calE}\Vert_{F}^{2}\leq C\epsilon\tau^{2}$.
\end{theorem}

Theorem~\ref{thm: NIECE_hd} is a general result that guarantees
that as long as we start with a reasonably accurate estimator $\hatmbM$,
we can estimate the envelope with the specified error bound. Next, we demonstrate that Theorem~\ref{thm: NIECE_hd} leads to the consistency of sparse NIECE under several important envelope models.

\subsection{Envelope in multivariate linear model}

In this section, we consider the multivariate linear model \eqref{lm} with jointly sub-Gaussian data \eqref{XYsubG}. To avoid redundancy, we only consider the response envelope. The predictor envelope in linear model is analogous to the response envelope and similar to the predictor envelope in the generalized linear model in the next section.

Recall that the NIECE estimator for response envelope is obtained by using the sample covariances $\hatmbM=\widehat\bolSigma_{\mbY}$ and $\hatmbU=\hatbolSigma_{\mbY\mbX}\hatbolSigma_{\mbX\mbY}$. In high-dimensional sparse setting, e.g.~the eigenvectors of $\mbM=\bolSigma_{\mbY}$ satisfies \eqref{sparsityPMD}, we have the following results.

\begin{theorem}\label{thm:res} 
For any $\epsilon$ such that $0<\epsilon\leq1/p$, assume that $\tau>\tau_{0}$
and $\Delta_{\mbU}>\sqrt{8c_{0}\nu^{2}\tau^{2}\epsilon}+(2c_{0}\nu+1)\tau^{2}\epsilon$
with constant $c_{0}$ defined in \eqref{eq:thmspc}, then there exists
a constant $C>0$ that does not depend on $n$, $r$, $p$ or $\epsilon$
such that
\begin{equation}
\Vert\mbP_{\hatcalE}-\mbP_{\calE}\Vert_{F}\leq C\epsilon\tau^{2}
\end{equation}
with a probability greater than $1-Cr^{2}\exp(-Cn\epsilon^{2})-Cpr\exp(-Cn\epsilon^{2})$.
\end{theorem}
The above non-asymptotic result leads to the following consistency of sparse NIECE. 

\begin{cy}\label{cy:hdResponse}
If $\sqrt{{\log p+\log r}}=o(\sqrt{n})$, $\sqrt{\log p+\log r}<\sqrt{n}/{p}$
and $\Delta_{\mbU}$ and $\tau$ satisfy that 
$$\max\left( \tau^{2}\sqrt{\log p+\log r},\ \Delta_{\mbU}^{-2}\tau^{2}\sqrt{\log p+\log r}\right) =o(\sqrt{n}),$$
then we have $\Vert\mbP_{\hatcalE}-\mbP_{\calE}\Vert_{F}\rightarrow0$  in probability as $n,p,r\rightarrow\infty$.
\end{cy}

The requirement that $\sqrt{\log p+\log r}<\sqrt{n}/{p}$
implies that the number of predictors can not diverge quickly ($p$ grow slower than $\sqrt{n}$), while the response envelope allows the number of responses to diverge much faster. To see this clearly, we let $p$ fixed in the following Corollary.

\begin{cy}\label{cy:hdResponse2}
If ${\log r}=o({n})$, and $\Delta_{\mbU}$ and
$\tau$ satisfy that $\max(\tau^{2}\sqrt{{\log r}},\ \Delta_{\mbU}^{-2}\tau^{2}\sqrt{{\log r}}) =o(\sqrt{n})$,
we have $\Vert\mbP_{\hatcalE}-\mbP_{\calE}\Vert_{F}\rightarrow0$  in probability as $n,r\rightarrow\infty$. \end{cy}

This justifies applying the NIECE approach for response envelope reduction in ultra high-dimensional settings, i.e.~$(\log r)/n\rightarrow0$. Similarly, for the predictor envelope in the linear model, if we fix the number of responses, consistency is achieved when $(\log p)/n\rightarrow0$. 

\subsection{Envelope in generalized linear model}
\citet{CookZhang2015foundation} extended the envelope model from multivariate linear regression context to generalized linear model with canonical link functions, Cox model, and general multivariate parameter estimation problems. We consider the envelope generalized linear model and the envelope Cox model in this and the following sections. The sparse NIECE procedure extends these non-standard envelope models from low dimensional settings to high-dimensional. 

We replace the joint sub-Gaussian tail distribution of $(\mbX\T,\mbY\T)\T$ with the following sub-Gaussian tail assumption on $\mbX$. Similar to \eqref{XYsubG}, this is weaker than assuming the normality and is widely used to prove the consistency of penalized methods \citep[e.g.,][]{NRWY12}. 
Suppose that there exists $\sigma>0$ such that for any $\mbw\in\mbbR^{p}$, $\Vert\mbw\Vert_{2}=1$, we have
\begin{equation}\label{XsubG}
P(|\mbw\T\mbX|\ge t)\le2\exp(-\dfrac{t^{2}}{\sigma^{2}}).
\end{equation}

In the generalized linear model with canonical link functions, the response $Y$ follows some exponential family distributions with
probability density (or mass) function as
$$f(y\mid\vartheta,\varphi)=\exp(\dfrac{y\vartheta-b(\vartheta)}{a(\varphi)}+c(y,\varphi)),
$$ where the canonical parameter $\vartheta$ is set to be $\vartheta=\bolbeta\T\mbX$
by the canonical link function, $\varphi$ is the dispersion parameter,
and functions $a$, $b$ and $c$ can be specified for different families
of the distribution of $Y$. For simplicity, we focus on one-parameter
family so that the dispersion parameter is not included (or, equivalently,
$\varphi$ is set to be 1). As such, we can write the log-likelihood
as
\begin{equation}
\ell_{n}(\bolbeta)=\sum_{i=1}^{n}\log f(Y_{i}\mid\bolbeta,\mbX_{i})=\sum_{i=1}^{n}\{Y_{i}(\bolbeta\T\mbX_{i})-b(\bolbeta\T\mbX_{i})\},
\end{equation}
where the function $b(t)=t^{2}/2$ for normally distributed $Y$;
it is $\exp(t)$ for Poisson regression and is $\log\{1+\exp(t)\}$
for Logistic regression. The envelope is $\calE_{\bolSigma_{\mbX}}(\bolbeta)\subseteq\mbbR^{p}$
where $\bolbeta\in\mbbR^{p\times1}$ is the regression coefficient
vector from above.

To avoid redundancy, we only show the results for high-dimensional
penalized Logistic regression where $Y=0$ or $1$, and $\ell_{n}(\bolbeta)=\sum_{i=1}^{n}\{Y_{i}(\bolbeta\T\mbX_{i})-\log(1+e^{\bolbeta\T\mbX_{i}})\}$.
For normally distributed $Y$, the GLM reduces to the predictor envelope
with a univariate response. We have shown that $\hatbolSigma_{\mbX}$
is an accurate estimator for $\bolSigma_{\mbX}$. For $\bolbeta$,
note that the sparsity of the envelope implies the sparsity of $\bolbeta$.
We can obtain $\hatbolbeta$ by $\ell_{1}$ logistic regression with
a tuning parameter $\lambda_{n}$, 
\begin{equation}
\hatbolbeta=\arg\min_{\bolbeta}\dfrac{1}{n}\sum_{i=1}^{n}\{-Y_{i}(\bolbeta\T\mbX_{i})+\log(1+e^{\bolbeta\T\mbX_{i}})\}+\lambda_{n}\Vert\bolbeta\Vert_{1},\label{betaGLM_l1}
\end{equation}
which is based on minimizing the negative log-likelihood plus an $\ell_{1}$-regularization
term. Then the NIECE procedure is based on $\hatmbM=\hatbolSigma_{\mbX}$
and $\hatmbU=\hatbolbeta\hatbolbeta\T$. 

\begin{theorem}\label{thm:glm} If $\tau>\tau_{0}$, $\lambda_{n}=4\epsilon/\sqrt{s}$,
and $\Delta_{\mbU}>\sqrt{8c_{0}\nu^{2}\tau^{2}\epsilon}+(2c_{0}\nu+1)\tau^{2}\epsilon$
with constant $c_{0}$ defined in \eqref{eq:thmspc}, then there exists
a constant $C>0$ that does not depend on $n$, $p$ or $\epsilon$
such that 
\begin{equation}
\Vert\mbP_{\hatcalE}-\mbP_{\calE}\Vert_{F}\le C\tau^{2}\epsilon
\end{equation}
with a probability greater than $1-Cp^{2}\exp(-Cn\epsilon^{2})-Cp\exp(-Cn\epsilon^{2}/s)$.
Furthermore, if ${{s\log p}}=o(n)$ and that $\lambda_{n},\tau$ and $\Delta_{\mbU}$
satisfy $\max\left\{ \tau^{2}\sqrt{\dfrac{s\log p}{n}},\ \Delta_{\mbU}^{-2}\tau^{2}\sqrt{\dfrac{s\log p}{n}},\ \lambda_{n}^{-1}\sqrt{\dfrac{\log p}{n}}\right\} =o(1)$,
we have $
\Vert\mbP_{\hatcalE}-\mbP_{\calE}\Vert_{F}\rightarrow0$ in probability as $n,p\rightarrow0$.\end{theorem}

Theorem \ref{thm:glm} thus summarizes both the non-asymptotic and asymptotic results for the sparse NIECE procedure in logistic regression.

\subsection{Envelope in Cox proportional hazards model}

We consider the Cox regression \citep{cox1972regression} that is
widely used in survival data analysis. The hazard function is assumed
to be $h(t\mid\mbX)=h_{0}(t)\exp(\bolbeta\T\mbX)$, where $h_{0}(t)$
is the baseline hazard function and $\bolbeta\in\mbbR^{p\times1}$
is the regression coefficients of the $p$-dimensional covariate vector
$\mbX\in\mbbR^{p}$. Let $Y$ and $C$ be the failure time and censoring
time, then we assume $Y$ and $C$ are conditionally independent given
$\mbX$. Define $\delta=I(Y\leq C)$ and $T=\min(Y,C)$, then the
observed data consists of independent copies of $\{T_{i},\delta_{i},\mbX_{i}\}$,
$i=1,\dots,n$. 

Estimation of $\bolbeta$ is typically achieved by maximizing the
Cox's partial likelihood \citep{cox1975partial}, or equivalently
by minimizing the following negative logarithm of that, 
\begin{equation}
\ell_{n}(\bolbeta)=-\dfrac{1}{n}\sum_{i=1}^{n}\delta_{i}\left[\bolbeta\T\mbX_{i}-\log\left\{ \sum_{j=1}^{n}I(T_{j}\geq T_{i})\exp(\bolbeta\T\mbX_{j})\right\} \right].
\end{equation}

When the covariates in $\mbX$ are highly correlated with each other,
\citet{CookZhang2015foundation} proposed to construct envelope estimator
\begin{equation}
\hatbolbeta_{\Env}=\arg\min_{\bolbeta=\bolGamma\boleta}\ell_{n}(\bolbeta)-\dfrac{1}{n}M_{n}(\bolGamma),
\end{equation}
where $M_{n}(\bolGamma)=-\dfrac{n}{2}(\log\vert\bolGamma\T\mbS_{\mbX}\bolGamma\vert+\log\vert\bolGamma\T\mbS_{\mbX}^{-1}\bolGamma\vert)$
is the partially maximized log likelihood of $\mbX\sim N(\bolmu_{\mbX},\bolSigma_{\mbX})$
under the envelope structure $\spn(\bolGamma)=\calE_{\bolSigma_{\mbX}}(\bolbeta)$.
It is shown to be more efficient than the partial least squares in
Cox regression \citet{nygaard2008partial}. However, the optimization is similar to (or even more challenging than) the Grassmannian optimization  \eqref{obj_pop} discussed in the Introduction section and is thus not feasible in the high-dimensional Cox model.

In the high-dimensional setting, to incorporate variable selection, we
consider the following $\ell_{1}$-penalized maximum (partial) likelihood
estimator,
\begin{equation}
\hatbolbeta=\arg\min_{\bolbeta}\ell_{n}(\bolbeta)+\lambda_{n}\Vert\bolbeta\Vert_{1},\label{betaCox_l1}
\end{equation}
where $\lambda_{n}>0$ is the tuning parameter. Similar to the envelope
GLM, the sparsity of the envelope $\calE_{\bolSigma_{\mbX}}(\bolbeta)$
implies the sparsity of $\bolbeta$. Then the NIECE procedure is based
on $\hatmbM=\hatbolSigma_{\mbX}$ and $\hatmbU=\hatbolbeta\hatbolbeta\T$.
The convergence of $\hatbolSigma_{\mbX}$ is guaranteed by sub-Gaussian
distribution of $\mbX$, i.e. \eqref{XsubG}. Properties of $\hatmbU$
is based on the estimation consistency of the Lasso estimator $\hatbolbeta$
that is derived in \citet{huang2013oracle}. Similar results for using
the SCAD penalty \citep{fan2001variable} instead of Lasso can be
found in \citet{bradic2011regularization}.

Following \citet{huang2013oracle}, we consider the following two conditions. 

(i) \emph{Uniformly bounded covariates.} There exists some constant $C$ such
that the covariates are bounded as $
\max_{i<i^{\prime}\leq n}\Vert\mbX_{i}-\mbX_{i^{\prime}}\Vert_{\infty}\le C$.

(ii) \emph{Restricted eigenvalue lower bound.} There exists positive constants $t$ and $M$ such that the smallest eigenvalue of the population matrix $\bolSigma(t,M)$ is greater than some constant $\rho^{\ast}$.

These conditions are derived from Theorem 4.1 of \citet{huang2013oracle},
where the introduction of $\bolSigma(t,M)$ is rather technical and
relegated to the Supplementary Materials. The lower bound $\rho^{\ast}$
on the smallest eigenvalue of $\bolSigma(t,M)$ provides a lower bound
for the compatibility and cone invertibility factors and the restricted
eigenvalue in high-dimensional Cox regression.

\begin{theorem}\label{thm:cox} If $\tau>\tau_{0}$, there exists
some constant $C_{\rho}$ and $C_{\lambda}$ such that if $s\sqrt{\dfrac{\log p}{n}}<C_{\rho}$
and $\lambda_{n}=C_{\lambda}\epsilon$, and $\Delta_{\mbU}>\sqrt{8c_{0}\nu^{2}\tau^{2}\epsilon}+(2c_{0}\nu+1)\tau^{2}\epsilon$
with constant $c_{0}$ defined in \eqref{eq:thmspc}, then there exists
a constant $C>0$ that does not depend on $n$, $r$, $p$ or $\epsilon$
such that 
\begin{equation}
\Vert\mbP_{\hatcalE}-\mbP_{\calE}\Vert_{F}\le C\tau^{2}\epsilon
\end{equation}
with a probability greater than $1-Cp^{2}\exp(-Cn\epsilon^{2})-Cp\exp(-Cn\epsilon^{2}/s)$.
Furthermore, if ${{s\log p}=o({n})}$ and  that $\lambda_{n},\tau$ and $\Delta_{\mbU}\}$
satisfy $\max\left\{ \tau^{2}\sqrt{{s\log p}},\ \Delta_{\mbU}^{-2}\tau^{2}\sqrt{{s\log p}},\ \lambda_{n}^{-1}\sqrt{{\log p}}\right\} =o(\sqrt{n})$,
we have $
\Vert\mbP_{\hatcalE}-\mbP_{\calE}\Vert_{F}\rightarrow0$ in probability as $n,p\rightarrow\infty$.
\end{theorem}

These asymptotic and non-asymptotic properties for sparse NIECE in the Cox model are analogous to Theorem \ref{thm:glm} for logistic regression. Hence, we have illustrated the widely applicable Theorem~\ref{thm: NIECE_hd} of NIECE in high-dimensional multivariate analysis.

\section{Practical considerations} \label{sec:imple}

The theory and methods presented so far in this paper are enough to justify NIECE procedures' applications in high-dimensional settings. When implementing the algorithms, however, several additional practical issues arise. This section considers these practical issues and how to deal with them in a sensible data-driven fashion. Because NIECE is a flexible framework for envelope estimation in various settings, we only provide general guidance on dealing with these practical issues, while additional information from specific models and problems would also be helpful.

First, we consider the constrained and the projected envelope estimators. Recall from Section~\ref{sub:intro2est} that the envelope estimator for $\boltheta\in\mbbR^p$ is the projected estimator $\hatboltheta_{\Env}=\hatbolGamma\hatbolGamma\T\hatboltheta$ that improves over the standard estimator $\hatboltheta$, where $\hatbolGamma\in\mbbR^{p\times u}$ is the estimated envelope basis. In many multivariate parameter estimation problems \citep[see][for more background]{CookZhang2015foundation}, this projected envelope estimator coincides with the constrained envelope estimator, which is obtained from optimizing a likelihood-based objective function under the constraint $\boltheta = \hatbolGamma\T\boleta$ for some $\boleta\in\mbbR^u$. In low-dimensional settings, \citet{CookZhang2015foundation} showed that the constrained and the projected envelope estimators are often asymptotically equivalent if not exactly the same. However, in the high-dimension estimation, the two estimators could be very different. In our experience (e.g.~simulations in Section~\ref{sec:sim}), the constrained envelope estimator is generally more robust and accurate in parameter estimation than the projected estimator. As such, we suggest using the constrained envelope estimator in practice by refitting the model (linear, generalized linear, or Cox proportional hazard model) on the reduced data $\hatbolGamma\T\mbX\in\mbbR^u$ to obtain the parameters $\hatboleta$ and $\hatboltheta_{\Env}=\hatbolGamma\hatboleta$. We implemented this procedure in all of our numerical studies.

Another important issue in practice is the tuning parameter selection. The sparsity level in high-dimensional NIECE procedure arises from the penalized matrix decomposition step \eqref{SPC}, where we follow \citet{witten2009pmd} to use cross-validation to select the parameter $c>0$, which is the maximal $L_1$ norm of the sparse singular vectors. Specifically, we choose $c$ that minimizes the cross-validated prediction mean squared error in the linear model; and in the logistic regression model and Cox model, we choose $c$ that minimizes the cross-validated negative log-likelihood.


Finally, the envelope dimension selection problem is still an open question in high dimensions. Selecting the envelope dimension $u$ is a crucial step in all envelope methods. \citet{EnvDim} recently proposed a versatile BIC-type criterion to select $u$ consistently without restricting to a specific model. However, the theory and optimization techniques are only applicable to low-dimensional problems. This paper assumes the envelope dimension $u$ is known and leaves the dimension selection problem for future studies.  
In practice, the envelope dimension $u$ and the number of principal components $d$ may be selected by cross-validation, which is widely used in practice \citep{bro2008cross,josse2012selecting}. 
In simulation studies (Section~\ref{sec:sim}), we use the true dimension $u$ for our methods and others and compare their subspace and parameter estimation accuracies, where the number of principal components $d$ is fixed at a much bigger number than $u$. In real data analysis (Section~\ref{sec:real}), we consider tuning $d$ and $u$ together as $d=2u$ to reduce the computational cost. The idea behind this practice is to view the envelope as an alternative to principal component regression: Instead of reducing the data into $k$ principal components un-supervised, we choose $k$ envelope components from the first $2k$ principal components.


\section{Simulation studies} \label{sec:sim}

In this section, we study the empirical performances of the sample NIECE procedure (Section~\ref{sec:NIECE}) and the high-dimensional sparse NIECE (SNIECE; Section~\ref{sec:SNIECE}) procedure through simulations. We consider four envelope models in high dimensions (response and predictor envelopes in linear regression, envelope in logistic regression, and envelope in Cox proportional hazards model), where we reduce the $p$-dimensional predictor (or $r$-dimensional response) onto the $u$-dimensional envelopes without loss of relevant information in regression. In all simulation models, we set the sample size $n=200$, the envelope dimension $u=3$, and the dimension $p=400$ or $1600$. Note that for response reduction in the linear models, we let $p$ be the number of response variables and $q$ be the predictor dimension (in contrast to the $r$ and $p$ notation in the previous sections). The main goal is to estimate a three-dimensional sparse subspace in high dimensions, where we set the sparsity level $s=10$. Specifically, the key parameter is $\bolGamma = (\bolGamma_s\T,\mathbf{0}\T)\T \in \mbbR^{p \times u}$, where entries in $\bolGamma_s \in \mbbR^{s \times u}$ are generated from $\mathrm{Uniform}[0,1]$ and then orthogonalized such that $\bolGamma\T\bolGamma=\bolGamma_s\T\bolGamma_s=\mbI_u$. To introduce complex correlations among the $s$ relevant variables, we consider the following three types of envelope covariance structures that will be used in each of the four models.
\begin{itemize}
\item $\bolSigma_1=\mbV\mbD\mbV\T$, where $\mbV = (\mbv_1,\dots,\mbv_s) \in \mbbR^{s \times s}$ is a randomly generated orthogonal matrix and $\mbD=\mathrm{diag}(D_{11},\dots,D_{ss}) \in \mbbR^{s \times s}$ is a diagonal matrix consists of distinct eigenvalues. In multivariate linear models,  the $D_{kk} = (k+1)^3$, $k=1,\dots,s$; In logistic regression and Cox  model, $D_{kk} = 3^{k+1}$, $k=1,\dots,s$. The envelope is constructed from $\bolGamma_s = (\mbv_7,\mbv_8,\mbv_9)$.

\item $\bolSigma_2=\bolGamma_s\bolOmega\bolGamma_s\T+\bolGamma_{0s}\bolOmega_0\bolGamma_{0s}\T$, where $\bolGamma_{0s}\in\mbbR^{s\times (s-u)}$ is the orthogonal completion of $\bolGamma_s$ and $\bolOmega,\bolOmega_0$ are symmetric positive definite matrices. In multivariate linear models, we generate $\bolOmega$ as $\mbO\mbD\mbO\T$, where $\mbO$ is a randomly generated orthogonal matrix and $\mbD$ is diagonal with entries $(k+1)^3$, $k=1,\dots,u$; $\bolOmega_0$ is a diagonal matrix with entries $50,1,1,\dots,1$. In logistic regression and Cox  model, we change the entries in $\mbD$ to $(k+1)^2$, $k=1,\dots,u$, and the diagonals of $\bolOmega_0$ to $50,0.01,\dots,0.01$.

\item $\bolSigma_3$ is generated similar to $\bolSigma_2$. We set the eigenvalues in $\bolOmega$ (i.e.~diagonals of $\mbD$) as $(k+1)^2$, $k=1,\dots,u$ and let $\bolOmega_0=0.01 \mbI_{s-u}$.
\end{itemize}

The covariance $\bolSigma_1$ contains eigenvalues that are very well-separated, where the largest one is many magnitudes larger than the smallest ones. Both the covariances $\bolSigma_1$ and $\bolSigma_2$ are created such that the largest eigenvalue associated eigenvector is orthogonal to the envelope. Therefore, as seen in the specific models, the first principal component of the data is irrelevant to the regression but only brings substantial amounts of estimative variability. In contrast, the covariance $\bolSigma_3$ is created in favor of principal component regression -- the leading $u$ eigenvalues are well-separated and more than a hundred times larger than the remaining eigenvalues.
Then these covariance structures will be used in the following four models.

\begin{itemize}
\item[{\bf M1:}] {\bf Response envelope model in multivariate linear regression.}
For predictor dimension $q=10$, we generate the sparse coefficient parameter $\bolbeta^* \in \mbbR^{p\times q}$ as $\bolbeta^* = \bolGamma\boleta$ and then standardized to $\bolbeta=10\cdot \bolbeta^*/ \Vert \bolbeta^* \Vert_{F}$, where the entries in $\boleta \in \mbbR^{u \times q}$ are sampled from $\mathrm{Uniform}[0,1]$. The response is generated as $\mbY_i =  \bolbeta \mbX_i + \bolepsilon_i$, $i=1,\dots,n$, where the error $\bolepsilon_i\sim N({0},\dg(\bolSigma_A,\mbI_{p-s}))$ for $A=1,2,3$ three different covariance structures. The predictors are generated from $N({0},30\mbI_q)$ when $\bolSigma_A=\bolSigma_1$ due to high variance in $\bolSigma_1$, and are generated from $N({0},\mbI_q)$ for $\bolSigma_A=\bolSigma_2$ and $\bolSigma_3$.

\item[{\bf M2:}] {\bf Predictor envelope model in multivariate linear regression.}
For response dimension $q=5$, we generate $\bolbeta\in\mbbR^{p\times q}$ in the same way as in Model M1. Then $\mbY_i =  \bolbeta\T \mbX_i + \bolepsilon_i$, $i=1,\dots,n$, 
where the predictor $\mbX_i\sim N({0},\dg(\bolSigma_A,0.01 \mbI_{p-s}))$ and the error $\bolepsilon_i\sim N( {0},\sigma^2  \mbI_q)$. The magnitude of error is controlled by $\sigma^2$ to be comparable with the predictor variability. Specifically, $\sigma = 200, 20, 10$ for $A=1,2,3$ respectively.

\item[{\bf M3:}] {\bf Envelope logistic regression.} The sparse coefficient vector $\bolbeta =\bolGamma\boleta \in \mbbR^{p}$ is generated in the same way as in Model M1 where we set $q=1$. The predictor $\mbX_i$ is also generated in the same way except that we change $\bolSigma_A$ to its scaled version $\bolSigma_A/\Vert\bolSigma_A\Vert$, so that the classification accuracy by envelope logistic regression is around 20\% (neither too challenging nor too easy). The $Y_i$ follows Bernoulli distribution with probability $\exp(\bolbeta\T\mbX_i)/(1+\exp(\bolbeta\T\mbX_i))$.

\item[{\bf M4:}] {\bf Envelope Cox proportional hazards model.} The procedure for generating  $\bolbeta\in\mbbR^p$ and $\mbX_i$ are the same as in Model M3. Then the failure time $Y_i$ and censoring time $C_i$ are generated from exponential distribution $\mathrm{Exp}(\bolbeta\T\mbX_i)$ and $\mathrm{Exp}(0.5)$, respectively. Finally, we let $T_i = \min(Y_i,C_i)$ and $\delta_i = I(Y_i \leq C_i)$.

\end{itemize}

We focus on the parameter estimation error $\Vert \bolbeta - \hatbolbeta \Vert_{F}$ and the subspace estimation error $\Vert \mbP_{\bolGamma}-\mbP_{\hatbolGamma} \Vert_{F}/\sqrt{2u}\in[0,1]$. For comparison purpose, we focus on principal component regression \citep[PCR,][]{jolliffe1986principal} and sparse principal component regression (SPCR) based on the same penalized matrix decomposition algorithm \citep{witten2009pmd} that was used in SNIECE. We also include partial least squares regression \citep[PLS,][]{wold1966estimation}, sparse partial least square regression \citep[SPLS,][]{chun2010sparse}, sparse reduced-rank regression \citep[SRRR,][]{chen2012sparse}, reduced rank stochastic regression \citep[RSSVD,][]{chen2012reduced}. In multivariate linear models, we also include ordinary least squares (OLS) and Lasso \citep{tibshirani1996regression} estimators as benchmarks for estimating regression parameter $\bolbeta$; in logistic regression model and Cox proportional hazards model, we include $\ell_1$-penalized MLE \citep[PMLE,][]{friedman2010regularization} for comparison while PLS and SPLS are not directly applicable. In all models, we set $d=10>u=3$ in the NIECE and SNIECE procedures. In our experience, moderately increase $d$ will not change the performance of our methods under these simulation settings.

\begin{table}[ht!]
	\ra{0.95}
	\setlength{\tabcolsep}{3.7pt}
	\centering
	\begin{footnotesize}
		\begin{tabular}[t]{@{}ccccccccccccccccccc@{}}
			\toprule
			&&\multicolumn{8}{c}{{\bf M1} (Response envelope in linear model)} &&\multicolumn{8}{c}{{\bf M2} (Predictor envelope in linear model)} \\
			\cmidrule{3-10} \cmidrule{12-19}
			\multicolumn{2}{c}{$\bolSigma_A$} &\multicolumn{2}{c}{$\bolSigma_1$} && \multicolumn{2}{c}{$\bolSigma_2$} &&\multicolumn{2}{c}{$\bolSigma_3$} &&\multicolumn{2}{c}{$\bolSigma_1$} && \multicolumn{2}{c}{$\bolSigma_2$} &&\multicolumn{2}{c}{$\bolSigma_3$}\\
			\cmidrule{3-4} \cmidrule{6-7} \cmidrule{9-10} \cmidrule{12-13} \cmidrule{15-16} \cmidrule{18-19}
			\multicolumn{2}{c}{$p$} &$400$ &$1600$ &&$400$ &$1600$ &&$400$ &$1600$ &&$400$ &$1600$ &&$400$ &$1600$ &&$400$ &$1600$ \\
			\cmidrule{3-10} \cmidrule{12-19}
			OLS &$\Delta_{\bolbeta}$ &2.89 &3.24 &&5.38 &9.63 &&4.72 &9.26 &&-- &-- &&-- &-- &&-- &--\\
			\cmidrule{3-10} \cmidrule{12-19}
			Lasso &$\Delta_{\bolbeta}$ &-- &-- &&-- &-- &&-- &-- &&26.18 &25.48 &&7.89 &7.79 &&12.88 &12.70\\
			\cmidrule{3-10} \cmidrule{12-19}
			\multirow{2}{*}{PLS} &$\Delta_{\bolbeta}$ &2.80 &2.90 &&3.69 &5.55 &&3.03 &5.19 &&2.96 &3.48 &&5.43 &8.02 &&6.06 &8.47\\
			&$\Delta_{\bolGamma}$ &0.28 &0.32 &&0.50 &0.67 &&0.51 &0.69 &&0.39 &0.43 &&0.44 &0.63 &&0.54 &0.79\\
			\cmidrule{3-10} \cmidrule{12-19}
			\multirow{2}{*}{SRRR} &$\Delta_{\bolbeta}$ &2.42 &2.54 &&3.61 &5.63 &&2.74 &5.11 &&-- &-- &&38.91 &23.23 &&21.06 &14.93\\
			&$\Delta_{\bolGamma}$ &0.26 &0.30 &&0.47 &0.65 &&0.47 &0.66 &&1.00 &1.00 &&0.97 &0.94 &&0.93 &0.89\\
			\cmidrule{3-10} \cmidrule{12-19}
			\multirow{2}{*}{RSSVD} &$\Delta_{\bolbeta}$ &2.41 &2.41 &&2.65 &2.65 &&1.36 &1.39 &&-- &-- &&8.58 &5.95 &&10.27 &9.85\\
			&$\Delta_{\bolGamma}$ &0.24 &0.24 &&0.26 &0.27 &&0.11 &0.12 &&-- &-- &&0.74 &0.61 &&0.71 &0.70\\
			\cmidrule{3-10} \cmidrule{12-19}
			\multirow{2}{*}{PCR} &$\Delta_{\bolbeta}$ &3.20 &3.23 &&3.80 &4.57 &&2.22 &3.91 &&6.98 &7.04 &&8.40 &8.66 &&6.49 &8.92\\
			&$\Delta_{\bolGamma}$ &0.58 &0.58 &&0.59 &0.62 &&0.31 &0.53 &&0.58 &0.59 &&0.61 &0.67 &&0.54 &0.79\\
			\cmidrule{3-10} \cmidrule{12-19}
			\multirow{2}{*}{NIECE} &$\Delta_{\bolbeta}$ &2.36 &2.40 &&2.99 &4.08 &&2.22 &3.91 &&2.86 &3.07 &&4.78 &7.16 &&6.49 &8.87\\
			&$\Delta_{\bolGamma}$ &0.20 &0.21 &&0.24 &0.41 &&0.31 &0.53 &&0.22 &0.24 &&0.38 &0.59 &&0.54 &0.79\\
			\cmidrule{3-10} \cmidrule{12-19}
			\multirow{2}{*}{SPLS} &$\Delta_{\bolbeta}$ &-- &-- &&-- &-- &&-- &-- &&3.02 &3.06 &&3.80 &2.93 &&2.65 &3.85\\
			&$\Delta_{\bolGamma}$ &-- &-- &&-- &-- &&-- &-- &&0.60 &0.71 &&0.31 &0.23 &&0.73 &0.74\\
			\cmidrule{3-10} \cmidrule{12-19}
			\multirow{2}{*}{SPCR} &$\Delta_{\bolbeta}$ &3.21 &3.21 &&3.50 &3.51 &&1.30 &1.34 &&7.00 &7.05 &&8.51 &8.80 &&1.75 &2.00\\
			&$\Delta_{\bolGamma}$ &0.58 &0.59 &&0.60 &0.65 &&0.06 &0.07 &&0.59 &0.59 &&0.63 &0.70 &&0.13 &0.15\\
			\cmidrule{3-10} \cmidrule{12-19}
			\multirow{2}{*}{SNIECE} &$\Delta_{\bolbeta}$ &2.35 &2.37 &&2.62 &2.63 &&1.30 &1.34 &&2.83 &2.88 &&2.10 &2.24 &&1.75  &1.89\\
			&$\Delta_{\bolGamma}$ &0.19 &0.19 &&0.15 &0.15 &&0.06 &0.07 &&0.20 &0.21 &&0.17 &0.17 &&0.13 &0.14\\
			\cmidrule{3-10} \cmidrule{12-19}
			\multirow{2}{*}{S.E.$\leq$} &$\Delta_{\bolbeta}$ &(0.01) &(0.02) &&(0.01) &(0.01) &&(0.02) &(0.01) &&(0.01) &(0.01) &&(0.02) &(0.01) &&(0.06) &(0.06)\\
			&$\Delta_{\bolGamma}$ &(0.01) &(0.01) &&(0.01) &(0.01) &&(0.01) &(0.01) &&(0.01) &(0.01) &&(0.01) &(0.01) &&(0.02) &(0.01)\\
			\bottomrule
		\end{tabular}
	\end{footnotesize}
	\centering
	\begin{footnotesize}
		\begin{tabular}[t]{@{}ccccccccccccccccccc@{}}
			&&\multicolumn{8}{c}{{\bf M3} (Logistic regression)} &&\multicolumn{8}{c}{{\bf M4} (Cox hazards model)} \\
			\cmidrule{3-10} \cmidrule{12-19}
			\multicolumn{2}{c}{$\bolSigma_A$} &\multicolumn{2}{c}{$\bolSigma_1$} && \multicolumn{2}{c}{$\bolSigma_2$} &&\multicolumn{2}{c}{$\bolSigma_3$} &&\multicolumn{2}{c}{$\bolSigma_1$} && \multicolumn{2}{c}{$\bolSigma_2$} &&\multicolumn{2}{c}{$\bolSigma_3$}\\
			\cmidrule{3-4} \cmidrule{6-7} \cmidrule{9-10} \cmidrule{12-13} \cmidrule{15-16} \cmidrule{18-19}
			\multicolumn{2}{c}{$p$} &$400$ &$1600$ &&$400$ &$1600$ &&$400$ &$1600$ &&$400$ &$1600$ &&$400$ &$1600$ &&$400$ &$1600$ \\
			\cmidrule{3-10} \cmidrule{12-19}
			PMLE &$\Delta_{\bolbeta}$ &7.13 &7.35 &&8.13 &8.20 &&9.25 &9.17 &&5.46 &5.72 &&7.96 &7.96 &&10.48 &10.69\\
			\cmidrule{3-10} \cmidrule{12-19}
			\multirow{2}{*}{PCR} &$\Delta_{\bolbeta}$ &5.80 &7.66 &&9.35 &9.48 &&3.17 &5.28 &&6.00 &6.61 &&9.66 &9.68 &&1.40 &2.61\\
			&$\Delta_{\bolGamma}$ &0.65 &0.76 &&0.63 &0.73 &&0.23 &0.42 &&0.59 &0.63 &&0.59 &0.62 &&0.10 &0.20\\
			\cmidrule{3-10} \cmidrule{12-19}
			\multirow{2}{*}{NIECE} &$\Delta_{\bolbeta}$ &5.27 &7.63 &&5.07 &7.73 &&3.19 &5.28 &&2.53 &4.36 &&2.48 &4.35 &&1.40 &2.61\\
			&$\Delta_{\bolGamma}$ &0.54 &0.75 &&0.39 &0.64 &&0.23 &0.42 &&0.24 &0.43 &&0.18 &0.33 &&0.10 &0.20\\
			\cmidrule{3-10} \cmidrule{12-19}
			\multirow{2}{*}{SPCR} &$\Delta_{\bolbeta}$ &3.92 &3.91 &&9.68 &9.53 &&1.81 &1.63 &&5.68 &5.66 &&9.71 &9.71 &&0.80 &0.76\\
			&$\Delta_{\bolGamma}$ &0.59 &0.59 &&0.58 &0.58 &&0.08 &0.08 &&0.58 &0.58 &&0.58 &0.58 &&0.05 &0.05\\
			\cmidrule{3-10} \cmidrule{12-19}
			\multirow{2}{*}{SNIECE} &$\Delta_{\bolbeta}$ &2.23 &2.44 &&2.08 &2.07 &&1.72 &1.56 &&1.32 &1.23 &&1.25 &1.13 &&0.83  &0.78\\
			&$\Delta_{\bolGamma}$ &0.17 &0.19 &&0.13 &0.14 &&0.07 &0.07 &&0.10 &0.10 &&0.07 &0.07 &&0.05 &0.05\\
			\cmidrule{3-10} \cmidrule{12-19}
			{S.E.$\leq$} & &(0.02) &(0.02) &&(0.01) &(0.01) &&(0.02) &(0.02) &&(0.01) &(0.01) &&(0.01) &(0.01) &&(0.01) &(0.01)\\
			\bottomrule
		\end{tabular}
	\end{footnotesize}
		\caption{The median estimation errors for the parameter $\Delta_{\bolbeta} = \Vert \bolbeta-\hatbolbeta \Vert_{F}$ and for the envelope subspace $\Delta_{\bolGamma} = \Vert \mbP_{\bolGamma}-\mbP_{\hatbolGamma} \Vert_{F}/\sqrt{2u}$. Results are based on 200 replications. The maximum standard error among all estimators in each setting (i.e.~each column of the Table) are included. In Model {M2} with covariance structure $\bolSigma_1$, SRRR had very big estimation errors ($\Delta_{\bolbeta} > 100$) and was excluded in comparison; similarly, RSSVD also failed due to extremely high variability in data.	\label{tb:linear}
}
\end{table}

Table~\ref{tb:linear} summarizes the results for all the four models across the three covariance structures. We have confirmed the following theoretical findings for linear, logistic, and Cox models through numerical experiments. (i) Under covariance structure $\bolSigma_3$, where the leading principal components span the envelope subspace, our NIECE procedure correctly identified these principal components as relevant information in regression. As a result, the NIECE estimator coincided with the PCR estimator; and the SNIECE coincided with the SPCR. (ii) Under covariance structures $\bolSigma_1$ and $\bolSigma_2$, the first principal component is orthogonal to the envelope, our NIECE procedure correctly identified the first principal component as irrelevant information in regression and eliminated it in subsequent estimation. As a result, the NIECE and SNIECE estimators substantially improved their unsupervised counterparts PCR and SPCR, respectively. The advantages were reflected in both subspace estimation and parameter estimation. (iii) In these high-dimensional sparse settings, both the SPCR (under covariance structure $\bolSigma_3$) and SNIECE converged much faster than their un-penalized counterparts (PCR and NIECE).

Although we did not establish the parameter estimation consistency of $\hatbolbeta_{\Env}$ in theoretical analysis, the overall simulation results showed that the NIECE estimators (penalized or un-penalized) had a promising performance. They outperformed the standard solutions such as OLS, Lasso in multiple linear regression, and $\ell_1$-penalized MLE in logistic and Cox models. The significant improvements are due to our design of highly correlated variables. On the other hand, while PCR and SPCR were widely used in such high correlation data applications, the NIECE and SNIECE can be viewed as practical and straightforward improvements. Finally, we have also seen significant improvements in SNIECE over popular multivariate regression methods such as SPLS, SRRR, and RSSVD. We want to remark that these simulations were designed to be very challenging and in favor of NIECE procedures to confirm their theoretical properties and potential advantages in a wide range of applications. The NIECE and those methods are not directly comparable because of their different focuses in the regression. For example, SRRR failed under Model M2 because it focused on imposing group sparsity on column vectors of the coefficient matrix $\bolbeta$ and failed to select the important predictors when there is a large variation in the error term. 


\section{Real data illustration} \label{sec:real}

In this section, we include three datasets to illustrate the various applications of the proposed high-dimensional sparse NIECE procedure in logistic regression and linear regression with either univariate response or multivariate response.  We focus on the predictive results of SNIECE in comparison to SPCA and $\ell_1$-penalized (generalized) linear model (i.e.~LASSO estimator). Following the practical suggestions in Section~\ref{sec:imple}, we use $d=2u$ principal components for each envelope dimension $u\in\{1,\dots,20\}$ in SNIECE. 

The first study is the meat property data from \citet{saebo2008st}, where they collected the Near-infrared (NIR) spectroscopy measurements and water, fat, protein compositions for $n=103$ meat samples. \citet{cook2013envelopes} took spectral measurements at every fourth wavelength between $850$ nm and $1050$ nm as predictors, yielding $p = 50$; and they showed promising performance of the predictor envelope in the linear regression model to predict the protein content. We use the spectral measurements at every two wavelengths as predictors, yielding a higher predictor dimension of $p=100$. We then use this to illustrate the envelope logistic regression model with NIECE to classify the $n=103$ meat samples into $49$ beef samples and $54$ pork samples. The binary response and highly correlated predictors brought additional challenges than the previous analysis. 
The second study is the riboflavin production data from \citet{buhlmann2014high}. It contains the riboflavin production of $n=71$ Bacillus subtilis samples. The univariate continuous response is log-transformed riboflavin production; the high-dimensional predictor variables measure the logarithm of the expression level of $p=4088$ genes. 
The third study is the music data from \citet{zhou2014predicting} and downloaded from the UCI machine learning repository (\url{https://archive.ics.uci.edu/ml/datasets/Geographical+Original+of+Music}). It includes $n=1059$ music tracks with $p=116$ audio features, and the $r=2$ bivariate response describes the origin location of a track, represented by standardized latitude and longitude.

\begin{figure}[t!]
	\begin{center}
		\centering
		\includegraphics[width=1\textwidth]{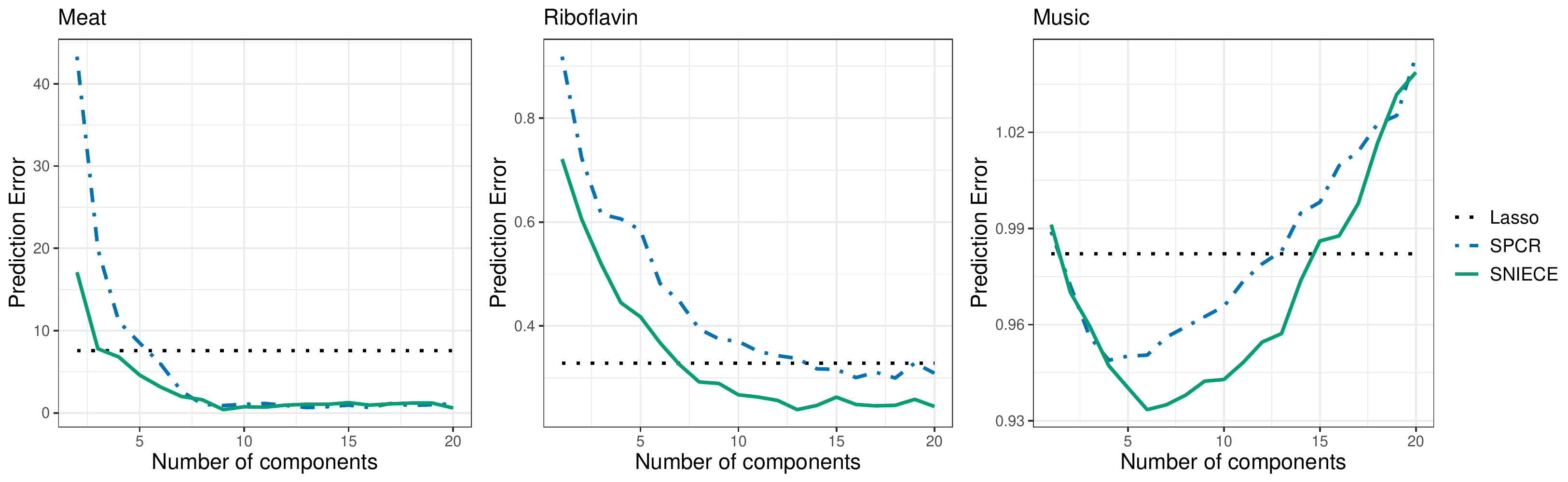}
		\caption{The average prediction errors using $100$ random data splittings where we vary the number of components in PCR and the envelope dimension in NIECE. For the meat data (left panel), the prediction error is defined as the misclassification rate in percentage; for the riboflavin data and the music data, the prediction error is the prediction mean squared error.}
		\label{fig:real_sparse}
	\end{center}
\end{figure}

For each dataset, we randomly split the data $100$ times into a training set of size $n_{\mathrm{train}}$ and a testing set of size $n_{\mathrm{test}}$, the averaged mis-classification error $\sum_{i=1}^{n_{\mathrm{test}}} I(Y_i\neq \widehat Y_i)/n_{\mathrm{test}}\times100\%$ or the averaged prediction mean squared error $\sum_{i=1}^{n_{\mathrm{test}}}\Vert \mbY_i-\hatmbY_i \Vert_2^2/n_{\mathrm{test}}$ is then recorded for each number of component $u$ ranges from $1$ to $20$. 
For the meat data and the riboflavin data, the training and testing sample size ratio is four to one (i.e.~five-fold cross-validation); for the music data, we set $n_{\mathrm{train}}=100$ to produce the high-dimension low-sample size scenario. 

Figure~\ref{fig:real_sparse} summarizes the prediction error. In the meat data, the NIR spectral measurements make the predictors extremely highly-correlated. Therefore, the classification errors of both SPCR and SNIECE decrease as we increase the number of components and eventually converge to nearly perfect classification. With a small number of components, the SNIECE is more effective than SPCR. This phenomenon can be explained by noticing that the first principal component is not useful for classification at all -- a situation resembles our simulations. For this data set, the $\ell_1$-penalized logistic regression ignores the predictor correlation and has an error rate of about $8\%$. 
In the riboflavin production data, the prediction error of SNIECE is uniformly smaller than that of SPCR for the whole range of dimensions. For this very high-dimensional data with much weaker correlations than in the meat data, the envelope approach still significantly improved over the lasso regression while SPCR fails to achieve so. These are very encouraging results and, to the best of our knowledge, the first real data application of envelope regression with $p$ in thousands. Finally, in the music data, the prediction error curves of SNIECE and SPCA in Figure~\ref{fig:real_sparse} is another typical situation: the SNIECE can achieve a much smaller error than SPCA and is also uniformly better than both SPCA and Lasso regression for a wide range of dimensions, $2\leq u\leq 14$. Based on the results in Figure~\ref{fig:real_sparse}, we have confirmed the potential advantages of NIECE over PCR in the high-dimensional setting. The proposed estimator can be widely adopted as a simple, unified, and effective alternative to SPCR.

\newpage

\vskip 0.5in
\appendix
\setcounter{equation}{0} 
\global\long\def\theequation{S\arabic{equation}}%
\global\long\def\thetheorem{S\arabic{theorem}}%
\global\long\def\thesection{S\arabic{section}}%
\global\long\def\thetable{S\arabic{table}}%
\global\long\def\thefigure{S\arabic{figure}}%
\renewcommand{\thelemma}{S\arabic{lemma}}
\renewcommand{\theproposition}{S\arabic{proposition}}

\begin{center}
	{\Large Supplementary Materials for\\ \centering ``Envelopes and Principal Component Regression''}
\end{center}

\section{Additional numerical results}
Furthermore, it is common in PCA to choose number of components that can explain a specific proportion of variability in data. Thus in meat property data, we also compare the classification performance by first applying PCA and sparse PCA on the original data and observe the number of components $u_1$ needed for PCA and $u_2$ needed for sparse PCA, to explain $90\%$, $95\%$ and $99\%$ of total variance. The comparison results are summarized in Table \ref{tb:meat}, to explain the same amount of variation in data, SNIECE achieves the lowest classification error.

\begin{table}[ht!]
	\ra{1}
	\centering
	\begin{footnotesize}
		\begin{tabular}[t]{@{}cccccccc@{}}
			\toprule
			Var. &$u_1$ &$u_2$ &PMLE &PCR &NIECE &SPCR &SNIECE\\ 
			\midrule
			\multirow{2}{*}{90\%} &\multirow{2}{*}{2} &\multirow{2}{*}{5}&7.55 &40.15 &34.90 &8.50 &5.90 \\
			&&&(0.61) &(1.06) &(1.64) &(0.61) &(0.51) \\
			\midrule
			\multirow{2}{*}{95\%} &\multirow{2}{*}{2} &\multirow{2}{*}{7}&7.55 &40.15 &34.90 &2.65 &2.15 \\
			&&&(0.61) &(1.06) &(1.64) &(0.39) &(0.37) \\
			\midrule
			\multirow{2}{*}{99\%} &\multirow{2}{*}{3} &\multirow{2}{*}{12}&7.55 &8.20 &8.20 &0.90 &0.55 \\
			&&&(0.61) &(0.51) &(0.51) &(0.22) &(0.17) \\				
			\bottomrule
		\end{tabular}
	\end{footnotesize}
	\caption{Envelope logistic regression on meat property data. Reported are the mean and standard error (in parenthesis) of mis-classification rates (\%) for 100 random data splits. {\textsl Var.} denotes total variance explained. $u_1$ is number of components for PCR and NIECE, $u_2$ is number of components for SPCR and SNIECE. \label{tb:meat}}
	
\end{table}

\section{Proofs for Lemmas \ref{lem: norm_vs_angle}--\ref{lemma: SPC}}

\subsection{Proof for Lemma \ref{lem: norm_vs_angle}}

For vectors, a natural measure of the distance between two vectors
is 
\begin{equation}
	\sin\Theta(\hatbolgamma_{j},\bolgamma_{j})=\sin\{\angle(\hatbolgamma_{j},\bolgamma_{j})\}=\sqrt{1-(\hatbolgamma_{j}\T\bolgamma_{j})^{2}},
\end{equation}
which in fact is equivalent to the Frobenius norm of the projection
matrices $\Vert\mbP_{\widehat{\bolgamma}_{j}}-\mbP_{\bolgamma_{j}}\Vert_{F}$.
To see this, 
\begin{eqnarray*}
	\Vert\mbP_{\widehat{\bolgamma}_{j}}-\mbP_{\bolgamma_{j}}\Vert_{F}^{2} & = & \tr\{(\mbP_{\widehat{\bolgamma}_{j}}-\mbP_{\bolgamma_{j}})\T(\mbP_{\widehat{\bolgamma}_{j}}-\mbP_{\bolgamma_{j}})\}\\
	& = & 2-2\tr\{\mbP_{\bolgamma_{j}}\mbP_{\hatbolgamma_{j}}\}\\
	& = & 2-2(\hatbolgamma_{j}\T\bolgamma_{j})^{2}\\
	& = & 2\{1-(\hatbolgamma_{j}\T\bolgamma_{j})^{2}\}.
\end{eqnarray*}
Therefore, for vectors, 
\begin{equation}
	\sqrt{2}\sin\Theta(\hatbolgamma_{j},\bolgamma_{j})=\Vert\mbP_{\widehat{\bolgamma}_{j}}-\mbP_{\bolgamma_{j}}\Vert_{F}.
\end{equation}

For matrices, a natural measure of the distance between two subspaces
is 
\begin{eqnarray*}
	\Vert\mbP_{\widehat{\calE}}-\mbP_{\calE}\Vert_{F} & = & \Vert\hatbolGamma\hatbolGamma\T-\bolGamma\bolGamma\T\Vert_{F}\\
	& = & \sqrt{\tr\{\hatbolGamma\hatbolGamma\T+\bolGamma\bolGamma\T-2\hatbolGamma\hatbolGamma\T\bolGamma\bolGamma\T\}}\\
	& = & \sqrt{2u-2\tr\{\hatbolGamma\T\bolGamma\cdot\bolGamma\T\hatbolGamma\}}\\
	& = & \sqrt{2\sum_{j=1}^{u}\{1-\lambda_{j}^{2}(\hatbolGamma\T\bolGamma)\}},
\end{eqnarray*}
which equals to $\sqrt{2}\sqrt{\sum_{j=1}^{u}\sin^{2}(\theta_{j})}$ with $\theta_{j}$ being the $j$-th principal angle between
the two subspace and the $j$-th diagonal element of $\bolTheta(\hatbolGamma,\bolGamma)$.
Therefore, we have $\Vert\mbP_{\widehat{\calE}}-\mbP_{\calE}\Vert_{F}=\sqrt{2}\Vert\sin\bolTheta(\hatbolGamma,\bolGamma)\Vert_{F}$.

\subsection{Proof for Lemma \ref{lem:assumption}}
The Lemma is straightforward to prove by applying Proposition~2.2 in \citet{Cook2010envelope}. We thus omit the proof.

\subsection{Proof for Lemma \ref{lemma: SPC}}

Since there is no penalty on $\mbu$, we can re-write the optimization
in \eqref{SPC} equivalently as,
\begin{eqnarray*}
	\hatmbu_{k} & = & \frac{\mbX^{k}\hatmbv_{k}}{\Vert\mbX^{k}\hatmbv_{k}\Vert_{2}},\\
	\hatmbv_{k} & = & \arg\max_{\mbv}\hatmbu_{k}\T\mbX^{k}\mbv\quad\mathrm{subject\ to}\ \Vert\mbv\Vert_{1}\leq c\ \mathrm{and}\ \Vert\mbv\Vert_{2}^{2}\leq1,
\end{eqnarray*}
from which we can plug in $\hatmbu_{k}$ into the optimization of
$\mbv$ and get
\begin{eqnarray*}
	\hatmbv_{k} & = & \arg\max_{\mbv}\frac{\mbv\T(\mbX^{k})\T\mbX^{k}\mbv}{\Vert\mbX^{k}\mbv\Vert_{2}}\quad\mathrm{subject\ to}\ \Vert\mbv\Vert_{1}\leq c\ \mathrm{and}\ \Vert\mbv\Vert_{2}^{2}\leq1\\
	& = & \arg\max_{\mbv}\mbv\T(\mbX^{k})\T\mbX^{k}\mbv\quad\mathrm{subject\ to}\ \Vert\mbv\Vert_{1}\leq c\ \mathrm{and}\ \Vert\mbv\Vert_{2}^{2}\leq1,
\end{eqnarray*}
where the last equality is because maximizing $\mbv\T(\mbX^{k})\T\mbX^{k}\mbv/\Vert\mbX^{k}\mbv\Vert_{2}$
is equivalent to maximizing $\left\{ \mbv\T(\mbX^{k})\T\mbX^{k}\mbv/\Vert\mbX^{k}\mbv\Vert_{2}\right\} ^{2}=\mbv\T(\mbX^{k})\T\mbX^{k}\mbv$.
Therefore, the optimization in \eqref{SPC} is equivalent to
\begin{equation}
	\hatmbv_{k}=\arg\max_{\mbv}\mbv\T\hatmbM_{k}\mbv\quad\mathrm{subject\ to}\ \Vert\mbv\Vert_{1}\leq c\ \mathrm{and}\ \Vert\mbv\Vert_{2}^{2}\leq1,
\end{equation}
where $\hatmbM_{k}=(\mbX^{k})\T\mbX^{k}$. 

Next, we have $\widehat{\sigma}_{k}=\hatmbu_{k}\T\mbX^{k}\hatmbv_{k}=\hatmbv_{k}\T(\mbX^{k})\T\hatmbu_{k}=\hatmbv_{k}\T(\mbX^{k})\T\mbX^{k}\hatmbv_{k}/\Vert\mbX^{k}\hatmbv_{k}\Vert_{2}$. Thus $\mbX^{k+1}$ can be written as,
\begin{eqnarray*}
	\mbX^{k+1} & = & \mbX^{k}-\widehat{\sigma}_{k}\hatmbu_{k}\hatmbv_{k}\T\\
	& = & \mbX^{k}-\dfrac{\hatmbv_{k}\T(\mbX^{k})\T\mbX^{k}\hatmbv_{k}\cdot\hatmbu_{k}\hatmbv_{k}\T}{\Vert\mbX^{k}\hatmbv_{k}\Vert_{2}}\\
	& = & \mbX^{k}-\dfrac{\hatmbv_{k}\T(\mbX^{k})\T\mbX^{k}\hatmbv_{k}\cdot\mbX^{k}\hatmbv_{k}\hatmbv_{k}\T}{\Vert\mbX^{k}\hatmbv_{k}\Vert_{2}^{2}}\\
	& = & \mbX^{k}(\mbI_{p}-\hatmbv_{k}\hatmbv_{k}\T).
\end{eqnarray*}
This implies that we can also write $\hatmbM_{k+1}$ as
\begin{equation}
	\hatmbM_{k+1}=(\mbX^{k+1})\T\mbX^{k+1}=(\mbI_{p}-\hatmbv_{k}\hatmbv_{k}\T)\hatmbM_{k}(\mbI_{p}-\hatmbv_{k}\hatmbv_{k}\T).
\end{equation}

\section{A brief discussion on common eigenvalues}

If the eigenvalues $\lambda_{\pi(j)}$'s, $j=1,\dots,u$, are all
distinct, we define the following population quantities
\begin{equation}
	\Delta=\min_{j=1,\dots,u}\Delta_{j},\quad\Delta_{j}\equiv\min\{\lambda_{\pi(j)-1}(\mbM)-\lambda_{\pi(j)}(\mbM),\ \lambda_{\pi(j)}(\mbM)-\lambda_{\pi(j)+1}(\mbM)\}.
\end{equation}
In presence of common eigenvalues for some $j=1,\dots,u$, denoted
as $\lambda_{\pi_{1}}=\cdots=\lambda_{\pi_{m}}$ with $\pi_{1}>\cdots>\pi_{m}$,
$m\leq u$, we re-define $\Delta_{j}$ as
\begin{equation}
	\Delta_{j}\equiv\min\{\lambda_{\pi_{1}-1}(\mbM)-\lambda_{\pi_{1}}(\mbM),\ \lambda_{\pi_{m}}(\mbM)-\lambda_{\pi_{m}+1}(\mbM)\}.
\end{equation}
Note that we need to assume these eigenvalues are distinct from the
eigenvalues associated with eigenvectors in $\calE_{\mbM}^{\perp}(\mbU)$
(an extreme case will be $\mbM=\mbI$, then the NIECE algorithm needs
modification), otherwise we need to modify/replace $\mbM$ by $\mbM+\mbU$.

\begin{lemma}\label{lem:eigenspace} Assume $\Delta>0$, and let $\epsilon>0$ be a constant
	such that $\Vert\hatmbM-\mbM\Vert_{\mathrm{op}}\leq\epsilon$, assume
	that $\{\widehat{\pi}(j)\mid j=1,\dots,u\}=\{\pi(j)\mid j=1,\dots,u\}$,
	then
	\begin{equation}
		\Vert\sin\bolTheta(\bolGamma,\hatbolGamma)\Vert_{F}\leq\frac{2\sqrt{u}\epsilon}{\Delta}.
	\end{equation}

\end{lemma}

\begin{proof}
	
	From Corollary 1 of \citet{YWS15}, we have that for distint eigenvalues,
	\begin{equation}
		\sin\Theta(\mbv_{j},\hatmbv_{j})\leq\frac{2\Vert\hatmbM-\mbM\Vert_{\mathrm{op}}}{\min(\lambda_{j-1}-\lambda_{j},\lambda_{j}-\lambda_{j+1})}\leq\frac{2\Vert\hatmbM-\mbM\Vert_{\mathrm{op}}}{\Delta_{j}}\leq\frac{2\Vert\hatmbM-\mbM\Vert_{\mathrm{op}}}{\Delta}.
	\end{equation}
	For common eigenvalues, $\lambda_{\pi_{1}}=\cdots=\lambda_{\pi_{m}}$
	with $\pi_{1}>\cdots>\pi_{m}$, we let $\mbW_{j}=(\mbv_{\pi_{1}},\dots,\mbv_{\pi_{m}})$
	and $\hatmbW_{j}=(\widehat{\mbv}_{\pi_{1}},\dots,\widehat{\mbv}_{\pi_{m}})$.
	Then we have,
	\begin{equation}
		\Vert\sin\bolTheta(\mbW_{j},\hatmbW_{j})\Vert_{F}\leq\frac{2\sqrt{m}\Vert\hatmbM-\mbM\Vert_{\mathrm{op}}}{\Delta_{j}}\leq\frac{2\sqrt{m}\Vert\hatmbM-\mbM\Vert_{\mathrm{op}}}{\Delta}.
	\end{equation}
	Since $\Vert\sin\bolTheta(\bolGamma,\hatbolGamma)\Vert_{F}^{2}=\sum_{j=1}^{u}\sin^{2}(\theta_{j})$,
	where $\theta_{j}$'s are the principal angles, we have
	\begin{equation}
		\Vert\sin\bolTheta(\bolGamma,\hatbolGamma)\Vert_{F}=\sqrt{\sum_{j\in\mathcal{J}_{1}}\sin^{2}\Theta(\mbv_{j},\hatmbv_{j})+\sum_{j\in\mathcal{J}_{2}}\Vert\sin\bolTheta(\mbW_{j},\hatmbW_{j})\Vert_{F}^{2}}\leq\frac{2\sqrt{u}\Vert\hatmbM-\mbM\Vert_{\mathrm{op}}}{\Delta},
	\end{equation}
	where the index sets $\mathcal{J}_{1}$ is for distinct eigenvalues
	and index set $\mathcal{J}_{2}$ is for common eigenvalues. The conclusion
	follows from $\Vert\hatmbM-\mbM\Vert_{\mathrm{op}}\leq\epsilon$.
	
\end{proof}

\section{Proofs for Theorem \ref{thm: basic_niece}}

Throughout this subsection, we assume that $\Vert\widehat{\mbM}-\mbM\Vert_{op}\le\epsilon,\Vert\widehat{\mbU}-\mbU\Vert_{op}\le\epsilon$. First, we introduce two technical Lemmas that are intermediate results for proving Theroem~\ref{thm: basic_niece}. Recall that we have defined the envelope scores $\widehat{\phi}_{j}=\hatmbv_{j}\T\hatmbU\hatmbv_{j}$
for $j=1,\dots,d$, and organize them in descending order $\widehat{\phi}_{(1)}\geq\cdots\geq\widehat{\phi}_{(d)}$
and define $\hatmbv_{(j)}$ correspondingly from $\hatphi_{(j)}=\hatmbv_{(j)}\T\hatmbU\hatmbv_{(j)}$.



\begin{lemma}\label{lem: sini} If $\Delta>0$, then $\sin\Theta(\hatmbv_{i},\mbv_{i})\leq\dfrac{2\epsilon}{\Delta},\quad i=1,\dots,d.$
\end{lemma}

\begin{proof} This result is a direct consequence of Corollary~1
	in \citet{YWS15}. 
\end{proof}
\begin{lemma}\label{lem: gammai} For $i=1,\ldots,d$, we have the
	following conclusions: 
	\begin{enumerate}
		\item If $\Vert\hatmbU-\mbU\Vert_{op}\le\epsilon$, $\epsilon\leq\Delta/4$,
		and $\sin\Theta(\hatmbv_{i},\mbv_{i})\le\dfrac{2\epsilon}{\Delta}$,
		then
		\begin{equation}
			\vert\widehat{\phi}_{i}-\phi_{i}\vert\leq\left(1+\dfrac{10\nu}{\Delta}\right)\epsilon,
		\end{equation}
		where $\nu=\Vert\mbU\Vert_{op}$ is the largest eigenvalue of $\mbU\geq0$.
		\item If $\Vert\hatmbU-\mbU\Vert_{\max}\le\epsilon$, $\sin^{2}\Theta(\hatmbv_{i},\mbv_{i})\leq c_{0}\tau^{2}\epsilon$
		and $\Vert\hatmbv_{i}\Vert_{1}\le\tau$, then
		\begin{equation}
			\vert\widehat{\phi}_{i}-\phi_{i}\vert\leq\left\{ (c_{0}+1)\tau^{2}+\dfrac{2\nu}{\Delta}\right\} \epsilon.
		\end{equation}
		
	\end{enumerate}
\end{lemma}

\begin{proof} First of all, note that, because $\sin\bolTheta(\hatmbv_{i},\mbv_{i})\le\frac{2\epsilon}{\Delta}$,
	we have
	\[
	\Vert\hatmbv_{i}-\mbv_{i}\Vert_{2}=\sqrt{2-2\cos\Theta(\hatmbv_{i},\mbv_{i})}\leq\sqrt{2-2\sqrt{1-\dfrac{4\epsilon^{2}}{\Delta^{2}}}}\leq\dfrac{2\sqrt{2}\epsilon}{\Delta}.
	\]
	Further notice that
	\begin{eqnarray}
		\vert\widehat{\phi}_{i}-\phi_{i}\vert & = & \vert\hatmbv_{i}\T\hatmbU\hatmbv_{i}-\mbv_{i}\T\mbU\mbv_{i}\vert\nonumber \\
		& \leq & |\hatmbv_{i}\T(\hatmbU-\mbU)\hatmbv_{i}|+|(\hat{\mbv}_{i}-\mbv_{i})\T\mbU(\hatmbv_{i}-\mbv_{i})|+2|\mbv_{i}\T\mbU(\hatmbv_{i}-\mbv_{i})|.\label{phi_diff_3terms}
	\end{eqnarray}
	For the last two terms, note that
	\begin{eqnarray}
		|(\hatmbv_{i}-\mbv_{i})\T\mbU(\hatmbv_{i}-\mbv_{i})|\leq\Vert\mbU\Vert_{op}\Vert\hatmbv_{i}-\mbv_{i}\Vert_{2}^{2} & \leq & 8\epsilon^{2}\nu/\Delta^{2}\leq2\epsilon\nu/\Delta,\\
		2|\mbv_{i}\T\mbU(\hatmbv_{i}-\mbv_{i})|\le2\sqrt{\mbv_{i}\T\mbU^{2}\mbv_{i}}\Vert\hatmbv_{i}-\mbv_{i}\Vert_{2} & \leq & 4\sqrt{2}\nu\epsilon/\Delta\leq8\epsilon\nu/\Delta,
	\end{eqnarray}
	where we have used $\epsilon/\Delta\leq1/4$.
	
	Similarly, if we assume that $\Vert\hatmbU-\mbU\Vert_{op}\le\epsilon$
	and $\sin^{2}\Theta(\hatmbv_{i},\mbv_{i})\leq c_{0}\tau^{2}\epsilon$.
	Then
	\[
	\Vert\hatmbv_{i}-\mbv_{i}\Vert_{\infty}^{2}\leq\Vert\hatmbv_{i}-\mbv_{i}\Vert_{2}^{2}=2-2\cos\Theta(\hatmbv_{i},\mbv_{i})\leq2-2\sqrt{1-c_{0}\tau^{2}\epsilon}=2\dfrac{c_{0}\tau^{2}\epsilon}{1+\sqrt{1-c_{0}\tau^{2}\epsilon}}\leq2c_{0}\tau^{2}\epsilon.
	\]
	Then the first term in \eqref{phi_diff_3terms} is
	\[
	|\hatmbv_{i}\T(\hatmbU-\mbU)\hatmbv_{i}|\le\Vert\hatmbU-\mbU\Vert_{\max}\Vert\hatmbv_{i}\Vert_{1}^{2}\leq\tau^{2}\epsilon.
	\]
	Then the last two terms in \eqref{phi_diff_3terms} become 
	\begin{eqnarray}
		|(\hatmbv_{i}-\mbv_{i})\T\mbU(\hatmbv_{i}-\mbv_{i})|\leq\Vert\mbU\Vert_{op}\Vert\hatmbv_{i}-\mbv_{i}\Vert_{2}^{2} & \leq & 2c_{0}\nu\tau^{2}\epsilon,\\
		2|\mbv_{i}\T\mbU(\hatmbv_{i}-\mbv_{i})|\le2\sqrt{\mbv_{i}\T\mbU^{2}\mbv_{i}}\Vert\hatmbv_{i}-\mbv_{i}\Vert_{2} & \leq & 2\nu\sqrt{2c_{0}\tau^{2}\epsilon}.
	\end{eqnarray}

	And we have the desired conclusions. 
\end{proof}

Next, to {prove Theorem \ref{thm: basic_niece}}, we apply these two Lemmas.
By Lemma~\ref{lem: gammai}, if $\Delta_{\mbU}>2(1+\dfrac{10\nu}{\Delta})\epsilon$,
we keep all the $\hatmbv_{i}$'s in the envelope. Combine this fact
with Lemma~\ref{lem: sini} and we we have $\vert\hatphi_{\pi(j)}-\phi_{\pi(j)}\vert<\Delta_{\mbU}/2$
and thus for $j=1,\dots,u,$ we have $\hatphi_{\pi(j)}>\phi_{\pi(j)}-\Delta_{\mbU}/2>\Delta_{\mbU}/2$,
and $0\leq\hatphi_{\pi(j)}<\Delta_{\mbU}/2$ for $j=u+1,\dots,d$.
Therefore we have selected the correct components $\widehat{\pi}(j)\in\{\pi(k)\vert k=1,\dots,u\}$
for all $j=1,\dots,u$. Therefore, 
\[
\Vert \mbP_{\widehat{\calE}}-\mbP_{\calE}\Vert_{F}=\sqrt{2}\sqrt{\sum_{i:\mbv_{i}\in\calE}\sin^{2}\bolTheta(\widehat{\mbv}_{i},\mbv_{i})}\le\dfrac{2\sqrt{2u}\epsilon}{\Delta}.
\]

\section{Proofs for Theorem \ref{thm.res.low} and Corollary \ref{cy.res.low}}

Note that the response and predictor envelopes are constructed in symmetry. The NIECE estimator is obtained by using the sample covariances $\hatmbM=\widehat\bolSigma_{\mbY}$ and $\hatmbU=\hatbolSigma_{\mbY\mbX}\hatbolSigma_{\mbX\mbY}$ for response reduction, and $\hatmbM=\widehat\bolSigma_{\mbX}$ and $\hatmbU=\hatbolSigma_{\mbX\mbY}\hatbolSigma_{\mbY\mbX}$ for predictor reduction. We therefore only lay out the proof for response envelope. For predictor envelope, we will need to interchange $p$ and $r$.

Recall from Theorem \ref{thm: basic_niece}, in order to show $\Vert\mbP_{\hatcalE}-\mbP_{\calE}\Vert_{F}\leq\dfrac{2\sqrt{2u}}{\Delta}\epsilon$, we need to verify
\begin{equation}
	\epsilon\leq\max\left\{ \dfrac{\Delta}{4},\!\dfrac{\Delta_{\mbU}}{2}\left(1+\dfrac{10\nu}{\Delta}\right)^{-1}\right\} ,\ \Vert\hatmbM-\mbM\Vert_{op}\leq\epsilon,\quad\Vert\hatmbU-\mbU\Vert_{op}\leq\epsilon.
\end{equation}
By the following Lemma~\ref{lem:res.low} and Theorem~\ref{thm: basic_niece},
and we have the desired conclusion.  Corollary~\ref{cy.res.low} follows from Theorem \ref{thm.res.low} by letting $\epsilon=pr\sqrt{\dfrac{\log p+\log r}{n}}\rightarrow0$.

\begin{lemma}\label{lem:res.low} 
	\begin{enumerate}
		\item For $\epsilon>0$, we have 
		\begin{equation}
			\Vert\hatbolSigma_{\mbY}-\bolSigma_{\mbY}\Vert_{op}\le\epsilon\label{res.low.eq1}
		\end{equation}
		with a probability greater than $1-Cr^{2}\exp(-C\dfrac{n\epsilon^{2}}{r^{2}})$. 
		\item For $0<\epsilon<\Vert\bolSigma_{\mbX\mbY}\Vert_{1}+\Vert\bolSigma_{\mbY\mbX}\Vert_{1}$,
		we have 
		\begin{equation}
			\Vert\hatbolSigma_{\mbY\mbX}\hatbolSigma_{\mbX\mbY}-\bolSigma_{\mbY\mbX}\bolSigma_{\mbX\mbY}\Vert_{op}\le2(\Vert\bolSigma_{\mbX\mbY}\Vert_{1}+\Vert\bolSigma_{\mbY\mbX}\Vert_{1})\epsilon\label{res.low.eq2}
		\end{equation}
		with a probability greater than $1-Cpr\exp(-C\dfrac{n\epsilon^{2}}{r^{2}})-Cpr\exp(-C\dfrac{n\epsilon^{2}}{p^{2}})$.
		In addition, if we assume $\Vert\bolSigma_{\mbX\mbY}\Vert_{1}+\Vert\bolSigma_{\mbY\mbX}\Vert_{1}<c_{1}$
		for some constant $c_{1}$, we can claim
		\begin{equation}
			\Vert\hatbolSigma_{\mbY\mbX}\hatbolSigma_{\mbX\mbY}-\bolSigma_{\mbY\mbX}\bolSigma_{\mbX\mbY}\Vert_{op}\le\epsilon\label{res.low.eq2-1}
		\end{equation}
		with a probability greater than $1-Cpr\exp(-C\dfrac{n\epsilon^{2}}{r^{2}})-Cpr\exp(-C\dfrac{n\epsilon^{2}}{p^{2}})$.
	\end{enumerate}
\end{lemma}

\begin{proof} We first show \eqref{res.low.eq1}. Note that 
	\begin{eqnarray}
		&  & \Pr(\Vert\hatbolSigma_{\mbY}-\bolSigma_{\mbY}\Vert_{op}>\epsilon)\le\Pr(\Vert\hatbolSigma_{\mbY}-\bolSigma_{\mbY}\Vert_{1}>\epsilon)\\
		& \le & \Pr(\cup_{i,j}|\hatsigma_{\mbY,ij}-\sigma_{\mbY,ij}|>\epsilon/r)\le\sum_{i,j}\Pr(|\hatsigma_{\mbY,ij}-\sigma_{\mbY,ij}|>\epsilon/r)\\
		& \le & r^{2}\exp(-Cn\dfrac{\epsilon^{2}}{r^{2}})
	\end{eqnarray}
	where the last inequality follows from Lemma \ref{lem:sampleCov}
	and joint sub-Gaussian of $(\mbX,\mbY)$ in \eqref{XYsubG}.
	
	Now we show \eqref{res.low.eq2}. Straightforward calculation shows
	that 
	\begin{eqnarray*}
		&  & \Vert\hatbolSigma_{\mbY\mbX}\hatbolSigma_{\mbX\mbY}-\hatbolSigma_{\mbY\mbX}\hatbolSigma_{\mbX\mbY}\Vert_{op}\\
		& \leq & \Vert\hatbolSigma_{\mbY\mbX}\hatbolSigma_{\mbX\mbY}-\hatbolSigma_{\mbY\mbX}\hatbolSigma_{\mbX\mbY}\Vert_{1}\\
		& \le & \Vert(\hatbolSigma_{\mbY\mbX}-\bolSigma_{\mbY\mbX})(\hatbolSigma_{\mbX\mbY}-\bolSigma_{\mbX\mbY})\Vert_{1}+\Vert(\hatbolSigma_{\mbY\mbX}-\bolSigma_{\mbY\mbX})\bolSigma_{\mbX\mbY}\Vert_{1}+\Vert\bolSigma_{\mbY\mbX}(\hatbolSigma_{\mbX\mbY}-\bolSigma_{\mbX\mbY})\Vert_{1}\\
		& \le & (\Vert\hatbolSigma_{\mbY\mbX}-\bolSigma_{\mbY\mbX}\Vert_{1}+\Vert\bolSigma_{\mbX\mbY}\Vert_{1}+\Vert\bolSigma_{\mbY\mbX}\Vert_{1})\max\{\Vert\hatbolSigma_{\mbX\mbY}-\bolSigma_{\mbX\mbY}\Vert_{1},\Vert\hatbolSigma_{\mbY\mbX}-\bolSigma_{\mbY\mbX}\Vert_{1}\}
	\end{eqnarray*}
	which is less than $2(\Vert\bolSigma_{\mbX\mbY}\Vert_{1}+\Vert\bolSigma_{\mbY\mbX}\Vert_{1})\epsilon$
	if we assume that 
	\begin{equation}
		\max\{\Vert\hatbolSigma_{\mbX\mbY}-\bolSigma_{\mbX\mbY}\Vert_{1},\Vert\hatbolSigma_{\mbY\mbX}-\bolSigma_{\mbY\mbX}\Vert_{1}\}\le\epsilon\leq\Vert\bolSigma_{\mbX\mbY}\Vert_{1}+\Vert\bolSigma_{\mbY\mbX}\Vert_{1}.
	\end{equation}
	Hence, it suffices to calculate the probability bound for $\max\{\Vert\hatbolSigma_{\mbX\mbY}-\bolSigma_{\mbX\mbY}\Vert_{1},\Vert\hatbolSigma_{\mbY\mbX}-\bolSigma_{\mbY\mbX}\Vert_{1}\}\le\epsilon$ as follows.
	\[
	\Pr(\Vert\hatbolSigma_{\mbX\mbY}-\bolSigma_{\mbX\mbY}\Vert_{1}>\epsilon)\le\Pr(\cup_{i,j}|\hatsigma_{\mbX\mbY,ij}-\sigma_{\mbX\mbY,ij}|>\epsilon/p)\le Cpr\exp(-Cn\dfrac{\epsilon^{2}}{p^{2}}).
	\]
	Similarly, $\Pr(\Vert\hatbolSigma_{\mbY\mbX}-\bolSigma_{\mbY\mbX}\Vert_{1}>\epsilon)\le Cpr\exp(-Cn\dfrac{\epsilon^{2}}{r^{2}})$.
	And the conclusion follows. \end{proof}

\section{Proofs for Theorems \ref{thm:SPC} and \ref{thm: NIECE_hd}}

We first introduce several technical lemmas, followed by proofs for
the two main theorems. 

We use $\lambda_{k}$ for the $k$-th eigenvalue of $\mbM$ and $\hatlambda_{k}=\hatmbv_{k}\T\hatmbM\hatmbv_{k}$,
which is not exactly the $k$-th eigenvalue of $\hatmbM$ but is a
good estimate for $\lambda_{k}$. We also define the following quantities
throughout this section,
\begin{eqnarray*}
	\tau_{0} & = & \max_{1\leq k\leq d}\Vert\mbv_{k}\Vert_{1}\\
	\eta_{k} & = & \Vert\hatmbv_{k}\Vert_{2}^{2},\quad\eta_{k}\leq1\\
	\tilmbv_{k} & = & \eta_{k}^{-1/2}\hatmbv_{k},\quad\Vert\tilmbv_{k}\Vert_{2}=1\\
	\eta_{0} & = & \min_{1\leq k\leq d}\eta_{k}.
\end{eqnarray*}
Recall that, for some $\epsilon>0$ and $\tau>0$, we assumed in Theorems
\ref{thm:SPC} and \ref{thm: NIECE_hd} that 
\begin{equation}
	\Vert\hatmbM-\mbM\Vert_{\max}\leq\epsilon,\quad\tau_{0}<\tau.
\end{equation}

\subsection{Technical Lemmas }

\begin{lemma} \label{lem:AngQuad} For any $\mbw\in\mbbR^{p}$ such
	that $\mbw\T\mbw=1$ we have 
	\begin{equation}
		\sin^{2}\Theta(\mbw,\mbv_{k})\leq\frac{\lambda_{k}-\mbw\T\mbM_{k}\mbw}{\Delta},\quad k=1,\dots,d.
	\end{equation}
	Morevoer, if $\sin^{2}\Theta(\mbw,\mbv_{k})=t$, then
	\begin{eqnarray*}
		\mbw\T\mbM_{k}\mbw & \leq & \lambda_{k}-\Delta t,\\
		\sin^{2}\Theta(\mbw,\mbv_{j}) & \geq & 1-2t,\quad j\neq k,\\
		(\mbw\T\mbv_{j})^{2} & \leq & 2t,\quad j\neq k,\\
		\mbw\T\mbM\mbw & \geq & \lambda_{k}(1-t).
	\end{eqnarray*}

\end{lemma}

\begin{proof} Recall that we can write $\mbM_{k}=\sum_{i=k}^{p}\lambda_{i}\mbv_{i}\mbv_{i}\T$,
	therefore,
	\begin{eqnarray*}
		\lambda_{k}-\mbw\T\mbM_{k}\mbw & = & \lambda_{k}-\sum_{i=k}^{p}\lambda_{i}(\mbw\T\mbv_{i})^{2}\\
		& = & \lambda_{k}-\sum_{i=k}^{p}\lambda_{i}\{1-\sin^{2}\Theta(\mbw,\mbv_{i})\}\\
		& = & \lambda_{k}-\lambda_{k}\{1-\sin^{2}\Theta(\mbw,\mbv_{k})\}-\sum_{i=k+1}^{p}\lambda_{i}\{1-\sin^{2}\Theta(\mbw,\mbv_{i})\}\\
		& = & \lambda_{k}\sin^{2}\Theta(\mbw,\mbv_{k})-\sum_{i=k+1}^{p}\lambda_{i}\{1-\sin^{2}\Theta(\mbw,\mbv_{i})\}.
	\end{eqnarray*}
	Because $1=\sum_{i=1}^{p}\{1-\sin^{2}\Theta(\mbw,\mbv_{i})\}$, $\sum_{i=k+1}^{p}\{1-\sin^{2}\Theta(\mbw,\mbv_{i})\}\leq\sin^{2}\Theta(\mbw,\mbv_{k})$.
	Therefore, we have,
	\begin{eqnarray*}
		\lambda_{k}-\mbw\T\mbM_{k}\mbw & = & \lambda_{k}\sin^{2}\Theta(\mbw,\mbv_{k})-\sum_{i=k+1}^{p}\lambda_{i}\{1-\sin^{2}\Theta(\mbw,\mbv_{i})\}\\
		& \geq & \lambda_{k}\sin^{2}\Theta(\mbw,\mbv_{k})-\lambda_{k+1}\sum_{i=k+1}^{p}\{1-\sin^{2}\Theta(\mbw,\mbv_{i})\}\\
		& \geq & \lambda_{k}\sin^{2}\Theta(\mbw,\mbv_{k})-\lambda_{k+1}\sin^{2}\Theta(\mbw,\mbv_{k})\\
		& \geq & \Delta\sin^{2}\Theta(\mbw,\mbv_{k}).
	\end{eqnarray*}
	Then we have $\sin^{2}\Theta(\mbw,\mbv_{k})\leq(\lambda_{k}-\mbw\T\mbM_{k}\mbw)/\Delta$
	and also $\mbw\T\mbM_{k}\mbw\leq\lambda_{k}-\Delta t,$ where $t=\sin^{2}\Theta(\mbw,\mbv_{k})$.
	
	For $j\neq k$, we write $\sin^{2}\Theta(\mbw,\mbv_{j})=1-(\mbw\T\mbv_{j})^{2}$,
	so it is equivalent to show that $(\mbw\T\mbv_{j})^{2}\leq2t$. Without
	loss of generality, suppose $\mbw\T\mbv_{k}\geq0$ (if $\mbw\T\mbv_{k}<0$
	then we can simply replace $\mbw-\mbv_{k}$ with $\mbw+\mbv_{k}$
	in the following derivation and reach the same conclusion). It follows
	from $\mbv_{k}\T\mbv_{j}=0$ that
	\begin{eqnarray*}
		(\mbw\T\mbv_{j})^{2} & = & \{(\mbw-\mbv_{k})\T\mbv_{j}\}^{2}\\
		& \leq & \{(\mbw-\mbv_{k})\T(\mbw-\mbv_{k})\}\{\mbv_{j}\T\mbv_{j}\}\\
		& = & (2-2\mbw\T\mbv_{k})\\
		& = & 2-2\sqrt{1-t}.
	\end{eqnarray*}
	It is then straightforward to verify $2-2\sqrt{1-t}\leq2t$ since
	$t=\sin^{2}\Theta(\mbw,\mbv_{k})\in[0,1]$, and the conclusion follows.
	
	To show that $\mbw\T\mbM\mbw\geq\lambda_{k}(1-t)$, we notice that
	\begin{eqnarray*}
		\mbw\T\mbM\mbw & = & \mbw\T(\sum_{j=1}^{p}\lambda_{j}\mbv_{j}\mbv_{j}\T)\mbw\\
		& \geq & \lambda_{k}(\mbw\T\mbv_{k})^{2}\\
		& = & \lambda_{k}\cos^{2}\Theta(\mbw,\mbv_{k})\\
		& = & \lambda_{k}(1-t).
	\end{eqnarray*}

\end{proof}

\begin{lemma}\label{lem: T0} For $k=1,\dots,d$, 
	\begin{equation}
		\vert\mbv_{k}\T\hatmbM\mbv_{k}-\lambda_{k}\vert\leq C\epsilon\tau_{0}^{2}\leq C\epsilon\tau^{2}.
	\end{equation}

\end{lemma}

\begin{proof}
	
	The conclusion follows immediately from the fact that $\vert\mbv_{k}\T\hatmbM\mbv_{k}-\lambda_{k}\vert=\vert\mbv_{k}\T(\hatmbM-\mbM)\mbv_{k}\vert\leq\Vert\mbv_{k}\Vert_{1}^{2}\Vert\hatmbM-\mbM\Vert_{\max}\leq\epsilon\tau_{0}^{2}\leq\epsilon\tau^{2}$.
	
\end{proof}

\begin{lemma}\label{lem: T1} If $\epsilon\tau^{2}<\min\{\dfrac{1}{4},\dfrac{\lambda_{1}}{4}\}$,
	then the following inequalities hold,
	\begin{eqnarray}
		\sin^{2}\Theta(\hatmbv_{1},\mbv_{1}) & \leq & 3\epsilon\tau^{2}/\Delta,\label{T11}\\
		\hatmbv_{1}\T\mbM\hatmbv_{1} & \geq & \lambda_{1}-2\epsilon\tau^{2},\label{T12}\\
		\eta_{1}=\Vert\hatmbv_{1}\Vert_{2}^{2} & \geq & 1-2\lambda_{1}^{-1}\epsilon\tau^{2},\label{T13}\\
		\vert\lambda_{1}-\hatlambda_{1}\vert & \leq & \epsilon\tau^{2},\label{T14}\\
		\Vert\hatmbv_{1}-\mbv_{1}\Vert_{2}^{2} & \leq & (2\lambda_{1}^{-1}+6)\epsilon\tau^{2}.\label{T15}
	\end{eqnarray}

\end{lemma}

\begin{proof}
	
	First, we prove \eqref{T13}.
	
	Since $\Vert\hatmbv_{1}\Vert_{2}^{2}=\eta_{1}\leq1$, we have 
	\begin{equation}
		\hatmbv_{1}\T\hatmbM\hatmbv_{1}=\hatmbv_{1}\T\mbM\hatmbv_{1}+\hatmbv_{1}\T(\hatmbM-\mbM)\hatmbv_{1}\leq\lambda_{1}\eta_{1}+\epsilon\tau^{2},
	\end{equation}
	where the inequality follows from $\hatmbv_{1}\T\mbM\hatmbv_{1}\leq\Vert\mbM\Vert_{\mathrm{op}}\Vert\hatmbv_{1}\Vert_{2}^{2}\leq\lambda_{1}\eta_{1}$
	and $\hatmbv_{1}\T(\hatmbM-\mbM)\hatmbv_{1}\leq\Vert\hatmbM-\mbM\Vert_{\max}\Vert\hatmbv_{1}\Vert_{1}^{2}\leq\epsilon\tau^{2}$.
	
	Analogously, we also have 
	\begin{equation}
		\mbv_{1}\T\hatmbM\mbv_{1}=\mbv_{1}\T\mbM\mbv_{1}+\mbv_{1}\T(\hatmbM-\mbM)\mbv_{1}\geq\lambda_{1}-\epsilon\tau_{0}^{2}.
	\end{equation}
	Since $\hatmbv_{1}$ is the maximizer of the optimization problem,
	where the population truth $\mbv_{1}$ is in the feasible set since
	$\tau\geq\tau_{0}$, we must have 
	\begin{equation}
		\lambda_{1}\eta_{1}+\epsilon\tau^{2}\geq\lambda_{1}-\epsilon\tau_{0}^{2},
	\end{equation}
	therefore, 
	\begin{equation}
		\eta_{1}\geq\frac{\lambda_{1}-\epsilon(\tau_{0}^{2}+\tau^{2})}{\lambda_{1}}=1-\frac{\epsilon(\tau_{0}^{2}+\tau^{2})}{\lambda_{1}}\geq1-\dfrac{2\epsilon\tau^{2}}{\lambda_{1}}.
	\end{equation}

	Second, we prove \eqref{T11}.
	
	From Lemma \ref{lem:AngQuad}, we have 
	\begin{equation}
		\sin^{2}\Theta(\hatmbv_{1},\mbv_{1})\leq\frac{\lambda_{1}-\eta_{1}^{-1}\hatmbv_{1}\T\mbM\hatmbv_{1}}{\Delta}.
	\end{equation}
	We can write 
	\begin{eqnarray*}
		\lambda_{1}-\eta_{1}^{-1}\hatmbv_{1}\T\mbM\hatmbv_{1} & = & \mbv_{1}\T\mbM\mbv_{1}-\eta_{1}^{-1}\hatmbv_{1}\T\mbM\hatmbv_{1}\\
		& = & \mbv_{1}\T\mbM\mbv_{1}-\eta_{1}^{-1}\hatmbv_{1}\T\hatmbM\hatmbv_{1}+\eta_{1}^{-1}\hatmbv_{1}\T(\hatmbM-\mbM)\hatmbv_{1}\\
		& \leq & \mbv_{1}\T\mbM\mbv_{1}-\eta_{1}^{-1}\hatmbv_{1}\T\hatmbM\hatmbv_{1}+\eta_{1}^{-1}\Vert\hatmbM-\mbM\Vert_{\max}\Vert\hatmbv_{1}\Vert_{1}^{2}\\
		& \leq & \mbv_{1}\T(\mbM-\hatmbM)\mbv_{1}+\mbv_{1}\T\hatmbM\mbv_{1}-\eta_{1}^{-1}\hatmbv_{1}\T\hatmbM\hatmbv_{1}+\eta_{1}^{-1}\epsilon\tau^{2}\\
		& \leq & \Vert\mbM-\hatmbM\Vert_{\max}\Vert\mbv_{1}\Vert_{1}^{2}+\mbv_{1}\T\hatmbM\mbv_{1}-\eta_{1}^{-1}\hatmbv_{1}\T\hatmbM\hatmbv_{1}+\eta_{1}^{-1}\epsilon\tau^{2}\\
		& \leq & \epsilon(\tau_{0}^{2}+\eta_{1}^{-1}\tau^{2})+\mbv_{1}\T\hatmbM\mbv_{1}-\eta_{1}^{-1}\hatmbv_{1}\T\hatmbM\hatmbv_{1}\\
		& \leq & \epsilon(\tau_{0}^{2}+\eta_{1}^{-1}\tau^{2}),
	\end{eqnarray*}
	where the last inequality follows from 
	\begin{equation}
		\mbv_{1}\T\hatmbM\mbv_{1}\leq\eta_{1}^{-1}\hatmbv_{1}\T\hatmbM\hatmbv_{1},
	\end{equation}
	because $\sqrt{\eta_{1}}\mbv_{1}$ is in the feasible set of the optimization
	where $\hatmbv_{1}$ is the maximizer. Hence, by \eqref{T13}, we
	have 
	\begin{equation}
		\lambda_{1}-\eta_{1}^{-1}\hatmbv_{1}\T\mbM\hatmbv_{1}\leq\epsilon\{\tau_{0}^{2}+\dfrac{\tau^{2}}{1-2\epsilon\tau^{2}\lambda_{1}^{-1}}\}\leq3\epsilon\tau^{2},
	\end{equation}
	if $2\epsilon\tau^{2}\lambda_{1}^{-1}\le1/2$. Therefore, 
	\begin{equation}
		\sin^{2}\Theta(\hatmbv_{1},\mbv_{1})\leq3\epsilon\tau^{2}/\Delta.
	\end{equation}

	Thirdly, we prove \eqref{T12}.
	
	Specifically, we have the following derivations to get \eqref{T12},
	\begin{eqnarray*}
		\hatmbv_{1}\T\mbM\hatmbv & = & \hatmbv_{1}\T\hatmbM\hatmbv_{1}+\hatmbv_{1}\T(\mbM-\hatmbM)\hatmbv_{1}\\
		& \geq & \hatmbv_{1}\T\hatmbM\hatmbv_{1}-\epsilon\tau^{2}\\
		& \geq & \mbv_{1}\T\hatmbM\mbv_{1}-\epsilon\tau^{2}\\
		& = & \lambda_{1}-\mbv_{1}\T(\mbM-\hatmbM)\mbv_{1}-\epsilon\tau^{2}\\
		& \geq & \lambda_{1}-\epsilon\tau_{0}^{2}-\epsilon\tau^{2}\\
		& \geq & \lambda_{1}-2\epsilon\tau^{2}.
	\end{eqnarray*}

	Fourth, we prove \eqref{T14}.
	
	We notice that, on one hand, 
	\begin{eqnarray*}
		\hatlambda_{1}-\lambda_{1} & = & \hatmbv_{1}\T\hatmbM\hatmbv_{1}-\mbv_{1}\T\mbM\mbv_{1}\\
		& \geq & \mbv_{1}\T\hatmbM\mbv_{1}-\mbv_{1}\T\mbM\mbv_{1}\\
		& \geq & -\epsilon\tau_{0}^{2},
	\end{eqnarray*}
	while on the other hand, 
	\begin{eqnarray*}
		\hatlambda_{1}-\lambda_{1} & = & \hatmbv_{1}\T\mbM\hatmbv_{1}+\hatmbv\T(\hatmbM-\mbM)\hatmbv_{1}-\mbv_{1}\T\mbM\mbv_{1}\\
		& \le & \hatmbv_{1}\T(\hatmbM-\mbM)\hatmbv_{1}\\
		& \leq & \epsilon\tau^{2}.
	\end{eqnarray*}

	Finally, we prove \eqref{T15}.
	
	\begin{eqnarray*}
		\Vert\hatmbv_{1}-\mbv_{1}\Vert_{2}^{2} & = & \hatmbv_{1}\T\hatmbv_{1}+\mbv_{1}\T\mbv_{1}-2\hatmbv_{1}\T\mbv_{1}\\
		& = & 1+\eta_{1}-2\sqrt{\eta_{1}\cos^{2}\Theta(\hatmbv_{1},\mbv_{1})}.
	\end{eqnarray*}
	Let $\beta\equiv\cos^{2}\Theta(\hatmbv_{1},\mbv_{1})$, then
	\begin{eqnarray*}
		\Vert\hatmbv_{1}-\mbv_{1}\Vert_{2}^{2} & = & 1-\eta_{1}+2\eta_{1}-2\sqrt{\eta_{2}\beta}\\
		& \leq & \dfrac{2\epsilon\tau^{2}}{\lambda_{1}}+2\sqrt{\eta_{1}}\dfrac{\eta_{1}-\beta}{\sqrt{\eta_{1}}+\sqrt{\beta}},\\
		& \leq & \dfrac{2\epsilon\tau^{2}}{\lambda_{1}}+2\cdot1\cdot\dfrac{3\epsilon\tau^{2}}{1},
	\end{eqnarray*}
	where the last inequality is because of $\eta_{1}\leq1$ and $\eta_{1}-\beta\leq1-\beta=\sin^{2}\Theta(\hatmbv_{1},\mbv_{1})\leq3\epsilon\tau^{2}$
	from \eqref{T11}, moreover, $\sqrt{\eta_{1}}+\sqrt{\beta}\geq1$
	if $\max(3\epsilon\tau^{2},\dfrac{2\epsilon\tau^{2}}{\lambda_{1}})\leq\dfrac{3}{4}$.
	
\end{proof}

\begin{lemma}\label{lem: Tk_upper} Assume that $\epsilon\tau^{2}\leq1$.
	If for all $j=1,\ldots,k-1$, we have 
	\begin{equation}
		\sin^{2}\Theta(\hatmbv_{j},\mbv_{j})\leq C\epsilon\tau^{2},\ \Vert\hatmbv_{j}-\mbv_{j}\Vert_{2}^{2}\leq C\epsilon\tau^{2},\label{Tk_upper1}
	\end{equation}
	then for any $\mbw\in\mbbR^{p}$, $\Vert\mbw\Vert_{2}^{2}=1$ and
	$\Vert\mbw\Vert_{1}\leq\tau$,
	\begin{eqnarray}
		\mbw\T\hatmbM_{k}\mbw & \leq & \lambda_{k}-\Delta\alpha+C\sqrt{\epsilon\tau^{2}\alpha}+C\epsilon\tau^{2},\label{Tk_upper}
	\end{eqnarray}
	where $\alpha=\sin^{2}\Theta(\mbw,\mbv_{k})$.
	
\end{lemma}

\begin{proof} From Lemma \ref{lem:AngQuad}, we have 
	\[
	\mbw\T\mbM_{k}\mbw\leq\lambda_{k}-\Delta\alpha,
	\]
	thus we can write 
	\begin{eqnarray*}
		\mbw\T\hatmbM_{k}\mbw & = & \mbw\T\mbM_{k}\mbw+\mbw\T(\hatmbM_{k}-\mbM_{k})\mbw\\
		& \leq & \lambda_{k}-\Delta\alpha+\mbw\T(\hatmbM_{k}-\mbM_{k})\mbw.
	\end{eqnarray*}
	It remains to show that
	\begin{equation}
		\mbw\T(\hatmbM_{k}-\mbM_{k})\mbw\leq C\sqrt{\epsilon\tau^{2}\alpha}+C\epsilon\tau^{2}.\label{Tk_upper_diff}
	\end{equation}
	First of all, we decompose the matrix $\hatmbM_{k}-\mbM_{k}$ into
	several terms as follows. We write $\mbM_{k}$ as
	\begin{eqnarray}
		\mbM_{k} & = & \mbM-\sum_{j<k}\lambda_{j}\mbv_{j}\mbv_{j}\T\nonumber \\
		& = & \mbM-\mbP_{k-1}\mbM\nonumber \\
		& = & \mbM-\mbP_{k-1}\mbM-\mbM\mbP_{k-1}+\mbP_{k-1}\mbM\mbP_{k-1},\label{Mk_4term}
	\end{eqnarray}
	where $\mbP_{k-1}$ is the projection matrix onto the subspace spanned
	by the first $(k-1)$-eigenvectors of $\mbM$, i.e. $\mbP_{k-1}=\sum_{j<k}\mbv_{j}\mbv_{j}\T$. 
	
	From Lemma \ref{lemma: SPC}, we have
	\begin{eqnarray*}
		\hatmbM_{k} & = & (\mbI_{p}-\hatmbv_{k-1}\hatmbv_{k-1}\T)\hatmbM_{k-1}(\mbI_{p}-\hatmbv_{k-1}\hatmbv_{k-1}\T)\\
		& = & \left\{ \prod_{j<k}(\mbI_{p}-\hatmbv_{j}\hatmbv_{j}\T)\right\} \hatmbM\left\{ \prod_{j<k}(\mbI_{p}-\hatmbv_{j}\hatmbv_{j}\T)\right\} .
	\end{eqnarray*}
	Similarly to the definition of $\mbP_{k-1}=\sum_{j<k}\mbv_{j}\mbv_{j}\T$,
	we define $\hatmbP_{k-1}=\sum_{j<k}\hatmbv_{j}\hatmbv_{j}\T$. It
	is worth mentioning that $\hatmbP_{k-1}$ is not a projection matrix
	because $\Vert\hatmbv_{j}\Vert_{2}^{2}$ may not be exactly 1 and
	also that $\hatmbv_{j}$ may not be orthogonal to $\hatmbv_{j-1}$.
	Therefore, we write
	\begin{eqnarray*}
		\prod_{j<k}(\mbI_{p}-\hatmbv_{j}\hatmbv_{j}\T) & = & \mbI_{p}-\sum_{j<k}\hatmbv_{j}\hatmbv_{j}\T+\sum_{s=2}^{k-1}\sum_{l_{1}<\cdots<l_{s}<k}(-1)^{s}(\hatmbv_{l_{1}}\hatmbv_{l_{1}}\T)\cdots(\hatmbv_{l_{s}}\hatmbv_{l_{s}}\T)\\
		& = & \mbI_{p}-\hatmbP_{k-1}+\hatmbbP_{k-1},
	\end{eqnarray*}
	where $\hatmbbP_{k-1}$ is not zero in general because $\hatmbv_{1},\dots,\hatmbv_{k-1}$
	are not orthogonal to each other. Also, $\hatmbP_{k-1}=0$ for $k=1$,
	and $\hatmbbP_{k-1}=0$ for $k=1,2$. Get back to $\hatmbM_{k}$,
	we have
	\begin{eqnarray}
		\hatmbM_{k} & = & \left\{ \mbI_{p}-\hatmbP_{k-1}+\hatmbbP_{k-1}\right\} \hatmbM\left\{ \mbI_{p}-\hatmbP_{k-1}+\hatmbbP_{k-1}\right\} \nonumber \\
		& = & \hatmbM-\hatmbP_{k-1}\hatmbM-\hatmbM\hatmbP_{k-1}+\hatmbP_{k-1}\hatmbM\hatmbP_{k-1}\nonumber \\
		& + & \hatmbbP_{k-1}\hatmbM(\mbI_{p}-\hatmbP_{k-1}+\hatmbbP_{k-1})+(\mbI_{p}-\hatmbP_{k-1}+\hatmbbP_{k-1})\hatmbM\hatmbbP_{k-1}.\label{hatMk_6term}
	\end{eqnarray}
	Comparing \eqref{Mk_4term} and \eqref{hatMk_6term}, we have
	\begin{eqnarray*}
		\mbw\T(\hatmbM_{k}-\mbM_{k})\mbw & = & \mbw\T(\hatmbM-\mbM)\mbw+\mbw\T(\hatmbP_{k-1}\hatmbM\hatmbP_{k-1}-\mbP_{k-1}\mbM\mbP_{k-1})\mbw\\
		& + & \mbw\T(\mbP_{k-1}\mbM-\hatmbP_{k-1}\hatmbM)\mbw+\mbw\T(\mbM\mbP_{k-1}-\hatmbM\hatmbP_{k-1})\mbw\\
		& + & \mbw\T\left\{ \hatmbbP_{k-1}\hatmbM(\mbI_{p}-\hatmbP_{k-1}+\hatmbbP_{k-1})\right\} \mbw+\mbw\T\left\{ (\mbI_{p}-\hatmbP_{k-1}+\hatmbbP_{k-1})\hatmbM\hatmbbP_{k-1}\right\} \mbw\\
		& \equiv & Q_{1}+Q_{2}+Q_{3}+Q_{4}+Q_{5}+Q_{6}.
	\end{eqnarray*}
	To show \eqref{Tk_upper_diff}, we need to show that 
	\[
	Q_{m}\leq C\sqrt{\epsilon\tau^{2}\alpha}+C\epsilon\tau^{2},\quad m=1,\ldots,6.
	\]
	First, we compute $Q_{1}$, 
	\[
	Q_{1}=\mbw\T(\hatmbM-\mbM)\mbw\leq\Vert\hatmbM-\mbM\Vert_{\max}\Vert\mbw\Vert_{1}^{2}\leq\epsilon\tau^{2}.
	\]
	Second, $Q_{3}=Q_{4}$ by symmetry, we only need to compute $Q_{3}$,
	\begin{eqnarray*}
		Q_{3} & = & \mbw\T(\mbP_{k-1}\mbM-\hatmbP_{k-1}\hatmbM)\mbw\\
		& = & \mbw\T(\mbP_{k-1}\mbM-\hatmbP_{k-1}\mbM)\mbw+\mbw\T(\hatmbP_{k-1}\mbM-\hatmbP_{k-1}\hatmbM)\mbw\\
		& \equiv & Q_{31}+Q_{32}.
	\end{eqnarray*}
	For $Q_{32}$, we have
	\begin{eqnarray*}
		Q_{32} & = & \mbw\T\hatmbP_{k-1}(\mbM-\hatmbM)\mbw=\sum_{j<k}(\mbw\T\hatmbv_{j})\{\hatmbv_{j}\T(\mbM-\hatmbM)\mbw\}\\
		& \leq & \sum_{j<k}(\mbw\T\hatmbv_{j})\cdot\epsilon\tau^{2}\leq k\epsilon\tau^{2}=C\epsilon\tau^{2}.
	\end{eqnarray*}
	For $Q_{31},$ we have 
	\begin{eqnarray*}
		Q_{31} & = & \mbw\T(\mbP_{k-1}-\hatmbP_{k-1})\mbM\mbw=\sum_{j<k}\mbw\T(\mbv_{j}\mbv_{j}\T-\hatmbv_{j}\hatmbv_{j}\T)\mbM\mbw\\
		& = & \sum_{j<k}\mbw\T(\mbv_{j}\mbv_{j}\T-\hatmbv_{j}\mbv_{j}\T)\mbM\mbw+\sum_{j<k}\mbw\T(\hatmbv_{j}\mbv_{j}\T-\hatmbv_{j}\hatmbv_{j}\T)\mbM\mbw\\
		& = & \sum_{j<k}\mbw\T(\mbv_{j}-\hatmbv_{j})\cdot(\mbv_{j}\T\mbM\mbw)+\sum_{j<k}(\mbw\T\hatmbv_{j})\cdot(\mbv_{j}-\hatmbv_{j})\T\mbM\mbw,
	\end{eqnarray*}
	where the two terms are computed as follows.
	\begin{eqnarray*}
		\sum_{j<k}\mbw\T(\mbv_{j}-\hatmbv_{j})\cdot(\mbv_{j}\T\mbM\mbw) & = & \sum_{j<k}\mbw\T(\mbv_{j}-\hatmbv_{j})\cdot(\lambda_{j}\mbv_{j}\T\mbw)\\
		& \leq & \sum_{j<k}\left(\Vert\mbw\Vert_{2}\Vert\mbv_{j}-\hatmbv_{j}\Vert_{2}\right)\cdot\left\{ \lambda_{j}\sqrt{1-(\mbv_{k}\T\mbw)^{2}}\right\} \\
		& \leq & Ck\sqrt{\epsilon\tau^{2}\alpha}=C\sqrt{\epsilon\tau^{2}\alpha}.
	\end{eqnarray*}
	\begin{eqnarray}
		\sum_{j<k}(\mbw\T\hatmbv_{j})\cdot(\mbv_{j}-\hatmbv_{j})\T\mbM\mbw & = & \sum_{j<k}\left\{ \mbw\T\mbv_{j}+\mbw\T(\hatmbv_{j}-\mbv_{j})\right\} \cdot(\mbv_{j}-\hatmbv_{j})\T\mbM\mbw\nonumber \\
		& \leq & \sum_{j<k}(\sqrt{\alpha}+C\sqrt{\epsilon\tau^{2}})\cdot\sqrt{(\mbv_{j}-\hatmbv_{j})\T\mbM(\mbv_{j}-\hatmbv_{j})}\sqrt{\mbw\T\mbM\mbw}\label{wvhat}\\
		& \leq & \sum_{j<k}(\sqrt{\alpha}+C\sqrt{\epsilon\tau^{2}})\cdot\lambda_{1}\sqrt{\Vert\mbv_{j}-\hatmbv_{j}\Vert_{2}^{2}\cdot\Vert\mbw\Vert_{2}^{2}}\label{vhatmvhat}\\
		& \leq & Ck(\sqrt{\alpha}+\sqrt{\epsilon\tau^{2}})\cdot(\sqrt{\epsilon\tau^{2}})\nonumber \\
		& = & C\epsilon\tau^{2}+C\sqrt{\epsilon\tau^{2}\alpha},\nonumber 
	\end{eqnarray}
	where the last inequality is because of \eqref{Tk_upper1}.
	
	Together, we have $Q_{3}=Q_{4}\leq C\epsilon\tau^{2}+C\sqrt{\epsilon\tau^{2}\alpha}$.
	
	Next, for $Q_{2}$, we have
	\begin{eqnarray*}
		Q_{2} & = & \mbw\T(\hatmbP_{k-1}\hatmbM\hatmbP_{k-1}-\mbP_{k-1}\mbM\mbP_{k-1})\mbw\\
		& = & \mbw\T\hatmbP_{k-1}(\hatmbM-\mbM)\hatmbP_{k-1}\mbw+\mbw\T\hatmbP_{k-1}\mbM(\hatmbP_{k-1}-\mbP_{k-1})\mbw\\
		& + & \mbw\T(\hatmbP_{k-1}-\mbP_{k-1})\mbM\mbP_{k-1}\mbw,\\
		& \equiv & Q_{21}+Q_{22}+Q_{23}
	\end{eqnarray*}
	where the three terms are computed as follows.
	\begin{eqnarray*}
		Q_{21} & = & \mbw\T\hatmbP_{k-1}(\hatmbM-\mbM)\hatmbP_{k-1}\mbw\\
		& \leq & \epsilon\Vert\hatmbP_{k-1}\mbw\Vert_{1}^{2}=\epsilon\Vert\sum_{j<k}(\mbw\T\hatmbv_{j})\hatmbv_{j}\Vert_{1}^{2}\\
		& \leq & k\epsilon\tau^{2}=C\epsilon\tau^{2}.
	\end{eqnarray*}
	\begin{eqnarray*}
		Q_{22} & = & \mbw\T\hatmbP_{k-1}\mbM(\hatmbP_{k-1}-\mbP_{k-1})\mbw\\
		& = & \mbw\T(\sum_{l<k}\hatmbv_{l}\hatmbv_{l}\T)\mbM\sum_{j<k}\left\{ \mbv_{j}(\hatmbv_{j}-\mbv_{j})\T\mbw+(\hatmbv_{j}-\mbv_{j})\hatmbv_{j}\T\mbw\right\} \\
		& = & \sum_{l,j<k}(\mbw\T\hatmbv_{l})\left\{ \hatmbv_{l}\T\mbM\mbv_{j}(\hatmbv_{j}-\mbv_{j})\T\mbw+\hatmbv_{l}\T\mbM(\hatmbv_{j}-\mbv_{j})\hatmbv_{j}\T\mbw\right\} \\
		& \leq & \sum_{l,j<k}(\sqrt{\epsilon\tau^{2}}+\sqrt{\alpha})\left\{ \hatmbv_{l}\T\mbM\mbv_{j}(\hatmbv_{j}-\mbv_{j})\T\mbw+\hatmbv_{l}\T\mbM(\hatmbv_{j}-\mbv_{j})\hatmbv_{j}\T\mbw\right\} ,
	\end{eqnarray*}
	where the last inequality is because $\mbw\T\hatmbv_{l}\leq\sqrt{\alpha}+\sqrt{\epsilon\tau^{2}}$
	as computed in \eqref{wvhat}. Moreover, $(\hatmbv_{j}-\mbv_{j})\T\mbw\leq C\sqrt{\epsilon\tau^{2}}$
	and $\hatmbv_{l}\T\mbM(\hatmbv_{j}-\mbv_{j})\leq C\sqrt{\epsilon\tau^{2}}$
	follows from previous calculation in \eqref{vhatmvhat}. Thus, 
	\begin{eqnarray*}
		Q_{22} & \leq & C(\sqrt{\epsilon\tau^{2}}+\sqrt{\alpha})\sqrt{\epsilon\tau^{2}}(\hatmbv_{l}\T\mbM\mbv_{j}+\hatmbv_{j}\T\mbw),\\
		& \leq & C(\sqrt{\epsilon\tau^{2}}+\sqrt{\alpha})\sqrt{\epsilon\tau^{2}}(\lambda_{1}+1)\\
		& = & C\epsilon\tau^{2}+C\sqrt{\epsilon\tau^{2}\alpha}.
	\end{eqnarray*}
	\begin{eqnarray*}
		Q_{23} & = & \mbw\T(\hatmbP_{k-1}-\mbP_{k-1})\mbM\mbP_{k-1}\mbw\\
		& = & \mbw\T\sum_{j<k}\left\{ (\hatmbv_{j}-\mbv_{j})\mbv_{j}\T+\hatmbv_{j}(\hatmbv_{j}-\mbv_{j})\T\right\} \mbM\mbP_{k-1}\mbw\\
		& = & \sum_{j<k}\left\{ \mbw\T(\hatmbv_{j}-\mbv_{j})\cdot(\mbv_{j}\T\mbM\mbP_{k-1}\mbw)+(\mbw\T\hatmbv_{j})\cdot(\hatmbv_{j}-\mbv_{j})\T\mbM\mbP_{k-1}\mbw\right\} \\
		& = & \sum_{j<k}\left\{ \mbw\T(\hatmbv_{j}-\mbv_{j})\cdot(\lambda_{j}\mbv_{j}\T\mbw)+(\mbw\T\hatmbv_{j})\cdot\sum_{l<k}\lambda_{l}(\hatmbv_{j}-\mbv_{j})\T\mbv_{l}\cdot\mbv_{l}\T\mbw\right\} \\
		& \leq & \sum_{j<k}\left\{ \sqrt{\epsilon\tau^{2}}\cdot(C\sqrt{\alpha})+(\mbw\T\mbv_{j}+\mbw\T\hatmbv_{j}-\mbw\T\mbv_{j})\cdot\sum_{l<k}C\sqrt{\epsilon\tau^{2}}\cdot\sqrt{\alpha}\right\} \\
		& \leq & C\sqrt{\epsilon\tau^{2}\alpha}+C(\sqrt{\alpha}+\sqrt{\epsilon\tau^{2}})\sqrt{\epsilon\tau^{2}\alpha}\\
		& = & C\sqrt{\epsilon\tau^{2}\alpha},
	\end{eqnarray*}
	where the last equality is because $\epsilon\tau^{2}\leq1$ and $\alpha\leq1$. 
	
	Finally, for $Q_{5}$ and $Q_{6}$, we have
	\[
	Q_{5}=Q_{6}=\mbw\T(\mbI_{p}-\hatmbP_{k-1}+\hatmbbP_{k-1})\hatmbM\hatmbbP_{k-1}\mbw,
	\]
	which is zero for $k\leq2$ since $\hatmbbP_{k-1}=0$ for $k\leq2$.
	For $k\geq2$, we have 
	\begin{eqnarray*}
		\hatmbbP_{k-1}\mbw & = & \sum_{s=2}^{k-1}\sum_{l_{1}<\cdots<l_{s}<k}(-1)^{s}(\hatmbv_{l_{1}}\hatmbv_{l_{1}}\T)\cdots(\hatmbv_{l_{s}}\hatmbv_{l_{s}}\T)\mbw\\
		& = & \sum_{s=2}^{k-1}\sum_{l_{1}<\cdots<l_{s}<k}(-1)^{s}(\hatmbv_{l_{3}}\hatmbv_{l_{3}}\T)\cdots(\hatmbv_{l_{s}}\hatmbv_{l_{s}}\T)\cdot(\hatmbv_{l_{1}}\hatmbv_{l_{1}}\T)(\hatmbv_{l_{2}}\hatmbv_{l_{2}}\T)\mbw,
	\end{eqnarray*}
	where 
	\[
	(\hatmbv_{l_{1}}\hatmbv_{l_{1}}\T)(\hatmbv_{l_{2}}\hatmbv_{l_{2}}\T)\mbw=(\hatmbv_{l_{1}}\T\hatmbv_{l_{2}})(\hatmbv_{l_{2}}\T\mbw)\hatmbv_{l_{1}}.
	\]
	Notice that $\hatmbv_{l_{2}}\T\mbw=\mbw\T\mbv_{l_{2}}+\mbw\T(\hatmbv_{l_{2}}-\mbv_{l_{2}})\leq\sqrt{\alpha}+\sqrt{\epsilon\tau^{2}}$
	for $l_{2}<k$. Also, notice that 
	\[
	\hatmbv_{l_{1}}\T\hatmbv_{l_{2}}=\hatmbv_{l_{1}}\T\mbv_{l_{2}}+\hatmbv_{l_{1}}\T(\hatmbv_{l_{2}}-\mbv_{l_{2}})\leq C\sqrt{\epsilon\tau^{2}},
	\]
	since $l_{1}<l_{2}<k$. Therefore, for any $\mbu\in\mbbR^{p}$, 
	\[
	\mbu\T\hatmbbP_{k-1}\mbw\leq(C\epsilon\tau^{2}+C\sqrt{\epsilon\tau^{2}\alpha})\left|\sum_{s=2}^{k-1}\sum_{l_{1}<\cdots<l_{s}<k}(-1)^{s}(\hatmbv_{l_{3}}\hatmbv_{l_{3}}\T)\cdots(\hatmbv_{l_{s}}\hatmbv_{l_{s}}\T)\mbu\T\hatmbv_{l_{1}}\right|.
	\]
	\begin{eqnarray*}
		\vert Q_{6}\vert & = & \vert\mbw\T(\mbI_{p}-\hatmbP_{k-1}+\hatmbbP_{k-1})\hatmbM\hatmbbP_{k-1}\mbw\vert\\
		& \leq & (C\epsilon\tau^{2}+C\sqrt{\epsilon\tau^{2}\alpha})\left|\sum_{s=2}^{k-1}\sum_{l_{1}<\cdots<l_{s}<k}(-1)^{s}(\hatmbv_{l_{3}}\hatmbv_{l_{3}}\T)\cdots(\hatmbv_{l_{s}}\hatmbv_{l_{s}}\T)\cdot\{\mbw\T(\mbI_{p}-\hatmbP_{k-1}+\hatmbbP_{k-1})\hatmbM\hatmbv_{l_{1}}\}\right|\\
		& \leq & (C\epsilon\tau^{2}+C\sqrt{\epsilon\tau^{2}\alpha})\{\mbw\T(\mbI_{p}-\hatmbP_{k-1}+\hatmbbP_{k-1})\hatmbM\hatmbv_{l_{1}}\}\\
		& = & (C\epsilon\tau^{2}+C\sqrt{\epsilon\tau^{2}\alpha}),
	\end{eqnarray*}
	where the last equality is because $\mbw\T\hatmbM\hatmbv_{j}$, $-\mbw\T\hatmbP_{k-1}\hatmbM\hatmbv_{j}$
	and $\mbw\T\hatmbbP_{k-1}\hatmbM\hatmbv_{j}$ are all less than or
	equals to $C(\lambda_{1}+C\epsilon\tau^{2})$. 
	
\end{proof}

\begin{lemma}\label{lem: Tk_lower} Assume that $\epsilon\tau^{2}\leq1$.
	If for all $j=1,\ldots,k-1$, we have 
	\begin{equation}
		\sin^{2}\Theta(\hatmbv_{j},\mbv_{j})\leq C\epsilon\tau^{2},\ \Vert\hatmbv_{j}-\mbv_{j}\Vert_{2}^{2}\leq C\epsilon\tau^{2},
	\end{equation}
	then we must have 
	\begin{eqnarray}
		\hatmbv_{k}\T\mbM_{k}\hatmbv_{k} & \geq & \lambda_{k}\cos^{2}\Theta(\hatmbv_{k},\mbv_{k}),\label{Tk_lower1}\\
		\mbv_{k}\T\hatmbM_{k}\mbv_{k} & \geq & \lambda_{k}-C\epsilon\tau^{2}.\label{Tk_lower2}
	\end{eqnarray}
\end{lemma}

\begin{proof} First, to obtain \eqref{Tk_lower1}, notice that 
	\begin{equation}
		\hatmbv_{k}\T\mbM_{k}\hatmbv_{k}\geq\hatmbv_{k}\T(\lambda_{k}\mbv_{k}\mbv_{k}\T)\hatmbv_{k}=\lambda_{k}\frac{\cos^{2}\Theta(\hatmbv_{k},\mbv_{k})}{\Vert\hatmbv_{k}\Vert_{2}^{2}},
	\end{equation}
	where $\Vert\hatmbv_{k}\Vert_{2}^{2}\leq1$. Thus $\hatmbv_{k}\T\mbM_{k}\hatmbv_{k}\geq\lambda_{k}\cos^{2}\Theta(\hatmbv_{k},\mbv_{k})$.
	
	Next, to get \eqref{Tk_lower2}, notice that 
	\begin{eqnarray*}
		\mbv_{k}\T\hatmbM_{k}\mbv_{k} & = & \mbv_{k}\T\mbM\mbv_{k}+\mbv_{k}\T(\hatmbM_{k}-\mbM)\mbv_{k}\\
		& = & \lambda_{k}-\mbv_{k}\T(\mbM-\hatmbM)\mbv_{k}-\mbv_{k}\T(\hatmbM-\hatmbM_{k})\mbv_{k}\\
		& \geq & \lambda_{k}-\epsilon\tau^{2}-\mbv_{k}\T(\hatmbM-\hatmbM_{k})\mbv_{k},
	\end{eqnarray*}
	which means that it remains to prove $\mbv_{k}\T(\hatmbM-\hatmbM_{k})\mbv_{k}\leq C\epsilon\tau^{2}$.
	Recall from \eqref{hatMk_6term} that,
	\begin{eqnarray}
		\hatmbM-\hatmbM_{k} & = & \hatmbM-\left\{ \mbI_{p}-\hatmbP_{k-1}+\hatmbbP_{k-1}\right\} \hatmbM\left\{ \mbI_{p}-\hatmbP_{k-1}+\hatmbbP_{k-1}\right\} \nonumber \\
		& = & \hatmbP_{k-1}\hatmbM+\hatmbM\hatmbP_{k-1}-\hatmbP_{k-1}\hatmbM\hatmbP_{k-1}\nonumber \\
		& - & \hatmbbP_{k-1}\hatmbM(\mbI_{p}-\hatmbP_{k-1}+\hatmbbP_{k-1})-(\mbI_{p}-\hatmbP_{k-1}+\hatmbbP_{k-1})\hatmbM\hatmbbP_{k-1}.\label{hatMk_5term}
	\end{eqnarray}
	Follow the same calculation in the proof for Lemma \ref{lem: Tk_upper}
	for $Q_{5}=Q_{6}$, we can see that
	\begin{eqnarray*}
		\mbv_{k}\T(\hatmbM-\hatmbM_{k})\mbv_{k} & \leq & \mbv_{k}\T(\hatmbP_{k-1}\hatmbM+\hatmbM\hatmbP_{k-1}-\hatmbP_{k-1}\hatmbM\hatmbP_{k-1})\mbv_{k}+C\epsilon\tau^{2}\\
		& \leq & \mbv_{k}\T(\hatmbP_{k-1}\hatmbM+\hatmbM\hatmbP_{k-1})\mbv_{k}+C\epsilon\tau^{2}\\
		& = & 2\mbv_{k}\T\hatmbP_{k-1}\hatmbM\mbv_{k}+C\epsilon\tau^{2},
	\end{eqnarray*}
	where $\mbv_{k}\T\hatmbP_{k-1}\hatmbM\mbv_{k}=\sum_{j<k}(\mbv_{k}\T\hatmbv_{j})\cdot(\hatmbv_{j}\T\hatmbM\mbv_{k})$.
	Because $\mbv_{k}\T\hatmbv_{j}=\mbv_{k}\T(\hatmbv_{j}-\mbv_{j})\leq\sqrt{\epsilon\tau^{2}}$
	and that
	\begin{eqnarray*}
		\hatmbv_{j}\T\hatmbM\mbv_{k} & = & \hatmbv_{j}\T\mbM\mbv_{k}+\hatmbv_{j}\T(\hatmbM-\mbM)\mbv_{k}\\
		& \leq & \lambda_{k}(\hatmbv_{j}\T\mbv_{k})+\epsilon\tau^{2}\\
		& = & C\sqrt{\epsilon\tau^{2}}+\epsilon\tau^{2}\\
		& \leq & (C+1)\sqrt{\epsilon\tau^{2}},
	\end{eqnarray*}
	we have $\mbv_{k}\T\hatmbP_{k-1}\hatmbM\mbv_{k}\leq C\epsilon\tau^{2}$.
	This completes the proof.
	
\end{proof}

\subsection{Proof for Theorem \ref{thm:SPC} }

We prove Theorem \ref{thm:SPC} by induction.

For $j=1$, the following results is obtained in Lemma \ref{lem: T1},
\[
\sin^{2}\Theta(\mbv_{1},\hatmbv_{1})\leq C\epsilon\tau^{2},\ \eta_{1}\geq1-C\epsilon\tau^{2},\ \Vert\mbv_{1}-\hatmbv_{1}\Vert_{2}^{2}\leq C\epsilon\tau^{2}.
\]

Suppose the following statements are true for all $j<k$, and we prove
it is also true for $j=k$,
\[
\sin^{2}\Theta(\mbv_{j},\hatmbv_{j})\leq C\epsilon\tau^{2},\ \eta_{j}\geq1-C\epsilon\tau^{2},\ \Vert\mbv_{j}-\hatmbv_{j}\Vert_{2}^{2}\leq C\epsilon\tau^{2}.
\]

First, we establish $\sin^{2}\Theta(\mbv_{k},\hatmbv_{k})\leq C\epsilon\tau^{2}$. 

From Lemma \ref{lem: Tk_lower} and the definition of $\hatmbv_{k}$,
we know that 
\[
\hatmbv_{k}\T\hatmbM_{k}\hatmbv_{k}\geq\mbv_{k}\T\hatmbM_{k}\mbv_{k}\geq\lambda_{k}-C\epsilon\tau^{2}.
\]
From Lemma \ref{lem: Tk_upper}, we let $\alpha=\sin^{2}\Theta(\hatmbv_{k},\mbv_{k})$
and have 
\begin{eqnarray*}
	\hatmbv_{k}\T\hatmbM_{k}\hatmbv_{k} & \leq & \tilmbv_{k}\T\hatmbM_{k}\tilmbv_{k}\\
	& \leq & \lambda_{k}-\Delta\alpha+C\sqrt{\epsilon\tau^{2}\alpha}+C\tau^{2}\epsilon.
\end{eqnarray*}
Comparing the lower and upper bound of $\hatmbv_{k}\T\hatmbM_{k}\hatmbv_{k}$,
we have
\[
\lambda_{k}-C\epsilon\tau^{2}\leq\lambda_{k}-\Delta\alpha+C\sqrt{\epsilon\tau^{2}\alpha}+C\tau^{2}\epsilon,
\]
or equivalently,
\[
0\geq\alpha-C\sqrt{\epsilon\tau^{2}}\sqrt{\alpha}-C\epsilon\tau^{2}=(\sqrt{\alpha}-\sqrt{C\epsilon\tau^{2}})^{2}-C\epsilon\tau^{2},
\]
which implies 
\[
\alpha\le C\epsilon\tau^{2}.
\]

Next, we establish $\eta_{k}=\Vert\hatmbv_{k}\Vert_{2}^{2}\geq1-C\epsilon\tau^{2}$. 

From Lemma \ref{lem: Tk_lower}, we have 
\[
\mbv_{k}\T\hatmbM_{k}\mbv_{k}\geq\lambda_{k}-C\epsilon\tau^{2}.
\]

From Lemma \ref{lem: Tk_upper}, we have
\[
\hatmbv_{k}\T\hatmbM_{k}\hatmbv_{k}=\eta_{k}\tilmbv_{k}\T\hatmbM_{k}\tilmbv_{k}\leq\eta_{k}(\lambda_{k}+C\epsilon\tau^{2}),
\]
where we have applied $\sin^{2}\Theta(\hatmbv_{k},\mbv_{k})\le C\epsilon\tau^{2}$
to obtain the last inequality. Comparing the two inequality and noticing
that $\mbv_{k}\T\hatmbM_{k}\mbv_{k}\leq\hatmbv_{k}\T\hatmbM_{k}\hatmbv_{k}$,
we have
\[
\eta_{k}\geq\dfrac{\lambda_{k}-C\epsilon\tau^{2}}{\lambda_{k}+C\epsilon\tau^{2}}=1-\dfrac{C\epsilon\tau^{2}}{\lambda_{k}+C\epsilon\tau^{2}}\geq1-C\epsilon\tau^{2}.
\]

Finally, we establish $\Vert\mbv_{k}-\hatmbv_{k}\Vert_{2}^{2}\leq C\epsilon\tau^{2}$.
By direct calculation, 
\begin{eqnarray*}
	\Vert\mbv_{k}-\hatmbv_{k}\Vert_{2}^{2} & = & 1+\eta_{k}-2\sqrt{\eta_{k}\cos^{2}\Theta(\hatmbv_{k},\mbv_{k})}\\
	& \leq & 1-\eta_{k}+2\eta_{k}-2\sqrt{\eta_{k}(1-C\epsilon\tau^{2})}\\
	& \leq & C\epsilon\tau^{2}+2\sqrt{\eta_{k}}(\sqrt{\eta_{k}}-\sqrt{1-C\epsilon\tau^{2}})\\
	& \leq & C\epsilon\tau^{2}+2\dfrac{\{\eta_{k}-(1-C\epsilon\tau^{2})\}}{\sqrt{\eta_{k}}+\sqrt{1-C\epsilon\tau^{2}}}\\
	& \leq & C\epsilon\tau^{2}.
\end{eqnarray*}

\subsection{Proof for Theorem \ref{thm: NIECE_hd}}

By Lemma~\ref{lem: gammai}, if $\Delta_{\mbU}>2(\tau^{2}+\dfrac{10\nu}{\Delta})\epsilon$,
we keep all the $\hatmbv_{i}$'s in the envelope because the high-dimensional
NIECE procedure keeps the correct ordering of directions $\hatmbv_{i}$.
Then the conclusion of Theorem \ref{thm: NIECE_hd} follows directly
from Theorem \ref{thm:SPC} and Lemma \ref{lem: gammai}.

\section{Proofs for Theorem \ref{thm:res} and Corollarys~\ref{cy:hdResponse} and \ref{cy:hdResponse2} }

Combine Lemma~\ref{lem:response} with Theorem~\ref{thm: NIECE_hd}
and we have the desired conclusion.  We have the Corollary~\ref{cy:hdResponse} by letting $\dfrac{1}{p}\geq\epsilon\gg\sqrt{\dfrac{\log p+\log r}{n}}\rightarrow0$
and $\tau^{2}\epsilon\rightarrow0$. Then Corollary~\ref{cy:hdResponse2} follows from Corollary~\ref{cy:hdResponse} by fix $p$ as constant and letting $n,r$ diverge.

\begin{lemma}\label{lem:sampleCov} Under sub-Gaussian condition, i.e.~\eqref{XsubG}, for any
	$\varepsilon>0$ and some constant $C>0$, we have
	\begin{equation}
		\Prob(\vert\hatsigma_{\mbX,ij}-\sigma_{\mbX,ij}\vert>\varepsilon)\leq C\exp(-Cn\varepsilon^{2}),
	\end{equation}
	where $n$ is the sample size $\hatsigma_{\mbX,ij}$ and $\sigma_{\mbX,ij}$
	are the $(ij)$-th element in $p\times p$ sample and population covariance
	of $\mbX\in\mbbR^{p}$.
	
	Under joint sub-Gaussian of $(\mbX,\mbY)$ in \eqref{XYsubG}, for any $\varepsilon>0$ and some constant $C>0$,
	we have
	\begin{equation}
		\Prob(\vert\hatsigma_{ij}-\sigma_{ij}\vert>\varepsilon)\leq C\exp(-Cn\varepsilon^{2}),
	\end{equation}
	where $n$ is the sample size $\hatsigma_{ij}$ and $\sigma_{ij}$
	are the $(ij)$-th element in $(p+r)\times(p+r)$ sample and population
	covariance of $\mathbf{Z}=(\mbX\T,\mbY\T)\T\in\mbbR^{p+r}$.
	
\end{lemma}

\begin{proof} Let $U$ and $V$ be the $i$-th and $j$-th elements
	in $\mbX$ or $\mbZ=(\mbX\T,\mbY\T)\T$. Then we want to show that
	\begin{equation}
		\Prob(\vert\hatsigma_{UV}-\sigma_{UV}\vert>\varepsilon)\leq\exp(-Cn\varepsilon^{2}),
	\end{equation}
	where $\hatsigma_{UV}=n^{-1}\sum_{i=1}^{n}U_{i}V_{i}$. We can write
	\[
	\sigma_{UV}=\Cov(U,V)=\{\Cov(U+V)-\Var(U)-\Var(V)\}/2\equiv(\sigma_{U+V}-\sigma_{U}-\sigma_{V})/2,
	\]
	and, similarly, 
	\[
	\hatsigma_{UV}=(\hatsigma_{U+V}-\hatsigma_{U}-\hatsigma_{V})/2.
	\]
	Therefore,
	\[
	\Prob(\vert\hatsigma_{UV}-\sigma_{UV}\vert>\varepsilon)\leq\Prob(\vert\hatsigma_{U+V}-\sigma_{U+V}\vert>\varepsilon)+\Prob(\vert\hatsigma_{V}-\sigma_{V}\vert>\varepsilon)+\Prob(\vert\hatsigma_{U}-\sigma_{U}\vert>\varepsilon).
	\]
	Because $U$, $V$ and $(U+V)$ are all sub-Gaussian, the random variables
	$\hatsigma_{U}$, $\hatsigma_{V}$ and $\hatsigma_{U+V}$ are sums
	of i.i.d. subexponential random variables. By equation (2.4) in \citet{vershynin2012close},
	we have
	\[
	\Prob(\vert\hatsigma_{W}-\sigma_{W}\vert>\varepsilon)\leq2\exp(-cn\varepsilon^{2}),
	\]
	where $W$ is either $U$, $V$ or $(U+V)$. This completes the proof.
	
\end{proof}

\begin{lemma}\label{lem:response} 
	\begin{enumerate}
		\item For $\epsilon>0$, we have 
		\begin{equation}
			\Vert\hatbolSigma_{\mbY}-\bolSigma_{\mbY}\Vert_{\max}\le\epsilon\label{res.high.eq1}
		\end{equation}
		with a probability greater than $1-Cr^{2}\exp(-Cn\epsilon^{2})$. 
		\item For $0<\epsilon\leq1/p$, we have 
		\begin{equation}
			\Vert\hatbolSigma_{\mbY\mbX}\hatbolSigma_{\mbX\mbY}-\bolSigma_{\mbY\mbX}\bolSigma_{\mbX\mbY}\Vert_{\max}\le\epsilon\label{res.high.eq2}
		\end{equation}
		with a probability greater than $1-Cpr\exp(-Cn\epsilon^{2})-Cpr\exp(-Cn/p^{2})$. 
	\end{enumerate}
\end{lemma}

\begin{proof} We first show \eqref{res.high.eq1}. Note that
	\begin{equation}
		\Pr(\Vert\hatbolSigma_{\mbY}-\bolSigma_{\mbY}\Vert_{\max}>\epsilon)=\Pr(\cup_{i,j}|\hat{\sigma}_{\mbY,ij}-\sigma_{\mbY,ij}|>\epsilon)\le Cr^{2}\exp(-Cn\epsilon^{2}),
	\end{equation}
	where the last inequality follows from joint sub-Gaussian of $(\mbX,\mbY)$ in \eqref{XYsubG} and Lemma \ref{lem:sampleCov}.
	
	Now we show\eqref{res.high.eq2}. Straightforward calculation shows
	that
	\begin{eqnarray*}
		&  & \Vert\hatbolSigma_{\mbY\mbX}\hatbolSigma_{\mbX\mbY}-\hatbolSigma_{\mbY\mbX}\hatbolSigma_{\mbX\mbY}\Vert_{\max}\\
		& \le & \Vert\bolSigma_{\mbY\mbX}(\hatbolSigma_{\mbX\mbY}-\bolSigma_{\mbX\mbY})\Vert_{\max}+\Vert(\hatbolSigma_{\mbY\mbX}-\bolSigma_{\mbY\mbX})\bolSigma_{\mbX\mbY}\Vert_{\max}\\
		& + & \Vert(\hatbolSigma_{\mbY\mbX}-\bolSigma_{\mbY\mbX})(\hatbolSigma_{\mbX\mbY}-\bolSigma_{\mbX\mbY})\Vert_{\max}\\
		& = & 2\Vert\bolSigma_{\mbY\mbX}(\hatbolSigma_{\mbX\mbY}-\bolSigma_{\mbX\mbY})\Vert_{\max}+\Vert(\hatbolSigma_{\mbY\mbX}-\bolSigma_{\mbY\mbX})(\hatbolSigma_{\mbX\mbY}-\bolSigma_{\mbX\mbY})\Vert_{\max}.
	\end{eqnarray*}
	Under the event that $\Vert\hatbolSigma_{\mbY\mbX}-\bolSigma_{\mbY\mbX}\Vert_{\max}\leq\epsilon$,
	we have
	\begin{eqnarray*}
		\Vert\bolSigma_{\mbY\mbX}(\hatbolSigma_{\mbX\mbY}-\bolSigma_{\mbX\mbY})\Vert_{\max} & \leq & \Vert\hatbolSigma_{\mbY\mbX}-\bolSigma_{\mbY\mbX}\Vert_{\max}\Vert\bolSigma_{\mbX\mbY}\Vert_{1}\leq\Vert\bolSigma_{\mbX\mbY}\Vert_{1}\epsilon,
	\end{eqnarray*}
	and if we further assume that $\Vert\hatbolSigma_{\mbY\mbX}-\bolSigma_{\mbY\mbX}\Vert_{\max}=\max_{ij}\vert\hatsigma_{\mbY\mbX,ji}-\sigma_{\mbY\mbX,ji}\vert\leq\dfrac{1}{p}$,
	\begin{eqnarray*}
		\Vert(\hatbolSigma_{\mbY\mbX}-\bolSigma_{\mbY\mbX})(\hatbolSigma_{\mbX\mbY}-\bolSigma_{\mbX\mbY})\Vert_{\max} & = & \sup_{j,k}|\sum_{i=1}^{p}(\hatsigma_{\mbY\mbX,ji}-\sigma_{\mbY\mbX,ji})(\hatsigma_{\mbY\mbX,ki}-\sigma_{\mbY\mbX,ki})|\\
		& \leq & \sup_{j,k}|\sum_{i=1}^{p}(\hatsigma_{\mbY\mbX,ji}-\sigma_{\mbY\mbX,ji})\cdot\dfrac{1}{p}|\\
		& \leq & \Vert\hatbolSigma_{\mbY\mbX}-\bolSigma_{\mbY\mbX}\Vert_{\max}\leq\epsilon.
	\end{eqnarray*}
	The probability of the event $\Vert\hatbolSigma_{\mbY\mbX}-\bolSigma_{\mbY\mbX}\Vert_{\max}\leq\min\{\epsilon,\dfrac{1}{p}\}$
	can be calculated as, 
	$$\Pr\left(|\hatsigma_{\mbY\mbX,jk}-\sigma_{\mbY\mbX,jk}|\le\min\{\epsilon,\dfrac{1}{p}\}\mbox{ for all \ensuremath{j,k}}\right)\ge1-Cpr\exp(-Cn\epsilon^{2})-Cpr\exp(-Cn/p^{2}).
	$$
	And we have the desired conclusion. \end{proof}

\section{Proofs for Theorem \ref{thm:glm}}

Theorem \ref{thm:glm} follows directly from the following
Lemma \ref{lem:glm} and Theorem \ref{thm: NIECE_hd}.

Recall from Theorem \ref{thm: NIECE_hd} that we need to verify the
following conditions,
\[
\Vert\hatmbM-\mbM\Vert_{\max}\leq\epsilon,\ \Vert\hatmbU-\mbU\Vert_{\max}\leq\epsilon,\ \Delta_{\mbU}>\sqrt{8c_{0}\nu^{2}\tau^{2}\epsilon}+(2c_{0}\nu+1)\tau^{2}\epsilon.
\]
From the sub-Gaussian distribution of $\mbX$, i.e.~\eqref{XsubG},
we have $\Vert\hatmbM-\mbM\Vert_{\max}\leq\epsilon$ holds with probability
at least $1-Cp^{2}\exp(-Cn\epsilon^{2})$. From the following Lemma
\ref{lem:glm}, we have $\Vert\hatmbU-\mbU\Vert_{\max}\leq\epsilon$
holds with probability at least $1-Cp\exp(-Cn\epsilon^{2}/s)$ if
$\lambda_{n}=4\epsilon/\sqrt{s}$. 

Finally, to achieve $\Vert\mbP_{\hatcalE}-\mbP_{\calE}\Vert_{F}\rightarrow0$
with probability going to 1, we could let 
\[
\dfrac{s\log p}{n\epsilon^{2}}\rightarrow0,\ \dfrac{n\epsilon^{2}}{s}\rightarrow\infty,\ \lambda_{n}=\dfrac{4\epsilon}{\sqrt{s}},\ \dfrac{\sqrt{\tau^{2}\epsilon}}{\Delta_{\mbU}}\rightarrow0,\ \tau^{2}\epsilon\rightarrow0,
\]
which are satisfied as we let 
\[
\sqrt{\dfrac{s\log p}{n}}\rightarrow0,\ \sqrt{\dfrac{s\log p}{n}}\ll\epsilon\ll\tau^{-2}\ll1,\ \sqrt{\dfrac{\log p}{n}}\ll\lambda_{n},\ \Delta_{\mbU}^{-2}\tau^{2}\sqrt{\dfrac{s\log p}{n}}\rightarrow0.
\]

\begin{lemma}\label{lem:glm} If $\lambda_{n}=4\epsilon$, with a
	probability greater than $1-Cp\exp(-Cn\epsilon^{2})$, we have 
	\begin{equation}
		\Vert\hatbolbeta-\bolbeta\Vert_{2}^{2}\le Cs\epsilon^{2},\quad\Vert\hatmbU-\mbU\Vert_{\max}\leq C(\sqrt{s\epsilon^{2}}+s\epsilon^{2}),
	\end{equation}
	where $s$ is the number of important predictors. \end{lemma} 

\begin{proof} By Corollary~1 in \citet{NRWY12} and the results
	in Section 4.4 in their paper, it suffices to verify that 
	\begin{equation}
		\lambda_{n}\ge2\Vert\nabla\{-\dfrac{1}{n}\ell_{n}(\bolbeta)\}\Vert_{\infty}=2\Vert\left(-\dfrac{1}{n}\sum_{i=1}^{n}\left\{ Y_{i}\mbX_{i}-\dfrac{e^{\mbX_{i}\T\bolbeta}}{1+e^{\mbX_{i}\T\bolbeta}}\mbX_{i}\right\} \right)\Vert_{\infty}
	\end{equation}
	with the specified probability.
	
	Note that $-\dfrac{1}{n}\sum_{i=1}^{n}\left\{ Y_{i}\mbX_{i}-\dfrac{e^{\mbX_{i}\T\bolbeta}}{1+e^{\mbX_{i}\T\bolbeta}}\mbX_{i}\right\} $
	is the derivative of the (negative) log likelihood function. It follows
	that its expectation is $0$. Hence, 
	\begin{eqnarray*}
		&  & \Pr(|-\dfrac{1}{n}\sum_{i=1}^{n}\left\{ Y_{i}\mbX_{i}-\dfrac{e^{\mbX_{i}\T\bolbeta}}{1+e^{\mbX_{i}\T\bolbeta}}\mbX_{i}\right\} |\ge\epsilon)\\
		& \le & \Pr(|\dfrac{1}{n}\sum_{i=1}^{n}Y_{i}\mbX_{i}-E(Y_{i}\mbX_{i})|\ge\epsilon/2)+\Pr(|\dfrac{1}{n}\sum_{i=1}^{n}\dfrac{e^{\mbX_{i}\T\bolbeta}}{1+e^{\mbX_{i}\T\bolbeta}}\mbX_{i}-\E\dfrac{e^{\mbX_{i}\T\bolbeta}}{1+e^{\mbX_{i}\T\bolbeta}}|\ge\epsilon/2)\\
		& \le & Cp\exp(-Cn\epsilon^{2})
	\end{eqnarray*}
	where we use the fact that, since $\mbX_{i}$ is sub-Gaussian, and
	$Y_{i}$ and $\dfrac{e^{\mbX_{i}\T\bolbeta}}{1+e^{\mbX_{i}\T\bolbeta}}$
	are bounded, we have $Y_{i}\mbX_{i}$ and $\dfrac{e^{\mbX_{i}\T\bolbeta}}{1+e^{\mbX_{i}\T\bolbeta}}\mbX_{i}$
	are also sub-Gaussian. 
	
	Since $\mbU=\bolbeta\bolbeta\T$ and $\hatmbU=\hatbolbeta\hatbolbeta\T$
	where $\hatbolbeta$ is the solution to the $\ell_{1}$-penalized
	maximum likelihood estimator \eqref{betaGLM_l1}. 
	\begin{eqnarray*}
		\Vert\hatmbU-\mbU\Vert_{\max} & = & \Vert(\hatbolbeta-\bolbeta)(\hatbolbeta-\bolbeta)\T\Vert_{\max}+2\Vert(\hatbolbeta-\bolbeta)\bolbeta\T\Vert_{\max}\\
		& \leq & \Vert\hatbolbeta-\bolbeta\Vert_{2}^{2}+2\Vert\hatbolbeta-\bolbeta\Vert_{\infty}\Vert\bolbeta\Vert_{\infty}\\
		& \leq & \Vert\hatbolbeta-\bolbeta\Vert_{2}^{2}+2\Vert\hatbolbeta-\bolbeta\Vert_{2}\Vert\bolbeta\Vert_{2}\\
		& = & \Vert\hatbolbeta-\bolbeta\Vert_{2}^{2}+2\sqrt{\nu}\Vert\hatbolbeta-\bolbeta\Vert_{2},
	\end{eqnarray*}
	where recall that $\nu=\Vert\mbU\Vert_{op}=\bolbeta\T\bolbeta$.\end{proof}

\section{Proofs for Theorem \ref{thm:cox}}

Analogous to Lemma \ref{lem:glm} for the GLM, we have the following
Lemma \ref{lem:cox} for the Cox model. Then the proof of Theorem
\ref{thm:cox} follows the same argument as in the proof for Theorem
\ref{thm:glm}.

\begin{lemma}\label{lem:cox} There
	exists some constant $C_{\rho}$ and $C_{\lambda}$ such that if $s\sqrt{\dfrac{\log p}{n}}<C_{\rho}$
	and $\lambda_{n}=C_{\lambda}\epsilon$, with a probability greater
	than $1-Cp\exp(-Cn\epsilon^{2})$, we have 
	\begin{equation}
		\Vert\hatbolbeta-\bolbeta\Vert_{2}^{2}\le Cs\epsilon^{2},\quad\Vert\hatmbU-\mbU\Vert_{\max}\leq C(\sqrt{s\epsilon^{2}}+s\epsilon^{2}),
	\end{equation}
	where $s$ is the number of important predictors. \end{lemma} 

\begin{proof} By Theorem 3.2. in \citet{huang2013oracle}, we have
	the following results for Lasso penalized Cox model under the uniformly bounded covariates and restricted eigenvalue lower bounds conditions. Replacing $\varepsilon$ in Theorem 3.2 of \citet{huang2013oracle}
	by $2p\exp(-cn\epsilon^{2})$, we have the following results. 
	
	Let $\lambda_{n}=C\epsilon$ for some constant that does not depend
	on $n$, $p$, or $s$. For any constant $C_{F}>0$, there exists
	some constant $C$ such that, 
	\[
	\Vert\hatbolbeta-\bolbeta\Vert_{2}\leq C\sqrt{s}\epsilon/C_{F},
	\]
	with probability at least $\Prob(\mathcal{A})-2p\exp(-Cn\epsilon^{2})$.
	The event $\mathcal{A}$ is the event that the compatibility and cone
	invertibility factors are bounded below by the constants $C_{F}$.
	Then by Theorem 4.1 of \citet{huang2013oracle}, we can have event
	$\mathcal{A}$ by setting $C_{F}=\rho^{\ast}/2$. More specifically,
	replacing $\varepsilon$ in Theorem 4.1 of \citet{huang2013oracle}
	by $p\exp(-cn\epsilon^{2})$, we have the following results: event
	$\mathcal{A}$ holds with probability at least $1-Cp\exp(-Cn\epsilon^{2})$
	if $\rho^{\ast}$ is a constant greater than 0 and $s\sqrt{\dfrac{\log p}{n}}<C_{\rho}$
	for some constant $C_{\rho}$. Finally, $\rho^{\ast}$ is defined
	as the smallest eigenvalue of the population matrix $\bolSigma(t,M)$
	for some positive constants $t$ and $M$. The accurate defintion
	of $\bolSigma(t,M)$ is given as follows.
	
	Following \citet{andersen1982cox}, let $\mbN^{(n)}(t)=(N_{1}(t),\dots,N_{n}(t)),$
	$t\geq0$, be the $n$-dimensional counting process where $N_{i}(t)$
	counts the number of observed events for the $i$-th individual and
	has predictable compensator $\bolLambda^{(n)}=(\Lambda_{1},\dots,\Lambda_{n})$
	defined as $d\Lambda_{i}(t)=Y_{i}(t)\exp(\bolbeta\T\mbX_{i})d\Lambda_{0}(t)$,
	where $\Lambda_{0}(t)$ is the unknown baseline cumulative hazard
	function and $Y_{i}(t)\in\{0,1\}$ is a predictable at risk indicator
	process. The matrix $\bolSigma(t,M)$ for some constants $t$ and
	$M$ is essentially a truncated approximation to the Hessian $\nabla^{2}\ell_{n}(\bolbeta)$
	up to time $t$, defined as,
	\begin{eqnarray*}
		\bolSigma(t,M)=\E\int_{0}^{t}\mbG_{n}(s;\ M)d\Lambda_{0}(s) &  & \mbG_{n}(t,M)=\dfrac{1}{n}\sum_{i=1}^{n}\{\mbX_{i}-\bolmu(t,M)\}\{\mbX_{i}-\bolmu(t,M)\}\T U_{i}(t)\\
		U_{i}(t)=Y_{i}(t)\min\{M,\exp(\bolbeta\T\mbX_{i})\} &  & \bolmu(t,M)=\E\{\mbX U(t)\}/\E\{U_{i}(t)\}.
	\end{eqnarray*}
	As noted in \citet{huang2013oracle}, since $\bolSigma(t,M)$ is a
	population matrix, it is reasonable to assume its smallest eigenvalue
	is a positive constants. Moreover, we only need this to be true for
	some $t$ and $M$ in the condition.
	
	Then follow the proof of Lemma \ref{lem:glm}, we have 
	\begin{eqnarray*}
		\Vert\hatmbU-\mbU\Vert_{\max} & \leq & \Vert\hatbolbeta-\bolbeta\Vert_{2}^{2}+2\sqrt{\nu}\Vert\hatbolbeta-\bolbeta\Vert_{2},
	\end{eqnarray*}
	which completes the proof.\end{proof}

\newpage

\baselineskip=5pt
\bibliographystyle{agsm}
\bibliography{nee_ref}
\end{document}